\documentclass[12pt,numbysec]{iopart}
\usepackage{amssymb,bm}
\usepackage{iopams}  
\usepackage{graphicx}
\usepackage{color} 
\usepackage{pstricks,pst-coil,xcolor}
\newcommand{\text}[1]{{\sf #1}}
\newcommand{\eqref}[1]{(\ref{#1})}

\renewcommand{\Re}{\rm{Re}} \renewcommand{\Im}{\rm{Im}}
\newcommand{\ub}{\mathring{u}}
\newcommand{\cb}{\mathring{c}}
\newcommand{\tb}{\mathring{\tau}}
\newcommand{\tcinf}{T_{\mathrm{c},\infty}}
\newcommand{\ket}[1]{|#1\rangle}
\newcommand{\normf}{\lambda}
\newcommand{\eigenf}{\upsilon}
\newcommand{\res}{\mathop{\mbox{Res}}\limits}
 \eqnobysec

\begin{document}

\title[]{Critical  Casimir effect in films for generic non-symmetry-breaking boundary conditions}
\author{H.~W. Diehl and Felix M. Schmidt}

\address{Fakult\"at f\"ur Physik, Universit{\"a}t Duisburg-Essen, 47048 Duisburg,
 Germany}

\begin{abstract}
 Systems described by an $O(n)$ symmetrical $\phi^4$ Hamiltonian are considered in a $d$-dimensional film geometry at their bulk critical points. 
A detailed renormalization-group (RG) study of the critical Casimir forces  induced between the film's boundary planes $\mathfrak{B}_j,\,j=1,2$, 
by  thermal fluctuations is presented for the case where the $O(n)$ symmetry remains unbroken by the surfaces. 
The boundary planes are assumed to cause short-ranged disturbances of the interactions that can be modelled by standard surface contributions $\propto \bm{\phi}^2$
 corresponding to subcritical or critical enhancement of the surface interactions. This translates into mesoscopic boundary conditions of the generic
 symmetry-preserving Robin type $\partial_n\bm{\phi}=\mathring{c}_j\bm{\phi}$. RG-improved perturbation theory and Abel-Plana
  techniques are used to compute the $L$-dependent   part $f_{\mathrm{res}}$ of the reduced excess free energy per film area $A\to\infty $ to two-loop order.
 When $d<4$, it takes the scaling form $f_{\mathrm{res}}\approx D(c_1L^{\Phi/\nu},c_2L^{\Phi/\nu})/L^{d-1}$ as $L\to\infty$, where $c_i$ are scaling fields
 associated with the surface-enhancement variables $\mathring{c}_i$, while $\Phi$ is a standard surface crossover exponent. The scaling function
 $D(\mathsf{c}_1,\mathsf{c}_2)$ and its analogue $\mathcal{D}(\mathsf{c}_1,\mathsf{c}_2)$ for the Casimir force $\mathcal{F}=-\partial f_{\mathrm{res}}/\partial L$ are determined via expansion in $\epsilon=4-d$ and extrapolated to $d=3$ dimensions. In the special case $\mathsf{c}_1=\mathsf{c}_2=0$,  the expansion becomes fractional.
 Consistency with the known fractional expansions of $D(0,0)$ and $\mathcal{D}(0,0)$ to order $\epsilon^{3/2}$ is achieved by appropriate
 reorganisation of RG-improved perturbation theory. For appropriate choices of $c_1$ and $c_2$, the Casimir forces can have either sign. Furthermore, crossovers from attraction to repulsion and vice versa may occur as $L$ increases.
\end{abstract}


\submitto{\NJP}

\contentsline {section}{\numberline {1}Introduction}{2}
\contentsline {section}{\numberline {2}Background}{5}
\contentsline {subsection}{\numberline {2.1}Model and boundary conditions}{5}
\contentsline {subsection}{\numberline {2.2}Free propagator}{6}
\contentsline {subsection}{\numberline {2.3}Many-point cumulants, their renormalization and RG equations}{9}
\contentsline {section}{\numberline {3}Free energy and Casimir force}{12}
\contentsline {subsection}{\numberline {3.1}Definitions}{12}
\contentsline {subsection}{\numberline {3.2}Renormalization of the residual free energy and Casimir force}{12}
\contentsline {subsection}{\numberline {3.3}Perturbation theory}{15}
\contentsline {subsection}{\numberline {3.4}Renormalized residual free-energy density and scaling functions}{17}
\contentsline {subsection}{\numberline {3.5}Modified RG-improved perturbation theory}{18}
\contentsline {subsubsection}{\numberline {3.5.1}Formulation and results}{18}
\contentsline {subsubsection}{\numberline {3.5.2}Consistency with fractional $\epsilon $ expansion for $\mathsf {c}_1=\mathsf {c}_2=0$ and Ginzburg-Levanyuk criterion}{25}
\contentsline {section}{\numberline {4}Discussion of results and extrapolation to $d=3$ dimensions}{26}
\contentsline {section}{\numberline {5}Summary and Conclusions}{32}
\contentsline {section}{\numberline {A}Calculation of the one-loop free-energy density $f^{[1]}$ }{35}
\contentsline {subsection}{\numberline {A.1}Calculation based on the Abel-Plana summation formula}{35}
\contentsline {subsection}{\numberline {A.2}Calculation based on the free propagator}{40}
\contentsline {section}{\numberline {B}Computation of two-loop free-energy terms}{41}
\contentsline {section}{\numberline {C}Calculation of the effective two-point vertex $\gamma ^{(2)}$}{42}
\contentsline {section}{\numberline {D}Behaviour of the shift $\delta \mathaccent "7017\relax {\tau }_L(\mathcal {C}_1,\mathcal {C}_2)$ for small $\mathcal {C}_j$}{43}
\contentsline {section}{\numberline {E}Calculation of $f_\psi -f$ to two-loop order}{43}

\section{Introduction}
When media exhibiting  low-energy fluctuations are confined by walls or interfaces, or macroscopic objects are immersed in them, their fluctuation spectrum changes \cite{BMM01,Kre94,KG99}. This may induce long-range effective forces between such objects and boundaries. A much studied example of such fluctuation-induced forces are the Casimir forces between a pair of parallel  and grounded metallic plates produced by their influence on the quantum vacuum fluctuations of the electromagnetic field \cite{Cas48}. These quantum electrodynamics (QED) Casimir forces are known to be very weak unless the separation of the bodies between which they act becomes very small. The latter condition is  met in micro-electromechanical devices (MEMS). As has recently become clear, Casimir forces of this kind must be taken into account in the design of such systems. Since they tend to be attractive for simple geometries, they may impair the functioning of MEMS, causing stiction \cite{SWM95, BR01, BR01b, CAKBC01}. Owing to the interest in --- and technological importance of --- such small-scale systems, this insight has triggered considerable new activity in the study of QED Casimir forces. Crucial issues are the knowledge and control of their strength, sign, and geometry dependence \cite{Emi10}.

Another example of fluctuation-induced forces that has attracted a great deal of attention recently are the so-called thermodynamic Casimir force caused by thermal fluctuations near critical points \cite{Kre94,Kre99,BDT00}. Predicted decades ago \cite{FdG78}, they were verified experimentally first in an indirect way through the thinning of ${}^4$He wetting layers near the lambda transition \cite{GC99} and their effects on wetting films of binary liquid mixtures \cite{FYP05,RBM07}. Subsequently they were measured directly in binary fluid mixtures near their critical points of demixing \cite{HHGDB08,Gam09}. 

In this paper we shall be concerned with thermodynamic Casimir forces in systems with a film geometry bounded by two planar free surfaces at a finite distance $L$.  We will consider systems that can be modelled by an $n$-component $|\bm{\phi}|^4$ Hamiltonian on sufficiently long length scales and study the Casimir forces directly at the bulk critical point. The internal $O(n)$ symmetry (or $Z_2$ symmetry if $n=1$) is presumed to be neither explicitly broken by surface contributions to the Hamiltonian nor spontaneously for all temperatures $T \ge T_{\mathrm{c},\infty}$, the bulk critical temperature. In other words, no ordered phase can occur for finite $L$ when $T \ge T_{\mathrm{c},\infty}$. Choosing a generic kind of such non-symmetry-breaking conditions, we will investigate the critical Casimir forces and demonstrate two important features of them. First, they can have either sign, depending on properties of the two surfaces --- an expected result since attractive and repulsive critical Casimir forces were found a long time ago for certain boundary conditions that ought to apply on long length scales  \cite{NI85,KD91,KD92a}.  Second, for appropriate surface properties, crossovers from attraction to repulsion of the Casimir forces and vice versa can occur as a function of  film thickness $L$. A brief account of parts of our work was given in \cite{SD08}. The purpose of the present paper is to provide details of the calculations and further results.

One motivation for our work is the obvious potential for cross-fertilisation between the fields of thermodynamic  and QED Casimir effects. Since both types of effects share a number of characteristic features, mutual benefits may be expected. On the other hand, one must be aware of certain essential differences. Common to both types is that they exhibit universal properties and depend on gross properties of the fluctuating media and confining objects, as well as on their shapes and geometry. Two important differences are: 
\begin{enumerate}
\item When studying QED Casimir forces, one frequently gets away with the analysis of effective \emph{Gaussian} theories in which the coupling between the electromagnetic field and matter is accounted for by proper choices of boundary conditions. This usually holds even in the case of polarisable and magnetisable media where material properties also enter via dielectric and permeability functions. By contrast, adequate treatment of critical Casimir forces normally involve the study of non-Gaussian theories such as $\phi^4$ theories or corresponding lattice models (e.g., Ising models) in bounded geometries. 
\item In studies of QED Casimir forces such as Casimir's original work \cite{Cas48}, the electromagnetic fields are taken to have zero averages. Hence the Casimir forces are  entirely due to fluctuations --- if there were no fluctuations, there would be no Casimir force. However, in the case of thermodynamic Casimir forces, the order-parameter densities $\bm{\phi}(\bm{x})$ have nonzero averages in ordered phases, where the order may either result from spontaneous symmetry breaking or else be imposed by symmetry-breaking bulk or boundary fields. If the medium undergoes a phase transition from a disordered high-temperature to an ordered low-temperature phase, then the order-parameter profile $\langle\bm{\phi}(\bm{x})\rangle$ does not vanish in the latter. Such a nontrivial profile contributes to the size-dependent part of the free energy  and hence, to the Casimir force. Consequently, a Casimir force will be obtained already at the level of Landau theory --- i.e., in the absence of fluctuations --- whenever it yields non-vanishing order-parameter profiles. Beyond Landau theory, the thermodynamic Casimir force will therefore consist of a part coming from a non-fluctuating background and a superimposed fluctuation-induced  contribution.
\end{enumerate}

The last statement applies in particular to confined binary fluid mixtures \cite{Gam09,Die86a,Die97}. These are known to generically involve symmetry-breaking boundary fields. Therefore, the thermodynamic Casimir forces they yield will normally have contributions from non-fluctuating backgrounds \emph{on either side of the order-disorder transitions}. Since even at the bulk critical temperature, $\langle\bm{\phi}(\bm{x})\rangle$  is not expected to vanish, this holds there as well.

It is well known that the boundary fields needed to describe binary fluid mixtures in contact with walls can have either sign, depending on which one of the two components gets preferentially adsorbed locally at the wall \cite{Kre99,FdG78,Die86a,Die97,Die94a}. Furthermore, there is clear evidence based on theoretical and experimental work as well as on Monte Carlo simulations \cite{Kre99,FYP05, RBM07,HHGDB08,Kre97,Huc07,VGMD07,VGMD09,GMHNB09,Has10b} that the critical Casimir force in binary fluid mixtures confined between two planar walls can be attractive or repulsive. Corresponding crossovers must occur and were investigated at the level of mean-field theory \cite{MMD10}.

Casimir forces in binary fluid mixtures are particularly well suited for direct experimental measurements. However, from our above remarks it is clear that their analogy with QED Casimir forces is limited: owing to the generic presence of a non-vanishing order parameter profile at, above and below the critical temperature, only a part of them is fluctuation induced. Since we focus here on the case in which the symmetry $\bm{\phi}\to-\bm{\phi}$ remains unbroken at $T_{\mathrm{c},\infty}$, we are dealing with critical Casimir forces that are entirely fluctuation induced and hence are more akin to QED Casimir forces, albeit due to thermal, rather than quantum, fluctuations.

The rest of this paper is organised as follows. In section~\ref{sec:Ham}  the model is  specified and  the boundary conditions are recapitulated. Next, the free propagator is determined in section~\ref{sec:freeprop} for general values of the surface interaction constants, its eigenfunction representation is given and the transcendental equations derived whose solutions yield its eigenvalues. In section~\ref{sec:cumren},  many-point cumulant functions are introduced,  their renormalization recapitulated and  their RG equations presented. These considerations are extended to the free energy and Casimir force in section~\ref{sec:freeen}. In the remainder of this section details of our approach  are explained and our analytical results for the scaling functions of the $L$-dependent part of the free-energy density and the Casimir force are given. In section~\ref{sec:consistency} it is shown how RG-improved perturbation theory can be reorganised to achieve consistency with the fractional series expansions in $\epsilon$ one encounters in the special case of two critically enhanced surface planes. Section~\ref{sec:results} discusses our results, deals with the issue of how to extrapolate them to $d=3$ dimensions and presents such extrapolation results for the scaling functions of the $L$-dependent part of the excess free-energy density and the Casimir force.
Section~\ref{sec:sumconcl} contains a brief summary and concluding remarks. Finally, there are 5 appendices describing  various calculational details. 

\section{Background}\label{sec:backg}

\subsection{Model and boundary conditions}\label{sec:Ham}

We consider a continuum model for a real-valued $n$-component  order-parameter field $\bm{\phi}(\bm{x})=(\phi_a(\bm{x}), a=1,\ldots,n)$ defined on the $d$-dimensional slab $\mathfrak{V}=\mathbb{R}^{d-1}\times[0,L]$. We write the position vector as $\bm{x}=(x_1,\ldots,x_d)=(\bm{y},z)$, where $\bm{y}=(y_1,\ldots,y_{d-1})$ denotes the $d-1$ Cartesian coordinates $y_j=x_j$ along the slab and $z=x_d$ the remaining one across it. The thickness of the slab, $L$, is taken to be finite. We choose periodic boundary conditions along all $y$-directions so that the boundary  $\partial\mathfrak{V}$ of $\mathfrak{V}$ consists of a pair of ($d-1$)-dimensional planes $\mathfrak{B}_1$ and $\mathfrak{B}_2$ at $z=0$ and $z=L$. Assuming the absence of long-range interactions, we can choose the Hamiltonian to have the local form
\begin{equation}
\mathcal{H}=\int_{\mathfrak{V}}\!\mathrm{d}^dx\;\mathcal{L}_{\mathrm{b}}(\bm{x})+\sum_{j=1}^2\int_{\mathfrak{B}_j}\!\mathrm{d}^{d-1}y\;\mathcal{L}_{j}(\bm{y})\;,
\end{equation}
where $\mathcal{L}_b$ and $\mathcal{L}_{j=1,2}$ are local bulk and surface densities depending on $\bm{\phi}(\bm{x})$ and its derivatives. 

In accordance with our considerations in the introduction we choose these densities to be $O(n)$ symmetric. For the bulk density we make the standard choice
\begin{equation}\label{eq:Lb}
\mathcal{L}_{\mathrm{b}}(\bm{x})=\frac{1}{2}[\nabla\bm{\phi}(\bm{x})]^2+\frac{\mathring\tau}{2}\bm{\phi}^2(\bm{x})+\frac{\mathring{u}}{4!}|\bm{\phi}^2(\bm{x})|^2
\end{equation}
where $(\nabla\bm{\phi})^2$ is the usual short-hand for $\sum_\alpha(\nabla\phi_\alpha)^2$. It has recently been emphasised that lattice systems lacking cubic symmetry, such as systems involving monoclinic or triclinic lattices, lead to $\phi^4$ bulk densities with a gradient-square term of the generalised form $\frac{1}{2}g^{ij}(\partial_i\bm{\phi})\cdot\partial_j\bm{\phi}$ where $\partial_i$ stands for the derivative $\partial/\partial x^i$ and  $g^{ij}$ is a non-diagonal, position-independent matrix \cite{Doh08}. We refrain from working with such generalised models because the matrix $g^{ij}$ is nothing but a constant metric \cite{DC09}. The geometric effects it has can be absorbed by a proper choice of variables which of course enters the way lattice quantities are related to those of the standard bulk $\phi^4$ model with action density~\eqref{eq:Lb}. For details the reader might want to consult reference~\cite[p.~14--17]{DC09}. 

The surface densities are given by
\begin{equation}
\mathcal{L}_{j}(\bm{r})=\frac{\mathring c_j}{2}\bm{\phi}^2(\bm{r}).
\end{equation}
The interaction constants $\cb_1$ and $\cb_2$ of the two boundary planes are allowed to differ. They are known to reflect the weakening or enhancement of the local pair interactions at $\mathfrak{B}_1$ and $\mathfrak{B}_2$, respectively \cite{Die86a,Die97}. In semi-infinite systems bounded by a plane $\mathfrak{B}_1$, a threshold value $\cb_{\mathrm{sp}}$ of the enhancement variable $\cb_1$  exists such that a phase with long-range surface order appears (in sufficiently high dimensions) in a temperature regime above $\tcinf$ when the enhancement $-\delta\cb_1\equiv\cb_{\mathrm{sp}}-\cb_1$ is positive (is ``supercritical'').  The transitions that occur at $\tcinf$ in such  semi-infinite systems are called ordinary, special and extraordinary, depending on whether the enhancement $\delta\cb_1$ is subcritical ($\delta\cb_1>0$), critical ($\delta\cb_1=0$) or supercritical ($\delta\cb_1<0$). Analogous statements hold for semi-infinite systems bounded by the plane $\mathfrak{B}_2$ with surface enhancement $\cb_2$.

In order to rule out spontaneous symmetry breakdown at $T=\tcinf$ for large $L$, we require that both enhancement variables $\delta\cb_j\equiv \cb_j-\cb_{\mathrm{sp}}$ satisfy the condition $\delta\cb_j\ge 0$, i.e.\ are  non-supercritical. 

From the boundary contribution to the classical equation of motion we get the boundary conditions of Landau theory 
\begin{equation}\label{eq:RobBC}
\partial_n\bm{\phi}(\bm{x})=\mathring{c}_j\bm{\phi}(\bm{x})\quad\mbox{for}\;\bm{x}\in \mathfrak{B}_j,
\end{equation}
where $\partial_n$ is the inner normal derivative on $\partial\mathfrak{V}=\mathfrak{B}_1\cup\mathfrak{B}_2$. Beyond Landau theory these boundary conditions still hold in an operator sense (inside of averages) \cite{Die86a,Die97,DD81a,DD81b,DD83a}.
 
\subsection{Free propagator}\label{sec:freeprop}

A quantity of central importance for the renormalization-group (RG) improved perturbation theory we are going to use below is the free propagator $G_L$ at $\tb=0$. It is given by the operator inverse 
$G_L=(-\triangle)^{-1}$ subject to the boundary conditions~\eqref{eq:RobBC}, where $\triangle= \nabla^2 $ is the Laplacian. Let $\ket{m}$ be the complete set of eigenfunctions of the operator $-\partial_z^2\equiv-\partial^2/\partial z^2$ on the interval $[0,L]$ satisfying the boundary conditions
\begin{eqnarray}
(\partial _z-\cb_1)|m\rangle\big|_{z=0}&=&0,\label{eq:BCef0}\\
(\partial _z+\cb_2)|m\rangle\big|_{z=L}&=&0.\label{eq:BCefL}
\end{eqnarray}
That is, we have 
\begin{equation}\label{eq:eigfunceq}
-\partial_z^2\ket{m}=k_m^2\ket{m},
\end{equation}
and the eigenstates fulfil the orthogonality and completeness relations
\begin{equation}\label{eq:completeness}
\langle m\ket{m'}=\delta_{mm'}
\end{equation}
and
\begin{equation}
\sum_{m=1}^\infty\langle z\ket{m}\langle m\ket{z'}=\delta(z-z'),
\end{equation}
where the label $m=1,2,\ldots,\infty$ enumerates the eigenvalues $k_m^2$ by size, starting with the smallest one. 

To exploit the translation invariance of the system along the direction parallel to the boundary planes $\mathfrak{B}_j$, we Fourier transform with respect to the difference of $\bm{y}$ coordinates, writing
\begin{equation}\label{eq:FTGLdef}
G_L(\bm{x},\bm{x}')=\int_{\bm{p}}^{(d-1)}\hat{G}_L(\bm{p};z,z')\,e^{i\bm{p}\cdot (\bm{y}'-\bm{y})},
\end{equation} 
where we have introduced the notation
\begin{equation}
\int_{\bm{p}}^{(d-1)}\equiv\int_{\mathbb{R}^{d-1}}\frac{\rmd^{d-1}p}{(2\pi)^{d-1}}.
\end{equation}
We thus arrive at the spectral representation 
\begin{equation}\label{eq:GLfreespecrep}
\hat{G}_L(\bm{p};z,z')=\sum_{m=1}^\infty\frac{\langle z\ket{m}\langle m\ket{z'}}{p^2+k_m^2}\,.
\end{equation}

Since our assumption that both enhancement variables $\cb_j$ are non-supercritical implies that the associated renormalized variables $c_j$  (whose definition will be recalled in the following section~\ref{sec:cumren}) are nonnegative, we shall need the eigenvalues $k_m^2$ and eigenfunctions $\ket{m}$ only for $\cb_j\ge 0$. In this case, there are infinitely many eigenvalues $k_m^2$ for finite $L$, all of which are nonnegative and non-degenerate (see reference~\cite[Appendix A]{RS02} and below). A zero eigenvalue $k_1^2=0$ occurs for arbitrary $L\in(0,\infty)$ only when $\cb_1=\cb_2=0$.

Let us briefly summarise the relevant properties of the eigensystem $\{\ket{m},k_m^2\}$ we shall need in our subsequent analysis. On dimensional grounds the eigenfunctions must have the form
\begin{equation}\label{eq:braketef}
\langle z\ket{m}=L^{-1/2}\,\eigenf_m(z/L|\cb_1L,\cb_2L).
\end{equation}
It is convenient to introduce the dimensionless variables
\begin{equation}\label{eq:dimvardef}
\zeta\equiv z/L,\quad\kappa_m\equiv Lk_m,\quad \mathcal{C}_j\equiv \cb_jL.
\end{equation}
Setting $L=1$ in equation~\eqref{eq:braketef} one concludes that the functions $\eigenf_m$ can be written as
\begin{equation}\label{eq:varphimform}
\eigenf_m(\zeta|\mathcal{C}_1,\mathcal{C}_2)=\normf^{1/2}_m\cos(\kappa_m\zeta-\vartheta_m).
\end{equation}
The analogue of the boundary condition~\eqref{eq:BCef0}
\begin{equation}\label{eq:BCvarphim0}
\eigenf_m'(0|\mathcal{C}_1,\mathcal{C}_2)=\mathcal{C}_1\eigenf_m(0|\mathcal{C}_1,\mathcal{C}_2)
\end{equation} 
implies that the phase shift $\vartheta_m$ is given by
\begin{equation}\label{eq:varthetam}
\tan\vartheta_m=\mathcal{C}_1/\kappa_m=\cb_1/k_m.
\end{equation}
We can choose it such that $0\le\vartheta_m\le\pi/2$, which leads to
\begin{equation}
\sin\vartheta_m=\frac{1}{\sqrt{1+(\kappa_m/\cb_1)^2}},\qquad \cos\vartheta_m=\frac{\kappa_m/\cb_1}{\sqrt{1+(\kappa_m/\cb_1)^2}}.
\end{equation}
The limiting values $\vartheta_m=0$ and $\vartheta_m=\pi/2$ are obtained when $\cb_1\to0$ and $\cb_1\to\infty$, respectively.

Using these results along with equation~\eqref{eq:varphimform}, the normalisation constant $\normf_m$ can be computed in a straightforward fashion. One obtains
\begin{equation}\label{eq:normf}
 \normf_m=2\left[1+\frac{\mathcal{C}_1}{\mathcal{C}^2_1+\kappa_m^2}+\frac{\mathcal{C}_2}{\mathcal{C}^2_2+\kappa_m^2}\right]^{-1}.
\end{equation}
To determine the spectrum $\{\kappa^2_m\}$, we use the analogue of the boundary condition~\eqref{eq:BCefL},
\begin{equation}\label{eq:BCvarphim1}
\eigenf_m'(1|\mathcal{C}_1,\mathcal{C}_2)=-\mathcal{C}_2\,\eigenf_m(1|\mathcal{C}_1,\mathcal{C}_2).
\end{equation} 
This yields the transcendental equation 
\begin{equation}\label{eq:transceq}
R_{\mathcal{C}_1,\mathcal{C}_2}(\kappa_m)=0
\end{equation}
for $\kappa_m$,
with
\begin{equation}\label{eq:Rdef}
R_{\mathcal{C}_1,\mathcal{C}_2}(\kappa)\equiv \big(\mathcal{C}_1+\mathcal{C}_2\big)\kappa\cos(\kappa)+\big(\mathcal{C}_1\mathcal{C}_2-\kappa^2\big)\sin\kappa.
\end{equation}

Equations~\eqref{eq:BCef0}--\eqref{eq:eigfunceq} specify a regular Sturm-Liouville problem for which the following mathematical properties are known  (see e.g.\ references~\cite{CH68}, \cite[Kap.~IX]{Jae01}, \cite[Kap.~IV.3]{Pey70b} and \cite[chapters 8, 9]{CC97}):  (i) The  eigenvalues $\kappa_m^2(\mathcal{C}_1,\mathcal{C}_2)$ are real, non-degenerate, countable and accumulate only at $\infty$. (ii) They can be ordered such that $\kappa_m^2<\kappa_{m'}^2$ for $m<m'$. There is a smallest eigenvalue $\kappa_1^2$ but no largest one, i.e. $\kappa_m\to \infty$ as $m\to\infty$.  (iii) The eigenfunctions $\upsilon_m$ corresponding to different eigenvalues are orthogonal with respect to the standard $L_2$ scalar product in $L_2(0,1)$.\footnote{In the general case of Sturm-Liouville problems, the scalar product involves a weight function \cite{Jae01}. This is unity in our case.} (iv) The normalised eigenfunctions are a complete orthonormal set (basis) in the Hilbert space  $L_2(0,1)$. (v) When $\mathcal{C}_1$ and $\mathcal{C}_2>0$, we can use the Robin boundary conditions~\eqref{eq:BCef0} and \eqref{eq:BCefL} in conjunction with the fact that the eigenfunctions vanish only for Dirichlet boundary conditions at the boundary planes $\mathfrak{B}_j$  to conclude that  $\upsilon(\rmd\upsilon/\rmd\zeta)|_{\zeta=0}^{\zeta=1}=-\mathcal{C}_2\upsilon^2|_{\zeta=1}-\mathcal{C}_1\upsilon^2|_{\zeta=0}<0$. Fulfilment of this condition guarantees (by theorem 7 of reference~\cite[p.~234]{Pey70b}) that $\kappa_1^2$ and hence all $\kappa_m^2$ are strictly positive.

The non-degeneracy of the eigenvalues $\kappa_m^2$ can be verified explicitly by showing that the positive zeros $\kappa_m$ of $R_{\mathcal{C}_1,\mathcal{C}_2}(\kappa)$  are simple. To this end, one can compute the derivative $R'_{\mathcal{C}_1,\mathcal{C}_2}(\kappa)\equiv\partial_\kappa R_{\mathcal{C}_1,\mathcal{C}_2}(\kappa)$ at $\kappa=\kappa_m$ and express the trigonometric functions $\cos\kappa_m$ and $\sin\kappa_m$  in terms of $\kappa_m$, $\mathcal{C}_1$ and $\mathcal{C}_2$ using equations~\eqref{eq:transceq} and \eqref{eq:varthetam} along with the analogue of the latter implied by the boundary condition at $\zeta=z/L=1$. This gives
\begin{equation}\label{eq:Rkappaprime}\fl
 R'_{\mathcal{C}_1,\mathcal{C}_2}(\kappa)=(-1)^m\,\frac{\kappa_m^4+\kappa_m^2(\mathcal{C}_1+\mathcal{C}_2+\mathcal{C}_1^2+\mathcal{C}_2^2)+\mathcal{C}_1\mathcal{C}_2(\mathcal{C}_1+\mathcal{C}_2+\mathcal{C}_1\mathcal{C}_2)}{\sqrt{(\kappa_m^2+\mathcal{C}_1^2)(\kappa_m^2+\mathcal{C}_2^2)}},
\end{equation}
which is nonzero for all $\mathcal{C}_1\ge0$, $\mathcal{C}_2\ge0$ and $\kappa_m>0$.

For the special choices $(\mathcal{C}_1,\mathcal{C}_2)=(0,0),(0,\infty),(\infty,0)$ and $(\infty,\infty)$, equation~\eqref{eq:transceq} can be solved explicitly for the $\kappa_m^2$. The respective results for Neumann-Neumann ($\mathrm{N},\mathrm{N}$) Neumann-Dirichlet ($\mathrm{N},\mathrm{D}$), Dirichlet-Neumann ($\mathrm{D},\mathrm{N}$), and Dirichlet-Dirichlet ($\mathrm{D},\mathrm{D}$) boundary conditions can be looked up in Appendix A of reference~\cite{KD92a}. However, for general values of the $\mathcal{C}_j$, the zeros $\kappa_m^2$ of the transcendental equation~\eqref{eq:transceq} cannot be determined in closed analytical form. This is a familiar difficulty, which makes the evaluation of the single and double mode sums $\sum_m$ one encounters in the calculation of one and two-loop Feynman graphs of the free energy a nontrivial problem. We shall turn to this issue in section~\ref{sec:freeen}.

Note also that in some of the above formulae we have tacitly assumed that $\mathcal{C}_1$ and $\mathcal{C}_2$ are both positive. However, these equations remain valid, firstly, for all $m$ when either one of the $\mathcal{C}_j$ vanishes while the other remains positive, and secondly, for all $m> 1$ when $\mathcal{C}_1=\mathcal{C}_2=0$, because the respective $\kappa_m$ then all approach positive values in the corresponding limits. Hence one can simply set $\mathcal{C}_1$ or $\mathcal{C}_2$ (or both) to zero in the respective equations.  The case of $m=1$ with $\mathcal{C}_1,\mathcal{C}_2\to 0$  is special in that $\kappa_1\to0$. It is an easy matter to check that the correct limiting eigenfunction $\upsilon(\zeta|0,0)=1$ results, independent of the order in which the limits $\mathcal{C}_1\to0$ and $\mathcal{C}_2\to0$ are taken.

It is also not difficult to determine how the eigenvalues $\kappa_m^2(\mathcal{C}_1,\mathcal{C}_2)$ approach the ($\mathrm{N},\mathrm{N}$)-values $\kappa_m^2(0,0)=(m-1)^2\pi^2$, $m=1,2,\ldots,\infty$, as both $\mathcal{C}_j$ approach zero. One can either determine the terms linear in $\mathcal{C}_j$ from equation~\eqref{eq:transceq} or else perform Rayleigh-Schr\"odinger perturbation theory in the potential $\mathcal{C}_1\,\delta(\zeta)+\mathcal{C}_2\,\delta(1-\zeta)$ to see that
\begin{equation}\label{eq:smallCkappa}
\kappa_m^2(\mathcal{C}_1,\mathcal{C}_2)\mathop{\approx}_{\mathcal{C}_1,\mathcal{C}_2\to 0}(m-1)^2\pi^2+(2-\delta_{m,1})\,(\mathcal{C}_1+\mathcal{C}_2),\;\;m=1,2,\ldots,\infty.
\end{equation}

\subsection{Many-point cumulants, their renormalization and RG equations}\label{sec:cumren}

We proceed by providing some of the necessary background on the renormalization of the theory defined above.\footnote{A more extensive discussion of this issue can be found in reference~\cite{Die86a}.} Let us begin by recalling the bulk and boundary counter-terms required to absorb the ultraviolet (UV) singularities of many-point cumulant functions.  To this end we consider the $(N{+}M_1{+}M_2)$-point cumulant functions 
\begin{equation}\eqalign{%
G^{(N;M_1,M_2)}_{\alpha_1,\dots,\beta_{M_1+M_2}}(\bm{x}_1,\ldots,\bm{y}_{M_1+M_2}) \cr =\left\langle\prod_{j=1}^N\phi_{\alpha_j}(\bm{x}_j)\prod_{k=1}^{M_1}\phi_{\beta_k}(\bm{y}_k,0)\prod_{l=M_1+1}^{M_1+M_2}\phi_{\beta_l}(\bm{y}_l,L)\right\rangle^{\mathrm{cum}}}
\end{equation}
involving $N$ fields $\phi_{\alpha_j}(\bm{x}_j)$ located at points in the interior of $\mathfrak{V}$ and $M_1{+}M_2$ fields $\phi_{\beta_k}$ on the boundary planes $\mathfrak{B}_1$  and $\mathfrak{B}_2$. As indicated, we take the first $M_1$ boundary points to lie on $\mathfrak{B}_1$ and the remaining ones on $\mathfrak{B}_2$.

Equations~\eqref{eq:GLfreespecrep}--\eqref{eq:varthetam} imply that the $\bm{p}$~transform of the free propagator for general nonnegative values of $\cb_1$ and $\cb_2$ can be written as
\begin{equation}\label{eq:GLc1c1}\fl
\eqalign{\hat{G}_L(\bm{p};z,z'|\tb,\cb_1,\cb_2)\cr =\frac{1}{L}\sum_m\frac{[k_m\cos(k_mz)+\cb_1\sin(k_mz)][k_m\cos(k_mz')+\cb_1\sin(k_mz')]}{(p^2+\tb+k_m^2)(\cb_1^2+k_m^2)}. }
\end{equation}
Its explicit form is known \cite{DC09,Sch08}. It is given by the expression into which the right-hand side of equation~(B5) of reference~\cite{DC09} turns upon replacement of its $\mathring{\kappa}_\omega$ by 
\begin{equation}\label{eq:varkappap}
\varkappa_{p}=\sqrt{p^2+\tb},
\end{equation}
i.e.\
\begin{eqnarray}\label{eq:GLPc1c2explform}
\fl \hat{G}_L(\bm{p};z,z'|\tb,\cb_1,\cb_2)\nonumber\\ \fl=\theta(z'-z)\frac{[\mathring{c}_1 \sinh(\varkappa_p z)+\varkappa_p\cosh
   (\varkappa_p z)  ]
   [\mathring{c}_2 \sinh [\varkappa_p(L-z'
   )]+\varkappa_p\cosh[\varkappa_p(L-z' )]}{\varkappa_p^2(\mathring{c}_1+\mathring{c}_2) \cosh
   (\varkappa_p L) +\varkappa_p (\varkappa_p^2+\mathring{c}_1
   \mathring{c}_2)\sinh(\varkappa_p L ) }
\nonumber\\ 
+(z\leftrightarrow z').
\end{eqnarray}

Both expressions \eqref{eq:GLc1c1} and \eqref{eq:GLPc1c2explform} are exceedingly difficult to work with. To understand the nature of the UV singularities of the theory, it is better to treat the boundary terms $\mathcal{L}_j$ as interactions and work with the free propagator for $\cb_1=\cb_2=0$. The latter is nothing but the free propagator $\hat{G}^{\mathrm{NN}}_L$ satisfying Neumann (N) boundary conditions on both boundary planes. It  can be written as \cite[section~IV.A]{Die86a}
\begin{equation}\label{eq:GNN}
\eqalign{\hat{G}^{\mathrm{NN}}_L(\bm{p};z,z')\equiv  \hat{G}_L(\bm{p};z,z'|\tb,0,0)\cr
 =\sum_{j=-\infty}^\infty[\hat{G}_{\mathrm{b}}(p,z-z'+j2L)+\hat{G}_{\mathrm{b}}(p,z+z'+j2L)],}
\end{equation}
where
\begin{equation}
\hat{G}_{\mathrm{b}}(\bm{p},z)=\frac{\exp(-\varkappa_p|z|)}{2\varkappa_p}
\end{equation}
is the free bulk propagator for $\tb=0$ in the $\bm{p}z$-representation. As expounded in reference~\cite{Die86a}, the $j=0$ contribution from the first term in square brackets in equation~\eqref{eq:GNN} is the origin of the familiar primitive bulk UV singularities. The $j=0$ and $j=-1$ contributions  from the second term in square brackets are singular at coinciding points on $\mathfrak{B}_1$ and $\mathfrak{B}_2$, respectively. The additional UV singularities they produce can be absorbed by  counter-terms with support on these boundary planes, i.e.\ the  surface counter-terms required for the corresponding semi-infinite systems \cite{Die86a,DD81b,DD83a}. All other contributions in the square brackets do not diverge at coinciding points and hence do not give rise to additional UV singularities. The upshot is that the usual bulk reparametrizations 
\begin{equation}\label{eq:bulkrep}
\eqalign{\bm{\phi}=Z_\phi^{1/2}\bm{\phi}_{\mathrm{R}}, \cr
\mathring\tau-\mathring\tau_{\mathrm{c},\infty}\equiv\delta\mathring\tau=Z_\tau\mu^2\tau,\cr
\mathring{u}N_d=\mu^{4-d}Z_u u,}
\end{equation}
of the field, temperature variable $\tb$ and interaction constant $\ub$ in conjunction with the reparametrizations 
\begin{equation}\label{eq:surfrep}
\eqalign{\mathring c_j-\mathring c_{\mathrm{sp}}\equiv\delta\mathring c_j=\mu Z_c\, c_j,\nonumber\\
\bm{\phi}^{\mathfrak{B}_j}=(Z_\phi Z_1)^{1/2}\bm{\phi}^{\mathfrak{B}_j}_{\mathrm{R}},}
\end{equation}
of the surface enhancement variables $\cb_j$ and the boundary operators $\bm{\phi}^{\mathfrak{B}_1}(\bm{y})\equiv \bm{\phi}(\bm{z},0)$ and $\bm{\phi}^{\mathfrak{B}_2}(\bm{y})\equiv \bm{\phi}(\bm{z},L)$ suffice to absorb the UV singularities of the functions $G^{(N;M_1,M_2)}$.

Following reference~\cite{GD08}, we choose the factor that is absorbed in the renormalized interaction constant as
\begin{equation}
 N_d=\frac{2\,\Gamma(3-d/2)}{(d-2)(4\pi)^{d/2}}=\frac{1}{16\pi^2}\left[1+
 \frac{1-\gamma_E+ \ln(4\pi)}{2}\epsilon+\Or(\epsilon^2)\right],
\end{equation}
where $\epsilon=4-d$ and $\gamma_E$ is the Euler-Mascheroni constant. This choice ensures that the bulk renormalization factors $Z_\phi$, $Z_\tau$ and $Z_u$ as well as the surface renormalization factors $Z_1$ and $Z_c$,  when determined by minimal subtraction of poles in $\epsilon$, up to second order in $u$ reduce to the two-loop results given in references~\cite{Die86a}, \cite{DD81a}, \cite{DD81b}, \cite{DD83a}, \cite{DD80} and used in Krech and Dietrich's work \cite{KD91,KD92a}. The quantity $\tb_{c,\infty}$ is the critical bulk value of $\tb$. Further, $\cb_{\mathrm{sp}}$ is the critical enhancement value associated with the special surface transition. In a theory whose UV singularities are regularised by means of a  large-momentum cut-off $\Lambda$, these quantities diverge   $\sim\Lambda^2$ and $\sim \Lambda$, respectively. In our calculations below we shall utilise dimensional regularisation.

Note also that the surface renormalization factors $Z_1$ and $Z_c$ depend exclusively on $u$ and $\epsilon$ but not on $c_j$ (nor on $L$) when fixed by the requirement that the UV poles in $\epsilon$ be minimally subtracted. In our calculations of the $L$-dependent part of the free energy at the bulk critical point and the critical Casimir force to be described in section~\ref{sec:freeen} we shall need $Z_c$ merely to first order in $u$. We therefore quote the result
\begin{equation}\label{eq:Z1orderu}
Z_c=1+\frac{n+2}{3\epsilon}\,u+\Or(u^2)
\end{equation}
for convenience.

Upon introducing the renormalized cumulants 
\begin{equation}\label{eq:GRsdef}
\eqalign{G_{\alpha_1,\ldots,\beta_{M_1+M_2};\mathrm{R}}^{(N;M_1,M_2)}(\bm{x}_1,\ldots;u,\tau,c_1,c_2,L,\mu)\cr =
Z_\phi^{-N/2}(Z_\phi Z_1)^{-(M_1+M_2)/2}\,G_{\alpha_1,\ldots,\beta_{M_1+M_2}}^{(N;M_1,M_2)}(\bm{x}_1,\ldots;\ub,\tb,\cb_1,\cb_2,L),}
\end{equation}
we can now derive RG equations for these functions by varying $\mu$ at fixed bare parameters $\ub$, $\tb$, $\cb_1$, $\cb_2$, and $L$. Let us define the exponent functions 
\begin{equation}
\eta_g\equiv\left.\mu\partial_\mu\right|_0Z_g,\qquad g=u,\tau,c_1,c_2,1,\phi,
\end{equation}
the beta function
\begin{equation}
\beta_u(u)\equiv \left.\mu\partial_\mu\right|_0 u=-[\epsilon+\eta_u(u)]u,
\end{equation}
(where $\partial_\mu|_0$ indicates a $\mu$-derivative at fixed bare parameters $\ub$, $\tb$, $\cb_1$, $\cb_2$, $L$)
and the operator 
\begin{equation}\label{eq:Dmu}
\mathcal{D}_\mu\equiv\mu\partial_\mu+\beta_u\partial_u-(2+\eta_\tau)\tau\partial_\tau-(1+\eta_c)(c_1\partial_{c_1}+c_2\partial_{c_2}).
\end{equation}
Then the RG equations can be written as
\begin{equation}\label{eq:RGEGs}
\left[\mathcal{D}_\mu+\frac{N}{2}\,\eta_\phi+\frac{M_1+M_2}{2}\,(\eta_\phi+\eta_1)\right]G_{\alpha_1,\ldots,\beta_{M_1+M_2};\mathrm{R}}^{(N;M_1,M_2)}=0.
\end{equation}
They are completely analogous to those for the renormalized cumulants of the respective semi-infinite geometries ($L=\infty$ with either $M_2= 0$ or $M_1=0$). We have given them here for general $\tau$, even though we will restrict ourselves to the critical case $\tau=0$ in our calculations in section~\ref{sec:freeen}. 

\section{Free energy and Casimir force}\label{sec:freeen}
\subsection{Definitions}\label{sec:freeenCFdef}
We next turn to the free energy and the Casimir force. Introducing the partition function $\mathcal{Z}$ and the total free energy $F$ via
\begin{equation}\label{eq:partfuncdef}
\mathcal{Z}\equiv\int\mathcal{D}[\bm{\phi}]\;\rme^{-\mathcal{H}[\bm{\phi}]}=\rme^{-F/k_{\mathrm{B}} T},
\end{equation}
we define the reduced free-energy density per hyper-surface area $A$,
\begin{equation}\label{eq:fdef}
f(L;\tb,\ub,\cb_1,\cb_2)\equiv \lim_{A\to \infty}\frac{F}{k_{\mathrm{B}}TA}.
\end{equation}
This quantity can be decomposed as 
\begin{equation}\label{eq:fdec}
f(L;\tb,\ub,\cb_1,\cb_2)=Lf_{\mathrm{b}}(\tb,\ub)+f_{\mathrm{s}}+f_{\mathrm{res}}(L;\tb,\ub,\cb_1,\cb_2),
\end{equation}
where $f_{\mathrm{b}}$ and $f_{\mathrm{s}}$ are the bulk free-energy density per $d$-dimensional volume $LA\to\infty$ and the excess  free-energy density per hyper-surface area $A\to\infty$, both of which are defined by appropriate thermodynamic limits (see e.g.\ references~\cite{Die86a} and \cite{Die82}). The former depends  only on the bulk interaction constants, the latter additionally on the boundary interaction constants. It is a sum of the reduced surface excess free-energy densities $f_{\mathrm{s},j}$ of the corresponding semi-infinite systems bounded by either $\mathfrak{B}_1$ or $\mathfrak{B}_2$:
\begin{equation}\label{eq:fsdec}
f_{\mathrm{s}}(\tb,\ub,\cb_1,\cb_2)=f_{\mathrm{s},1}(\tb,\ub,\cb_1)+f_{\mathrm{s},2}(\tb,\ub,\cb_2).
\end{equation}
The remaining term in the decomposition~\eqref{eq:fdec}, the reduced residual free-energy density $f_{\mathrm{res}}$, contains the full $L$-dependence of $f-Lf_{\mathrm{b}}$. The effective (``Casimir'') force per hyper-surface area to which it gives rise is
\begin{equation}\label{eq:CFdef}
\frac{\mathcal{F}(L;T,\ldots)}{k_{\mathrm{B}}T}=-\frac{\partial f_{\mathrm{res}}}{\partial L},
\end{equation}
where the ellipsis stands for all other (bulk and surface) variables on which it depends.\footnote{Since the right-hand side of equation~\eqref{eq:CFdef} depends on the variables $\tb$, $\ub$, $\cb_1$ and  $\cb_2$, so does its left-hand side.}

\subsection{Renormalization of the residual free energy and Casimir force}\label{eq:renfresCF}

Inspection of the perturbation series of the reduced free energy $F/k_{\mathrm{B}}T$ reveals that the bulk reparametrizations~\eqref{eq:bulkrep} and surface reparametrizations~\eqref{eq:surfrep} are not sufficient to absorb all UV singularities. This is because both the bulk free-energy density $f_{\mathrm{b}}$ as well as the surface free-energy densities $f_{\mathrm{s},j}$ have additional primitive UV singularities \cite{Die86a}. To cure these, additive bulk and surface counter-terms are required. Since these counter-terms must merely absorb primitive UV singularities of $f_{\mathrm{b}}$ and $f_{\mathrm{s},j}$, respectively, they can be chosen independent of $L$. One convenient way to fix them is by subtracting from $f_{\mathrm{b}}$ the Taylor expansion in $\delta\tb$, and from $f_{s,j}$ that in $\delta\tb$ and $\delta\cb_j$ about non-vanishing reference values to the appropriate orders (see e.g.\ \cite[section~III.C.12]{Die86a} or \cite[section~II.E]{GD08}). Owing to the $L$-independence of these additive counter-terms, their contributions cancel out in the difference of free-energy densities that $f_{\mathrm{res}}$ involves. Using the reparametrizations~\eqref{eq:bulkrep} and ~\eqref{eq:surfrep} therefore gives us the UV-finite renormalized residual free-energy density
\begin{equation}\label{eq:fresRdef}
f_{\mathrm{res},\mathrm{R}}(L;\tau,u,c_1,c_2,\mu)=f_{\mathrm{res}}(L;\tb,\ub,\cb_1,\cb_2).
\end{equation}
Hence this quantity satisfies a homogeneous RG equation, namely
\begin{equation}\label{eq:RGfres}
\mathcal{D}_\mu f_{\mathrm{res},\mathrm{R}}=0.
\end{equation}

The latter can be solved in a standard fashion by means of characteristics. Let us define the running variables $\bar{u}(\ell)$ and $\bar{c}_j(\ell)$ as solutions to the initial value problems
\begin{eqnarray}\label{eq:ubardef}
\ell \frac{\rmd}{\rmd \ell}\bar{u}(\ell)=\beta_u[\bar{u}(\ell)],\qquad &\bar{u}(1)=u,\\
\label{eq:cjbardef}
\ell \frac{\rmd}{\rmd \ell}\bar{c}_j(\ell)=-\big\{1+\eta_c[\bar{u}(\ell)]\big\}\bar{c}_j(\ell),\qquad&\bar{c}_j(1)=c_j.
\end{eqnarray}
Let  $u^*$ be the infrared-stable zero of the beta function $\beta_u$ for $d=4-\epsilon<4$. We shall need its $\epsilon$ expansion
\begin{equation}\label{eq:ustarepsexp}
u^*=\frac{3\epsilon}{n+8}+\frac{9(3n+14)\epsilon^2}{(n+8)^3}+\Or(\epsilon^3)
\end{equation}
below only to $\Or(\epsilon)$. Solving the flow equations~\eqref{eq:cjbardef} for $\bar{c}_j$ yields 
\begin{equation}
\eqalign{\bar{c}_j(\ell)=E_c[\bar{u}(\ell),u]\,\ell^{-y_c}\,c_j,\quad y_c\equiv\Phi/\nu,\cr
 E_c(\bar{u},u)=\exp\left\{-\int_u^{\bar{u}}\rmd u'[1+\eta_c(u')-y_c]\right\}},
\end{equation}
where $\Phi$ is the surface crossover exponent whose $\epsilon$ expansion is known to $\Or(\epsilon^2)$ \cite{DD81b,DD83a}. Since we shall need the $\epsilon$ expansion of the RG eigenexponent $y_c$ to first order in $\epsilon$, we recall the result
\begin{equation}\label{eq:Phinuepsexp}
y_c=1-\frac{n+2}{n+8}\,\epsilon+\Or(\epsilon^2).
\end{equation} 

In the large-length-scale limit $\ell\to 0$, the running variables $\bar{u}$ and $\bar{c}_j$ behave as $\bar{u}(\ell)\approx u^*$ and
\begin{equation}\label{eq:scalvarc}
\bar{c}_j(\ell)\mathop{\approx}_{\ell\to 0}E_c^*(u)\,\ell^{-y_c}\,c_j,\quad E_c^*(u)\equiv E_c(u^*,u),
\end{equation}
respectively, where $E_c^*(u)$ is a non-universal amplitude. We now set $\tau=0$, solve the RG equation~\eqref{eq:RGfres}, choose the flow parameter as $\ell\mu L=1$ and use the above limiting expressions for $\bar{u}$ and $\bar{c}_j$. This gives the asymptotic behaviour
\begin{equation}\label{eq:fressf}
f_{\mathrm{res},\mathrm{R}}(L;0,u,c_1,c_2,\mu)/n\mathop{\approx}\limits_{L\to\infty} \frac{D(\mathsf{c}_1,\mathsf{c}_2)}{L^{d-1}}
\end{equation}
with
\begin{equation}\label{eq:Dfres}
D(\mathsf{c}_1,\mathsf{c}_2)=f_{\mathrm{res}}(1;0,u^*,\mathsf{c}_1,\mathsf{c}_2,1)/n
\end{equation}
where
\begin{equation}\label{eq:cscalvar}
\mathsf{c}_j=E_c^*(u)\,c_j\,{(\mu L)}^{y_c}.
\end{equation}

The analogous scaling form of the reduced critical Casimir force per hyper-surface area follows by differentiation with respect to $L$. It reads
\begin{equation}\label{eq:CFscf}
\frac{\mathcal{F}(L;T_{c,\infty},u,c_1,c_2,\mu)}{k_{\mathrm{B}}T}\approx n\, L^{-d}\,\mathcal{D}(\mathsf{c}_1,\mathsf{c}_2)
\end{equation}
with
\begin{equation}\label{eq:mathcalD}
\mathcal{D}(\mathsf{c}_1,\mathsf{c}_2)=\big[d-1+y_c\big(\mathsf{c}_1\partial_{\mathsf{c}_1}+ \mathsf{c}_2\partial_{\mathsf{c}_2}\big)\big]D(\mathsf{c}_1,\mathsf{c}_2).
\end{equation}

To appreciate these results, recall that the large-$L$ form of the residual free-energy density at the bulk critical point $T=T_{c,\infty}$,  for given asymptotic large-scale boundary conditions BC, is conventionally written as
\begin{equation}\label{eq:critfresBC}
f^{\mathrm{BC}}_{\mathrm{res},\mathrm{R}}(L;T_{c,\infty})\mathop{\approx}\limits_{L\to\infty}\Delta^{\mathrm{BC}}\,L^{-(d-1)},
\end{equation}
where $\Delta^{\mathrm{BC}}$ is an $L$-independent universal, but BC dependent, number, called ``Casimir amplitude''. According to our result~\eqref{eq:Dfres}, a scaling function $D$ appears for general values of $c_1$ and $c_2$ in place of the Casimir amplitude --- that is, the Casimir amplitude becomes \emph{scale-dependent}.

On the other hand, various Casimir amplitudes $\Delta^{\mathrm{BC}}$ can be recovered from the knowledge of $D(\mathsf{c}_1,\mathsf{c}_2)$. Since we assumed both surface enhancement variables $c_j$ to be non-supercritical, and also ruled out the breaking of the $O(n)$ symmetry via boundary terms, the case of Robin boundary conditions we consider includes the four cases $\mathrm{BC}=(\mathrm{O},\mathrm{O})$, $(\mathrm{O},\mathrm{sp})$, $(\mathrm{sp},\mathrm{O})$ and $(\mathrm{sp},\mathrm{sp})$ of large-scale boundary conditions that are associated with the fixed-point values $c_j=\infty$ and $c_j=0$ of the respective ordinary ($\mathrm{O}$) and special ($\mathrm{sp}$) fixed points of the semi-infinite systems bounded by $\mathfrak{B}_j$. Hence we have
\begin{equation}\label{eq:DelD}
\eqalign{\Delta^{(\mathrm{O},\mathrm{O})}/n&=D(\infty,\infty),\cr
\Delta^{(\mathrm{O},\mathrm{sp})}/n&=D(\infty,0)=\Delta^{(\mathrm{sp},\mathrm{O})}/n=D(0,\infty), \cr
\Delta^{(\mathrm{sp},\mathrm{sp})}/n&=D(0,0).}
\end{equation}

The first two of these amplitudes, $\Delta^{(\mathrm{O},\mathrm{O})}$ and $\Delta^{(\mathrm{O},\mathrm{sp})}$, are known to have  expansions in integer powers of $\epsilon$. The leading two terms of these series expansions were determined in reference~\cite{KD92a}. The corresponding results can be written as
\begin{eqnarray}\label{eq:DelOO}
\Delta^{(\mathrm{O},\mathrm{O})}/n&=&a_0+a_1(n)\,\epsilon+\Or(\epsilon^{2}),
\end{eqnarray}
and
\begin{eqnarray}\label{eq:DelOsp}
\Delta^{(\mathrm{O},\mathrm{sp})}/n&=&\frac{-7a_0}{8}+\left[\frac{\pi^2}{1024}\left(\frac{n+2}{n+8}-\frac{4\ln 2}{45}\right)-\frac{7a_1(n)}{8}\right]\epsilon+\Or(\epsilon^{2})
\end{eqnarray}
with 
 \begin{equation}
a_0=-\frac{\pi^{2}}{1440}
\end{equation}
and
\begin{equation}\label{eq:a1}
a_1(n)=\frac{\pi^{2}}{2880}\!\left[1-\gamma_E-\ln(4\pi) +\frac{2\zeta^{\prime}(4)}{\zeta(4)}+\frac{5}{2}\,\frac{n+2}{n+8}\right].
\end{equation}
By contrast, the Casimir amplitude $\Delta^{(\mathrm{sp},\mathrm{sp})}$ does not have a power-series expansion in $\epsilon$. As shown in references~\cite{GD08} and \cite{DGS06}, the presence of the $k_1=0$ mode leads to a breakdown of the $\epsilon$ expansion of $\Delta^{(\mathrm{sp},\mathrm{sp})}$ and produces additional half-integer powers $\epsilon^{j/2}$  with $j=3,5,\ldots$, which are modulated by powers of  $\ln \epsilon$ when $j\ge 5$. 
The expansion is known to order $\epsilon^{3/2}$; it reads \cite{GD08,DGS06}
\begin{eqnarray}\label{eq:Delspsp}
\Delta^{(\mathrm{sp},\mathrm{sp})}/n &=& a_0+a_1(n)\,\epsilon+a_{3/2}(n)\, \epsilon^{3/2}
+ \mathrm{o}\big(\epsilon^{3/2}\big)
\end{eqnarray}
with
\begin{equation}\label{eq:a32}
a_{3/2}(n)=-\frac{\pi^{2}}{72\sqrt{6}} \bigg(\frac{n+2}{n+8}\bigg)^{3/2}.
\end{equation}

The $\epsilon$-expansion results for the scaling function $D(\mathsf{c}_1,\mathsf{c}_2)$ we are going to derive below must  be consistent with the series expansions~\eqref{eq:DelOO}--\eqref{eq:Delspsp} and equation~\eqref{eq:DelD}. This requirement provides nontrivial checks for these results.

\subsection{Perturbation theory}\label{sec:pt}

We next turn to the loop expansion 
\begin{equation}\label{eq:floopexp}
f(L;\tb,\ub,\cb_1,\cb_2)=\sum_{l=0}^\infty f^{[l]}(L;\tb,\ub,\cb_1,\cb_2)
\end{equation}
of the reduced free-energy density and its computation for $\tb=0$ to two-loop order. The zero-loop contribution $f^{[0]}$ vanishes in the disordered phase. The next two terms can be written as
\begin{equation}\label{eq:f1loop}
\eqalign{f^{[1]}(L;\tb,\ub,\cb_1,\cb_2)&=\frac{-1}{A}\;\raisebox{-0.8em}{\includegraphics[width=2em]{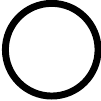}}
=\frac{n}{2A}\Tr\ln{\left(\tb-\triangle\right)}\cr
&=\frac{n}{2}\sum_{m=1}^\infty\int_{\bm{p}}^{(d-1)}\ln(\tb+p^2+k_m^2)}
\end{equation}
and
\begin{equation}\label{eq:f2loop}\fl
\eqalign{
f^{[2]}(L;\tb,\ub,\cb_1,\cb_2)=\frac{-1}{A}
\;\raisebox{-1.5em}{\includegraphics[width=1.75em]{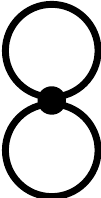}}
=
\frac{n(n+2)\ub}{4!}\int_{\mathfrak{V}}\frac{\rmd^dx}{A}\left[G_L(\bm{x},\bm{x}|\mathring{c}_1,\mathring{c}_2)\right]^2\cr
=\frac{\ub n(n+2)}{4!}\sum_{m_1,m_2=1}^\infty\int_{\bm{p}_1}^{(d-1)}\int_{\bm{p}_2}^{(d-1)}\frac{\Delta_{m_1,m_1,m_2,m_2}(\cb_1L,\cb_2L)}{\left(p_1^2+k_{m_1}^2+\mathring{\tau}\right)\left(p_2^2+k_{m_2}^2+\mathring{\tau}\right)}}
\end{equation} 
for general $\tb\ge 0$. Here we have introduced the functions
\begin{equation}
\Delta_{m_1,m_2,m_3,m_4}(\mathcal{C}_1,\mathcal{C}_2)\equiv\int_0^1\rmd{\zeta}\prod_{j=1}^4\eigenf_{m_j}(\zeta|\mathcal{C}_1,\mathcal{C}_2).
\end{equation}
For the parameter values for which they are needed in equation~\eqref{eq:f2loop}, a lengthy but straightforward calculation yields 
\begin{equation}\label{eq:delta}\fl\eqalign{%
\Delta_{m_1,m_1,m_2,m_2}(\mathcal{C}_1,\mathcal{C}_2)\cr =\frac{1}{8}\sum_{\sigma,\rho=0}^2\frac{\normf_{m_1}\kappa_{m_1}^{2\sigma}}{\left(\mathcal{C}_1^2+\kappa_{m_1}^2\right)\left(\mathcal{C}_2^2+\kappa_{m_1}^2\right)}\frac{\normf_{m_2}\kappa_{m_2}^{2\rho}}{\left(\mathcal{C}_1^2+\kappa_{m_2}^2\right)\left(\mathcal{C}_2^2+\kappa_{m_2}^2\right)}P^{(\sigma,\rho)}_{\mathcal{C}_1,\mathcal{C}_2}
+\frac{1}{4}\normf_{m_1}\delta_{m_1,m_2}}
\end{equation}
with the polynomials
\begin{eqnarray} \label{eq:Ps}
  P^{(0,0)}_{\mathcal{C}_1,\mathcal{C}_2}&=&
    2\mathcal{C}_1^3\mathcal{C}_2^3(\mathcal{C}_1+\mathcal{C}_2+\mathcal{C}_1\mathcal{C}_2),\nonumber\\
  P^{(1,1)}_{\mathcal{C}_1,\mathcal{C}_2}&=&2[\mathcal{C}_1^3+\mathcal{C}_2^3
  +2\mathcal{C}_1^2\mathcal{C}_2+2\mathcal{C}_1\mathcal{C}_2^2+(\mathcal{C}_1^2+\mathcal{C}_2^2)^2],\nonumber\\ 
P^{(2,2)}_{\mathcal{C}_1,\mathcal{C}_2}&=&2,\nonumber\\
P^{(1,0)}_{\mathcal{C}_1,\mathcal{C}_2}&=&2\mathcal{C}_1\mathcal{C}_2(\mathcal{C}_1^2+\mathcal{C}_2^2)
(\mathcal{C}_1+\mathcal{C}_2+\mathcal{C}_1\mathcal{C}_2) =P^{(0,1)}_{\mathcal{C}_1,\mathcal{C}_2},\nonumber\\
P^{(2,1)}_{\mathcal{C}_1,\mathcal{C}_2}&=&
2(\mathcal{C}_1+\mathcal{C}_2+\mathcal{C}_1^2+\mathcal{C}_2^2)=P^{(1,2)}_{\mathcal{C}_1,\mathcal{C}_2},
\nonumber\\
P^{(2,0)}_{\mathcal{C}_1,\mathcal{C}_2}&=&
2\mathcal{C}_1\mathcal{C}_2(\mathcal{C}_1+\mathcal{C}_2+\mathcal{C}_1\mathcal{C}_2)=P^{(0,2)}_{\mathcal{C}_1,\mathcal{C}_2}.
\end{eqnarray}

The $\bm{p}$-integrals in the above equations~\eqref{eq:f1loop} and \eqref{eq:f2loop} can be handled in a standard way using dimensional regularisation. The main difficulty one then is faced with is the calculation of the resulting single and double sums over the not explicitly known eigenvalues $\kappa^2_m$. We have done this by means of complex integration, modifying the Abel-Plana techniques described in reference~\cite{RS02} for our purposes.%
\footnote{As we show in \ref{app:f1alt}, it can also be derived from the result~\eqref{eq:GLc1c1} for the free propagator.} 
The technical details are given in \ref{app:f1}--\ref{app:2loops}. Here we just present our results. 
 
As before, we use the notation $Q^{[l]}$ to specify the $l$-loop term of a quantity $Q$. We choose the additive constant in the free-energy density such that the dimensionally regularised bulk free-energy density $f_{\mathrm{b}}$ vanishes for $\tb=0$. Hence
 \begin{equation}\label{eq:fbc}
 f_{\mathrm{b}}^{[1]}(\tb=0)= f_{\mathrm{b}}^{[2]}(\tb=0,\ub)=0.
 \end{equation}
Our results for the surface free-energy densities read
\begin{equation}\label{eq:fsc1}
f_{\mathrm{s},j}^{[1]}(\tb=0,\cb_j)=-\frac{n\pi K_{d-1}}{2(d-1)\sin(d\pi)}\,\cb_j^{d-1}
\end{equation}
and
\begin{equation}\label{eq:fsc2}
f_{\mathrm{s},j}^{[2]}(\tb=0, \ub,\cb_j)=\frac{n(n+2)}{2}\,\frac{\ub}{4!}\,\frac{\Gamma^2[(3-d)/2]}{(4\pi)^{d-1}}\,\frac{\cb_j^{2d-5}}{\cos(d\pi)-1},
\end{equation}
where we have introduced the familiar quantity
 \begin{equation}\label{eq:Kddef}
K_d\equiv\int_{\bm{q}}^{(d)}\delta(q-1)=2^{1-d}\pi^{-d/2}/\Gamma(d/2).
\end{equation}

In order to write the one- and two-loop residual free-energy densities in a compact fashion, it is helpful to define the functions
\begin{equation}
h_{\mathcal{C}_1,\mathcal{C}_2}(t)\equiv \ln\left[1-\frac{(t-\mathcal{C}_1)(t-\mathcal{C}_2)}{(t+\mathcal{C}_1)(t+\mathcal{C}_2)}\,\rme^{-2t}\right],
\end{equation}
\begin{equation}\label{eq:Xdef}
X^{(d,\sigma)}_{\mathcal{C}_1,\mathcal{C}_2}=\frac{1}{2}\,\csc\left(\frac{d+2\sigma}{2}\pi\right)\,\frac{\mathcal{C}_1^{2\sigma+d-4}-\mathcal{C}_2^{2\sigma+d-4}}{\mathcal{C}_1^2-\mathcal{C}_2^2},
\end{equation}
\begin{equation}\label{eq:Ydef}\fl
Y^{(d,\sigma)}_{\mathcal{C}_1,\mathcal{C}_2}=-\frac{2}{\pi}\cos(d\pi/2)\int_0^\infty\rmd{t}\, \frac{(-1)^\sigma \,t^{2\sigma+d-3}}{(t+\mathcal{C}_1)^2(t+\mathcal{C}_2)^2\,\rme^{2t}-(t^2-\mathcal{C}_1^2)(t^2-\mathcal{C}_2^2)}
\end{equation}
and
\begin{equation}\label{eq:Zdef}
Z^{(d)}_{\mathcal{C}_1,\mathcal{C}_2}=\frac{2}{\pi}\sin(d\pi)\int_0^\infty\rmd{t}\,t^{2d-6}\left[1-\frac{(t-\mathcal{C}_1)(t-\mathcal{C}_2)}{(t+\mathcal{C}_1)(t+\mathcal{C}_2)}\,\rme^{-2t}\right]^{-1}.
\end{equation}
In terms of these, our results can be written as
\begin{equation}\label{eq:fres1}
f_{\mathrm{res}}^{[1]}(L;0,\cb_1,\cb_2)=\frac{nK_{d-1}}{2L^{d-1}}\int_0^\infty\rmd{t}\,t^{d-2}\,h_{L\cb_1,L\cb_2}(t)
\end{equation}
and
\begin{equation}\label{eq:fres2}\fl
\eqalign{f_{\mathrm{res}}^{[2]}(L;0,\ub,\cb_1,\cb_2)=&\frac{n(n+2)}{2}\,\frac{\ub}{4!} \,\frac{\Gamma^2[(3-d)/2]}{(4\pi)^{d-1}}\,L^{5-2d}\bigg[Z^{(d)}_{L\cb_1,L\cb_2}\cr
&+\sum_{\sigma,\rho=0}^2P^{(\sigma,\rho)}_{L\cb_1,L\cb_2}\Big(Y^{(d,\sigma)}_{L\cb_1,L\cb_2}\,Y^{(d,\rho)}_{L\cb_1,L\cb_2}+2X^{(d,\sigma)}_{L\cb_1,L\cb_2}Y^{(d,\rho)}_{L\cb_1,L\cb_2}\Big)
\bigg].}
\end{equation}

\subsection{Renormalized residual free-energy density and scaling functions}\label{sec:renfresscfcts}

The functions $X_{\cb_1L,\cb_2L}^{(d,1)}$ and $X_{\cb_1L,\cb_2L}^{(d,2)}$ have  simple poles at $d=4$, caused by the behaviour  of the respective pre-factor of the integral in equation~\eqref{eq:Xdef}. These UV singularities imply that  the two-loop term~\eqref{eq:fres2} is not regular at $\epsilon=0$. It has a simple pole originating from the terms proportional to $X_{\cb_1L,\cb_2L}^{(d,1)}$ and $X_{\cb_1L,\cb_2L}^{(d,2)}$. One finds
\begin{equation}\label{eq:f2respole}\fl
\eqalign{\frac{f_{\mathrm{res}}^{[2]}(L;0,\ub,\cb_1,\cb_2)}{L^{-(d-1)}}=&\frac{n(n+2)\ub L^{-\epsilon} N_d}{12\pi\epsilon}\,(\cb_1L+\cb_2L)\Big(\cb_1\cb_2L^2\,Y_{\cb_1L,\cb_2L}^{(4,1)}\cr&+Y_{\cb_1L,\cb_2L}^{(4,2)}\Big)+\Or(\epsilon^0),}
\end{equation}
where $\ub N_d=\mu^\epsilon u+\Or(u^2)$. It is an easy matter to check that this pole gets cancelled by the contribution $\propto u/\epsilon$ one obtains from $f_{\mathrm{res}}^{[1]}(L;0,\cb_1,\cb_2)$ upon expressing the bare variables $\cb_j$ in terms of their renormalized analogues $c_j$  via equations~\eqref{eq:surfrep} and \eqref{eq:Z1orderu}. Substitution of the one- and two-loop terms~\eqref{eq:fres1} and \eqref{eq:fres2} into equation~\eqref{eq:fresRdef} therefore  yields indeed a UV-finite  $\Or(u)$ result for the renormalized residual free-energy density $f_{\mathrm{res,R}}(L;0,u,c_1,c_2,\mu)$, namely
\begin{equation}\label{eq:fresRres}\fl
\eqalign{\frac{f_{\mathrm{res,R}}(L;0,u,c_1,c_2,\mu)}{nL^{-(d-1)}}=&D_0(c_1 L\mu,c_2 L\mu)+\epsilon\Bigg[\left(1-\frac{\gamma_E-\ln\pi}{2}\right)D_0(c_1 L\mu,c_2 L\mu)\cr&\strut
-\frac{1}{4\pi^2}\int_0^\infty \rmd{t}\,
h_{c_1 L\mu,c_2 L\mu}(t) \,t^2\ln t\Bigg]\cr&\strut+u\Bigg\{\frac{n+2}{3}\sum_{j=1}^2\bigg[-\frac{1}{2}+\ln\left(2c_j\right)\bigg]c_j\partial_{c_j}D_0(c_1 L\mu,c_2 L\mu)\cr&\strut+
\frac{n+2}{12\pi^2}\sum_{\sigma,\rho=0}^2
P^{(\sigma,\rho)}_{c_1 L\mu,c_2 L\mu}\,
J^{(\sigma)}_{c_1 L\mu,c_2 L\mu}\,
J^{(\rho)}_{c_1 L\mu,c_2 L\mu}\Bigg\}+\Or(u^2),}
\end{equation}
where
\begin{equation}\label{eq:D0}
  D_0(\mathcal{C}_1,\mathcal{C}_2)\equiv f^{[1]}_{\mathrm{res}}(1;0,\mathcal{C}_1,\mathcal{C}_2)/n=\frac{1}{4\pi^2}\int_0^\infty
  \rmd{t}\,t^2\,h_{\mathcal{C}_1,\mathcal{C}_2}(t)\;,
\end{equation}
and
\begin{equation}\label{eq:Jdef}\fl
  J^{(\sigma)}_{\mathcal{C}_1,\mathcal{C}_2}\equiv -\frac{\pi}{2}\,Y^{(4,\sigma)}_{\mathcal{C}_1,\mathcal{C}_2}=\int_0^\infty\rmd{t}\,
  \frac{(-1)^\sigma\,
    t^{1+2\sigma}}{(t+\mathcal{C}_1)^2(t+\mathcal{C}_2)^2\,\rme^{2t}
    -(t^2-\mathcal{C}_1^2)(t^2-\mathcal{C}_2^2)}\;.
\end{equation}
To  obtain the $\epsilon$ expansion of the scaling function $D$, we set $u=u^*$. Using 
\begin{equation}
(\mu L)^{y_c}=\mu L\left[1-\frac{n+2}{n+8}\,\epsilon\ln(\mu L)\right]+\Or(\epsilon^2),
\end{equation}
one sees that the result is consistent with the predicted scaling form~\eqref{eq:fressf} and yields the expansion
\begin{equation}\label{eq:Depsexp}
D(\mathsf{c}_1,\mathsf{c}_2)=D_0(\mathsf{c}_1,\mathsf{c}_2)+\epsilon D_1(\mathsf{c}_1,\mathsf{c}_2)+\Or(\epsilon^2)
\end{equation}
with
\begin{equation}\label{eq:D1}\fl
\eqalign{D_1(\mathsf{c}_1,\mathsf{c}_2)=&
  \bigg(1-\frac{\gamma_E-\ln\pi}{2}\bigg)\,D_0(\mathsf{c}_1,\mathsf{c}_2) 
-\frac{1}{4\pi^2}\int_0^\infty \rmd{t}\,
h_{\mathsf{c}_1,\mathsf{c}_2}(t) \,t^2\ln t
\cr&\strut
 +\frac{n+2}{n+8}\Bigg\{\sum_{j=1}^2\left[-\frac{1}{2} +\ln
(2\mathsf{c}_j)\right]\,\mathsf{c}_j\partial_{\mathsf{c}_j}
D_0(\mathsf{c}_1,\mathsf{c}_2)\cr &
+ \frac{1}{4\pi^2}\sum_{\sigma,\rho=0}^2
P^{(\sigma,\rho)}_{\mathsf{c}_1,\mathsf{c}_2}\,
J^{(\sigma)}_{\mathsf{c}_1,\mathsf{c}_2}\,
J^{(\rho)}_{\mathsf{c}_1,\mathsf{c}_2}\Bigg\} .}
\end{equation}

For the special values $\mathsf{c}_j=0$ and $\infty$, the functions $D_0$ and $D_1$ can be analytically computed in a straightforward manner. One finds
\begin{equation}
\eqalign{D_0(\infty,\infty)&=D_0(0,0)=\frac{-8}{7}\,D_0(0,\infty)=-\frac{\pi^2}{1440},\cr
D_1(\infty,\infty)&=D_1(0,0)=a_1(n),\cr
D_1(\infty,0)&=\frac{\pi^2}{1024}\left(\frac{n+2}{n+8}-\frac{4\ln 2}{45}\right)-\frac{7a_1(n)}{8},
}
\end{equation}
where $a_1(n)$ is the coefficient defined in equation~\eqref{eq:a1}. These results confirm the validity of the relations~\eqref{eq:DelD} to first order in $\epsilon$. However, the present form~\eqref{eq:Depsexp} of our result does not yield the contribution $\sim\epsilon^{3/2}$ to $D(0,0)=\Delta^{(\mathrm{O},\mathrm{sp})}/n$. The reason is that we assumed that $c_1+c_2>0$  so that the lowest eigenvalue $k_1^2$ of $-\partial_z^2$ [cf.\ equation~\eqref{eq:eigfunceq}] is strictly positive. Since $k_1^2=0$ when $c_1=c_2$, there is a zero mode in the free propagator. As already mentioned, this causes a breakdown of the conventional RG-improved perturbation theory at $T_{c,\infty}$ \cite{GD08,DGS06,DG09}. In the following we will improve our results in such a manner that they fully comply with all of the small-$\epsilon$ expansions of  Casimir amplitudes given in equations~\eqref{eq:DelOO}--\eqref{eq:Delspsp}, including the fractional expansion~\eqref{eq:Delspsp} of $\Delta^{(\mathrm{sp},\mathrm{sp})}$ to $\Or(\epsilon^{3/2})$.

\subsection{Modified RG-improved perturbation theory}\label{sec:consistency}
\subsubsection{Formulation and results}

 It is known from references~\cite{GD08}, \cite{DGS06} and \cite{DG09} how one can cope with the mentioned zero-mode problem one encounters at $\tau=0$  when $c_1=c_2=0$: one can reorganise RG-improved perturbation theory such that it becomes well-defined at $\tau=0$. This suggests an obvious strategy to ensure consistency with the small-$\epsilon$ expansions of $\Delta^{(\mathrm{sp},\mathrm{sp})}$ to $\epsilon^{3/2}$. We should reorganise RG-improved perturbation theory for general $c_1>0$ and $c_2>0$ in a way that reduces to the one used in previous work \cite{GD08,DGS06} for the case $c_1=c_2=0$.

This can be done as follows. We decompose the order-parameter field into an ($m=1$)-component and an orthogonal remainder, writing
 \begin{equation}\label{eq:phidec}
\eqalign{ \bm{\phi}(\bm{y},z)&=L^{-1/2}\bm{\varphi}(\bm{y})\,\eigenf_1(z/L)+\bm{\psi}(\bm{y},z),\cr
 \bm{\psi}(\bm{y},z)&=L^{-1/2}\sum_{m=2}^\infty\bm{\phi}^{(m)}(\bm{y})\,\eigenf_m(z/L)}
 \end{equation}
 with
 \begin{equation}
 \bm{\phi}^{(m)}(\bm{y})=\int_0^1\rmd{\zeta}\,\eigenf_m(\zeta)\,\bm{\phi}(\bm{y},\zeta L)
 \end{equation}
 and $\bm{\varphi}(\bm{y})\equiv\bm{\phi}^{(1)}(\bm{y})$.
The main difference compared to the previously studied cases of periodic and $(\mathrm{sp},\mathrm{sp})$ BC \cite{GD08,DGS06}, as well as  generalisations of the latter \cite{DG09}, is that the ($m=1$)-mode is not $z$-independent unless $c_1=c_2=0$. Thus the orthogonality of the $\psi_\alpha$ and the eigenfunction $\eigenf_1$ does no longer translate into the vanishing of the integral $\int_0^L\rmd{z}\,\bm{\psi}(\bm{y},z)$. The $\phi^4$~term of the Hamiltonian therefore now generates also a vertex $\propto(\bm{\psi}\cdot\bm{\varphi})\varphi^2$ (which vanishes when  $c_1=c_2=0$), in addition to four-point vertices involving $l=0,1,3$ and $4$ fields $\psi_\alpha$ and $4-l$ powers of $\varphi_\alpha$. 

Let us introduce $f_\psi$, the free-energy density associated with the $\psi$-field, by
 \begin{equation}\label{eq:fpsi}
 Af_\psi=-\ln\Tr_{\bm{\psi}}\rme^{-\mathcal{H}[\bm{\psi}]}.
 \end{equation}
 Its loop expansion 
 \begin{equation}\label{eq:fpsiloopexp}\fl
 \eqalign{f_\psi(L;\tb,\ub,\cb_1,\cb_2) &= f^{[1]}_\psi(L;\tb,\ub,\cb_1,\cb_2)+f^{[2]}_\psi(L;\tb,\ub,\cb_1,\cb_2)+\Or(3\mbox{-loop})\cr &=\frac{-1}{A}\;\raisebox{-0.8em}{\includegraphics[width=2em]{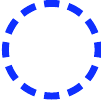}}-\frac{1}{A}\;\raisebox{-1.5em}{\includegraphics[width=1.75em]{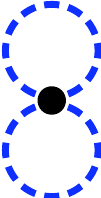}}+\Or(3\mbox{-loop})}
 \end{equation}
 is analogous to that of $f(L;\tb,\ub,\cb_1,\cb_2)$  specified in equation~\eqref{eq:floopexp}. The  one- and two-loop terms $f_\psi^{[1]}$ and $f_\psi^{[2]}$ are given by the series in the second lines of  equations~\eqref{eq:f1loop} and \eqref{eq:f2loop} excluding the summands with $m=1$ and $m_1,m_2=1$, respectively. In the corresponding Feynman graphs (second line), we depict free $\psi$-propagators as dashed blue lines.
 
\begin{figure}[htbp]
\centerline{\includegraphics[scale=0.85]{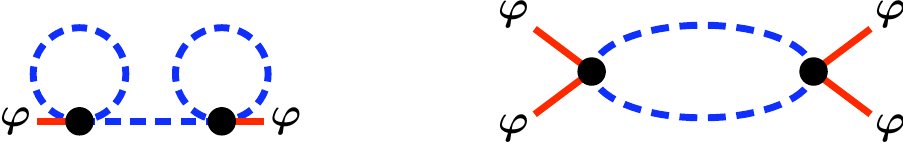}}
\caption{Examples of graphs producing  nonlocal two- and four-point vertices of $\mathcal{H}_{\mathrm{eff}}$.
\label{fig:nonlocalint}}
\end{figure}

We now set $\tb=0$ and define an effective $(d-1)$-dimensional field theory  with  Hamiltonian $\mathcal{H}_{\mathrm{eff}}[\bm{\varphi}]$ by integrating out $\bm{\psi}$:
  \begin{equation}
 \rme^{-Af_\psi+\mathcal{H}_{\mathrm{eff}}[\bm{\varphi}]}=\Tr_{\bm{\psi}}\rme^{-\mathcal{H}[\bm{\phi}]}.
 \end{equation}
To determine $\mathcal{H}_{\mathrm{eff}}[\bm{\varphi}]$, we use perturbation theory. The contribution to the two-point vertex caused by the coupling between $\bm{\varphi}$ and $\bm{\psi}$, to first order in $\ub$, originates from the graph
\begin{equation}
\raisebox{-0.5em}{\includegraphics[width=3em]{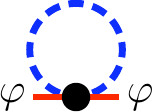}} =-\frac{1}{2}\,\delta\tb_L(\cb_1,\cb_2)\int\rmd^{d-1}{y}\,\varphi^2,
 \end{equation}
 where
 \begin{equation}\label{eq:deltauLc1c2}
 \delta\tb_L(\cb_1,\cb_2)=\frac{n+2}{3}\,\frac{\ub}{2L^{d-2}}\sum_{m=2}^\infty\int_{\bm{p}}^{(d-1)}\frac{\Delta_{1,1,m,m}(\mathcal{C}_1,\mathcal{C}_2)}{p^2+\kappa_m^2(\mathcal{C}_1,\mathcal{C}_2)}.
 \end{equation}
It represents a local interaction corresponding to a shift $\delta\tb_L$ of $\tb$. 
The Hamiltonian $\mathcal{H}_{\mathrm{eff}}[\bm{\varphi}]$ becomes
\begin{equation}\label{eq:Heff}
\eqalign{ \mathcal{H}_{\mathrm{eff}}[\bm{\varphi}]=&\int\rmd^{d-1}y\left\{\frac{1}{2}(\nabla_{\bm{y}}\bm{\varphi})^2+\frac{1}{2}\left[\delta\tb_L(\cb_1,\cb_2)+k_1^2\right]\bm{\varphi}^2\right.\cr &\left.+\frac{\ub}{4!L}\Delta_{1,1,1,1}(\cb_1 L,\cb_2 L)\,|\bm{\varphi}|^4 +\Or(\ub^2)
\right\}.}
 \end{equation}
 It should be clear that beyond first order in $\ub$ also nonlocal two-point and $2s$-point vertices with $s\ge 2$ appear in $\mathcal{H}_{\mathrm{eff}}[\bm{\varphi}]$. Two examples of such graphs are shown in figure~\ref{fig:nonlocalint}.
 %

To obtain the free-energy density $f(L;0,\ub,\cb_1,\cb_2)$ in the present modified perturbation scheme, we must add to  $f_\psi(L;0,\ub,\cb_1,\cb_2)$  the contribution associated with the $\bm{\varphi}$-field. We denote it as $f_\varphi$ and define it via 
\begin{equation}
 Af_\varphi=-\ln\Tr_{\bm{\varphi}}\rme^{-\mathcal{H}_{\mathrm{eff}}[\bm{\varphi}]}
\end{equation}
by analogy with equation~\eqref{eq:fpsi}.
Hence we have
\begin{equation}\label{eq:fpsivarphi}
f(L;0,\ub,\cb_1,\cb_2) =f_\varphi(L;0,\ub,\cb_1,\cb_2)+f_\psi(L;0,\ub,\cb_1,\cb_2),
\end{equation}
where $f_\psi(L;\tb=0,\ub,\cb_1,\cb_2)$ is given by equation~\eqref{eq:fpsiloopexp} and $f_\varphi(L;0,\ub,\cb_1,\cb_2)$ has the loop expansion
\begin{equation}\label{eq:fvarphi}
\eqalign{f_\varphi(L;0,\ub,\cb_1,\cb_2)&=\frac{-1}{A}\;\raisebox{-0.8em}{\includegraphics[width=2em]{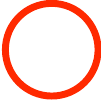}}-\frac{1}{A}\;\raisebox{-1.5em}{\includegraphics[width=1.75em]{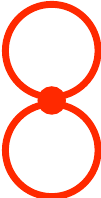}}+\ldots\cr&=f^{[1]}_\varphi(L;0,\ub,\cb_1,\cb_2)+f^{[2]}_\varphi(L;0,\ub,\cb_1,\cb_2)+\ldots.}
\end{equation}
Here $f_\varphi^{[1]}$ and $f_\varphi^{[2]}$ are the one- and two-loop terms of a $(d-1)$-dimensional $\phi^4$~theory with quadratic and $\varphi^4$ interaction constants $\delta\tb_L+k_1^2$ and  $\ub\Delta_{1,1,1,1}/L$, respectively. Their explicit expressions may be inferred from equation~(4.28) of reference \cite{GD08}. They read
\begin{equation}\label{eq:fvarphi1}
f^{[1]}_\varphi(L;0,\ub,\cb_1,\cb_2)=-\frac{nA_{d-1}}{d-1}\left[k_1^2+\delta\tb_L(\cb_1,\cb_2)\right]^{(d-1)/2}
\end{equation}
and 
\begin{equation}\label{eq:fvarphi2}\fl
f^{[2]}_\varphi(L;0,\ub,\cb_1,\cb_2)=\frac{\ub}{L}\,\frac{n(n+2)}{4!}\,A^2_{d-1}\left[k_1^2+\delta\tb_L(\cb_1,\cb_2)\right]^{d-3}\,\Delta_{1,1,1,1}(\cb_1L,\cb_2L)
\end{equation}
with
\begin{equation}\label{eq:Addef}
A_d\equiv\frac{2N_d}{4-d}=-(4\pi)^{-d/2}\,\Gamma(1-d/2).
\end{equation}

Building on the reorganisation of perturbation theory just described, we now wish to compute the residual free-energy density $f_{\mathrm{res}}(L;0,\ub,\cb_1,\cb_2)$ and  express it in terms of renormalized variables  to obtain improved results for its scaling function $D(\mathsf{c}_1,\mathsf{c}_2)$ and that of the Casimir force. Clearly, from these results all Taylor expansions in $\epsilon$ reported in sections~\ref{eq:renfresCF} and \ref{sec:renfresscfcts} must be recovered, namely, the $\epsilon$~expansions~\eqref{eq:Depsexp}--\eqref{eq:D1} of the scaling function $D(\mathsf{c}_1,\mathsf{c}_2)$ when $\min(\mathsf{c}_1+\mathsf{c}_2)>0$, as well as those of the Casimir amplitudes $\Delta^{(\mathrm{O},\mathrm{O})}$ and $\Delta^{(\mathrm{O},\mathrm{sp})}$ given in equations~\eqref{eq:DelOO} and \eqref{eq:DelOsp}, respectively. In addition, it must yield the fractional $\epsilon$~expansion~\eqref{eq:Delspsp} of $\Delta^{(\mathrm{sp},\mathrm{sp})}$ to $\Or(\epsilon^{3/2})$ since it reduces to the scheme used in references~\cite{DGS06} and \cite{GD08} when $c_1=c_2=\tau=0$.

What makes this approach technically difficult to implement is that the eigenvalues $k_m^2$, whose behaviours for small $\cb_1+\cb_2$ we gave in equation~\eqref{eq:smallCkappa}, are \emph{not explicitly known} for general values of $\cb_1$ and $\cb_2$. Hence we shall have to resort again to numerical means.

Let us start by taking a look at the effective two-point vertex
\begin{equation}\label{eq:gamma2eff}
\gamma^{(2)}(\bm{y}-\bm{y}')\,\delta_{\alpha\beta} \equiv\frac{\delta^2\mathcal{H}_{\mathrm{eff}}[\varphi]}{\delta\varphi_\alpha(\bm{y})\delta\varphi_\beta(\bm{y}')}\Bigg|_{\bm{\varphi}=\bm{0}}\,,
\end{equation}
whose one-loop approximation appears in the expressions~\eqref{eq:fvarphi1} and \eqref{eq:fvarphi2} for $f_\varphi^{[1]}$ and $f_\varphi^{[2]}$. Its renormalized counterpart is given by $\gamma^{(2)}_{\mathrm{R}}=Z_\phi\gamma^{(2)}$. Thus its Fourier $\bm{p}$-transform $\hat{\gamma}^{(2)}_{\mathrm{R}}$ satisfies the RG $(\mathcal{D}_\mu-\eta)\hat{\gamma}^{(2)}_{\mathrm{R}}$, by analogy with equation~\eqref{eq:RGEGs}. Solving this at $\bm{p}=\bm{0}$ and $\tau=0$ yields the scaling form
\begin{equation}\label{eq:scfgammma2}
\hat{\gamma}^{(2)}_{\mathrm{R}}(\bm{p}=\bm{0},L,\tau=0,u,c_1,c_2,\mu)\approx \mu^{\eta}[E^*_h(u)]^2L^{-2+\eta}\,\Omega(\mathsf{c}_1,\mathsf{c}_2),
\end{equation}
where $\mathsf{c}_j$ denote the scaling variables defined in equation~\eqref{eq:cscalvar}. Further,
$E_h^*$ is a non-universal amplitude whose representation as a trajectory integral can be found in equation~(3.85c) of reference~\cite{Die86a} but will not be needed in the remainder. Since this amplitude drops out of the Casimir force, we need not keep track of it and may therefore set it to $1$ henceforth.

In \ref{app:gamma2eff} we determine $\hat{\gamma}_{\mathrm{R}}^{(2)}$ to one-loop order, show that it is UV finite and compute the scaling function $\Omega(\mathsf{c}_1,\mathsf{c}_2)$ to $\Or(\epsilon)$. The result is 
\begin{equation}\label{eq:Omegares}\fl
\eqalign{\Omega(\mathsf{c}_1,\mathsf{c}_2)&=\kappa_1^2+\epsilon\normf_1\kappa_1\frac{n+2}{n+8}\Bigg[\sum_{\sigma,\rho=1}^2\frac{\kappa_1^{2\sigma-1}P_{\mathsf{c}_1,\mathsf{c}_2}^{(\sigma,\rho)}J_{\mathsf{c}_1,\mathsf{c}_2}^{\rho}}{(\mathsf{c}_1^2+\kappa_1^2)(\mathsf{c}_2^2+\kappa_1^2)}+\kappa_1\sum_{j=1}^2\frac{\mathsf{c}_j\ln\mathsf{c}_j}{\mathsf{c}_j^2+\kappa_1^2}-\frac{\pi}{2}\cr
& +\frac{2\pi}{\normf_1}\Delta_{1,1,1,1}(\mathsf{c}_1,\mathsf{c}_2)-(1-2\ln 2)\kappa_1\frac{(\mathsf{c}_1+\mathsf{c}_2)(\mathsf{c}_1\mathsf{c}_2+\kappa_1^2)}{2(\mathsf{c}_1^2+\kappa_1^2)(\mathsf{c}_2^2+\kappa_1^2)}+\Or\big(\epsilon^2\big),}
\end{equation}
where $\kappa_1$ and $\normf_1$  represent the eigenvalue $\kappa_1(\mathsf{c}_1,\mathsf{c}_2)$ and the normalization factor $\normf_1(\mathsf{c}_1,\mathsf{c}_2)$, respectively.

\begin{figure}[htbp]
\begin{center}
\includegraphics[width=0.75\textwidth]{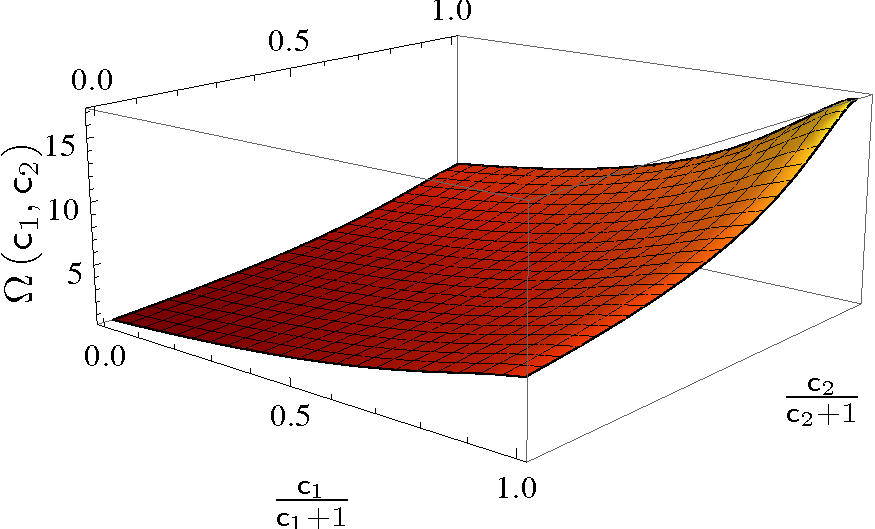}
\caption{Scaling function $\Omega(\mathsf{c}_1,\mathsf{c}_2) $ for $n=1$ and $d=3$, obtained by extrapolation of the $\Or(\epsilon)$ result~\eqref{eq:Omegares}. To cover the full range $(0,\infty)$ of the scaling variables $\mathsf{c}_j$, we plotted $\mathsf{c}_j/(1+\mathsf{c}_j)$ along two axes.\label{fig:Omega}}
\end{center}
\end{figure}

The extrapolation one obtains for $\Omega(\mathsf{c}_1,\mathsf{c}_2;d=3,n=1)$ by evaluating the above $\Or(\epsilon)$ result at $\epsilon=1$ is depicted in 
figure~\ref{fig:Omega}.
%
%
Since the eigenvalues $\kappa_1(\mathsf{c}_1,\mathsf{c}_2)$ are analytically known for all combinations of $(\mathsf{c}_1,\mathsf{c}_2)$ with $\mathsf{c}_j=0,\infty$, $j=1,2$, the $\epsilon$ expansions of the corresponding limiting values of $\Omega$ can be determined exactly. The same applies to the term linear in $\mathsf{c}_1+\mathsf{c}_2$ of the Taylor series expansion of $\Omega$ in $\mathsf{c}_j$ at $(\mathsf{c}_1,\mathsf{c}_2)=(0,0)$. One obtains
\begin{equation}
 \Omega(\infty,\infty)=\pi^2+\epsilon \frac{13\pi^2}{6}\frac{n+2}{n+8}+\Or\big(\epsilon^2\big),
 \end{equation}
 \begin{equation}
 \Omega(0,\infty)=\Omega(\infty,0)=\frac{\pi^2}{4}+\epsilon \frac{11\pi^2}{12}\frac{n+2}{n+8}+\Or\big(\epsilon^2\big)
 \end{equation}
 and
 \begin{equation}\label{eq:Omegalin}
\eqalign{\Omega(\mathsf{c}_1,\mathsf{c}_2) = &\; \mathsf{c}_1+\mathsf{c}_2+\Or(\mathsf{c}_1^2,\mathsf{c}_2^2,\mathsf{c}_1\mathsf{c}_1) \cr
&+\epsilon\frac{n+2}{n+8}\Bigg\{\frac{\pi^2}{6}+\big(\mathsf{c}_1+\mathsf{c}_2\big)\Bigg[\frac{3}{2}-\gamma_{\mathrm{E}}+\ln (2\pi)\Bigg]\cr
&\strut+\Or(\mathsf{c}_1^2,\mathsf{c}_2^2,\mathsf{c}_1\mathsf{c}_2)\Bigg\}+\Or\big(\epsilon^2\big).}
\end{equation}

The first term on the right-hand side reflects the small-$\mathsf{c}_j$ behaviour $\kappa_1^2(\mathsf{c}_1,\mathsf{c}_2)\approx \mathsf{c}_1+\mathsf{c}_2$ implied by equation~\eqref{eq:smallCkappa}. The remaining terms originate from the  expansion  of the shift $\delta\tb$ to linear order,
\begin{equation}\label{eq:deltauLsamllc}
\delta\tb_L(\cb_1,\cb_2)=\delta\tb_L^{\mathrm{sp}}+(\cb_1+\cb_2)\,\delta\tb_L^{\prime,\mathrm{sp}}+\Or(\cb_1^2,\cb_2^2,\cb_1\cb_2),
\end{equation}
whose zeroth-order term is
\begin{equation}\label{eq:deltauLsp}
\delta\tb_L^{\mathrm{sp}}\equiv\delta\tb_L(0,0)=\frac{(n+2)\ub N_d}{3L^{d-2}}\,\frac{\Gamma(d/2)\,\zeta(d-2)}{\Gamma(3-d/2)},
\end{equation}
according to equation~(14) of reference~\cite{DGS06} (where it was denoted as $\delta\tb_{\mathrm{sp},\mathrm{sp}}^{(L)}$). Note that this result can easily be recovered from equation~\eqref{eq:deltauLc1c2} using the fact that $\kappa_m(0,0)=(m-1)\pi$ and $\Delta_{1,1,m,m}(0,0)=1+\delta_{m,1}/2$. The coefficient $\delta\tb_L^{\prime,\mathrm{sp}}$ is computed in \ref{app:shiftprime}. It reads
\begin{equation}\label{eq:shiftprime}
\delta\tb_L^{\prime,\mathrm{sp}}=\frac{(n+2)\ub N_d}{3L^{d-3}}\,\frac{(d-2)(2d-7)\,\Gamma(3/2-d/2)}{4\pi^{9/2-d}\,\Gamma(3-d/2)}\,\zeta(5-d).
\end{equation}

The above results, equations~\eqref{eq:deltauLsamllc}--\eqref{eq:shiftprime}, are consistent with equation~\eqref{eq:Omegares}. To see this note that 
$\delta\tb_L^{\prime,\mathrm{sp}}$ has the Laurent expansion 
\begin{equation}\label{eq:tauLprimeLaurent}
\delta\tb_L^{\prime,\mathrm{sp}}=-\frac{n+2}{3L}\,\frac{u}{\epsilon}\left[1+\epsilon\left(\gamma_E-\frac{3}{2}+\ln\frac{\mu L}{2\pi}\right)+\Or(\epsilon^2)\right].
\end{equation}
Just as in our calculation of $\gamma^{(2)}_{\text{R}}$, the pole term gets cancelled by the counter-term provided by the contribution linear in $\cb_1+\cb_2=\mu Z_c\,(c_1+c_2)$ to $k_1^2$. If we substitute equation~\eqref{eq:deltauLsamllc} into $k_1^2+\delta\tb_L$ along with equations~\eqref{eq:deltauLsp} and \eqref{eq:shiftprime}, express the result in terms of renormalized variables and set $u=u^*$, we recover the expansion of $\Omega(\mathsf{c}_1,\mathsf{c}_2)$ to linear order in $\mathsf{c}_j$ given on the right-hand-side of equation~\eqref{eq:Omegalin}. 

We now insert the scaling form~\eqref{eq:scfgammma2} of the effective two-point vertex into the expressions~\eqref{eq:fvarphi1} and \eqref{eq:fvarphi2} for $f_\varphi^{[1]}$ and $f_\varphi^{[2]}$. Expressing the sum in terms of renormalized variables yields 
\begin{equation}\label{eq:Dvarphidef}\fl
\eqalign{f_{\varphi,\text{R}}(L;0,u^*,c_1,c_2,\mu)\,\frac{L^{d-1}}{n}&\equiv D_\varphi(\mathsf{c}_1,\mathsf{c}_2)
\cr &=-\frac{ A_{d-1}}{d-1}\big[ \Omega(\mathsf{c}_1,\mathsf{c}_2)\big]^{(d-1)/2}\cr
&\quad+\frac{u^*(n+2)}{4! N_d}\,A_{d-1}^2\big[\Omega(\mathsf{c}_1,\mathsf{c}_2)\big]^{d-3}\Delta_{1,1,1,1}(\mathsf{c}_1,\mathsf{c}_2)+\ldots.
}
\end{equation}
Here the ellipsis stands for contributions that are $\Or(u^*\epsilon,\epsilon^2)$ as long as $\Omega(\mathsf{c}_1,\mathsf{c}_2)=\Or(\epsilon^0)$. This condition is fulfilled whenever $\mathsf{c}_1+\mathsf{c}_2>0$. Depending on whether this is the case or not, the function $D_\varphi(\mathsf{c}_1,\mathsf{c}_2)$ has an expansion in integer powers of $\epsilon$, namely
\begin{equation}
\eqalign{%
D_\varphi(\mathsf{c}_1,\mathsf{c}_2)=&-\frac{\kappa_1^3}{12\pi}+\epsilon\Bigg[\frac{\kappa_1^3}{72\pi}\bigg(3\gamma_E-8+3\ln\frac{\kappa_1^2}{\pi}\bigg)\cr & \strut +\frac{n+2}{n+8}\, \frac{\kappa_1^2}{8}\,\Delta_{1,1,1,1}-\frac{\kappa_1\,\Omega'_0}{8\pi}\Bigg]+\Or(\epsilon^2) \qquad (\mathsf{c}_1+\mathsf{c}_2>0),}
\end{equation}
while
\begin{equation}\label{eq:Dvarphi00}
D_\varphi(0,0)=a_{3/2}(n)\,\epsilon^{3/2}+\mathrm{o}(\epsilon^{3/2}).
\end{equation}
Here $\Omega'_0\equiv\partial\Omega(\mathsf{c}_1,\mathsf{c}_2;\epsilon)/\partial\epsilon|_{\epsilon=1}$ and $a_{3/2}(n)$ are the coefficient of the $\Or(\epsilon)$ term in equation~\eqref{eq:Omegares} and the quantity given in equation~\eqref{eq:a32}, respectively.

One caveat should be noted: Although the right-hand side of equation~\eqref{eq:Dvarphidef} is UV finite for $\epsilon<1$, it is not so at $d=3$ since $A_{d-1}=[2\pi(d-3)]^{-1}+\Or(1)$. The origin of these UV singularities is obvious. The free-energy density $f_\varphi$ of the effective ($d-1$)-dimensional theory at $d=3$ involves  UV singularities of the form $\Lambda^2$ and $\Omega\ln\Lambda $. The bulk and surface counter-terms included so far do not cure these; additional subtractions (of the kind needed for a super-renormalizable effective bulk theory in $d-1=2$ dimensions, cf.~\cite{Sac97}) would be needed. Evaluated at $d=3$, the first term on the right-hand side of equation~\eqref{eq:Dvarphidef} would then yield a contribution $(\Omega/8\pi)\big[1-\gamma_E-\ln(\Omega/4\pi)\big]$ to $f_{\varphi,\text{res},\text{R}}$ and hence to $D_\varphi$. Such logarithmic terms are encountered  at $d=3$  also in Gaussian film  models; see e.g.\ reference~\cite{KD10}. 

Let us postpone any further discussion of the behaviour of $D_\varphi$ at $d=3$ for the moment and first compute the free-energy contributions $f^{[1]}_\psi$ and $f^{[2]}_\psi$. The former differs from $f^{[1]}$ through the lowest-mode contribution. This is given by equation~\eqref{eq:fvarphi1} with the shift $\delta\tau_L$ set to zero. Hence we have
\begin{equation}\label{eq:fpsi1l}\fl
f^{[1]}_\psi(L;0,\ub,\cb_1,\cb_2)-f^{[1]}(L;0,\ub,\cb_1,\cb_2)=nL^{1-d}\,\frac{A_{d-1}}{d-1}\kappa_1^{d-1}.
\end{equation}
To determine the two-loop graph of $f_\psi^{[2]}$, we must subtract from the two-loop graph of  $f^{[2]}$ displayed in equation~\eqref{eq:f2loop} the contributions 
for which the mode indices $m_1$ and $m_2$  of the upper or lower loops take the value $1$, i.e.
\begin{eqnarray}\label{eq:fdiff2l}
-A\,f_\psi^{[2]}=\raisebox{-1.5em}{\includegraphics[width=1.75em]{./figure4}}=\raisebox{-1.5em}{\includegraphics[width=1.75em]{./figure2}}-\left[\raisebox{-1.5em}{\includegraphics[width=1.75em]{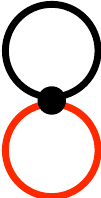}}-\raisebox{-1.5em}{\includegraphics[width=1.75em]{./figure8}}\right]_{\delta\tau_L=0}.
\end{eqnarray}
Details of the calculation are presented in \ref{app:fpsidiff}. They yield
\begin{equation}\label{eq:fpsiresdiff}\fl
\eqalign{f_{\psi,\text{res},\text{R}}(L;0,u^*,c_1,c_2,\mu)\,\frac{L^{d-1}}{n}&\equiv D_{\psi}(\mathsf{c}_1,\mathsf{c}_2) \cr &=D_{\psi,0}(\mathsf{c}_1,\mathsf{c}_2)+\epsilon\,D_{\psi,1}(\mathsf{c}_1,\mathsf{c}_2)+\Or(\epsilon^2)}
\end{equation}
with
\begin{equation}\label{eq:Dpsi0}
D_{\psi,0}(\mathsf{c}_1,\mathsf{c}_2)=\frac{\kappa_1^3}{12\pi}+D_{0}(\mathsf{c}_1,\mathsf{c}_2)
\end{equation}
and 
\begin{equation}\label{eq:Dpsi1}\fl
\eqalign{D_{\psi,1}(\mathsf{c}_1,\mathsf{c}_2)&=D_{1}(\mathsf{c}_1,\mathsf{c}_2)
 +
\frac{1}{72\pi}\kappa_1^3\left(8-3\gamma_{\mathrm{E}}-3\ln(\kappa_1^2/\pi)\right)\cr &\quad\strut +\frac{\normf_1\kappa_1^2}{8\pi}\frac{n+2}{n+8}\Bigg(\sum_{\sigma,\rho=0}^2\frac{\kappa_1^{2\sigma-1}P_{\mathsf{c}_1,\mathsf{c}_2}^{(\sigma,\rho)}J_{\mathsf{c}_1,\mathsf{c}_2}^{\rho}}{(\mathsf{c}_1^2+\kappa_1^2)(\mathsf{c}_2^2+\kappa_1^2)}+\kappa_1\sum_{j=1}^2\frac{\mathsf{c}_j\ln\mathsf{c}_j}{\mathsf{c}_j^2+\kappa_1^2}\cr &\quad\strut+\frac{\pi}{\normf_1}\Delta_{1,1,1,1}(\mathsf{c}_1,\mathsf{c}_2)-\frac{\pi}{2} -\frac{1-\ln 4}{2}\kappa_1\frac{(\mathsf{c}_1+\mathsf{c}_2)(\mathsf{c}_1\mathsf{c}_2+\kappa_1^2)}{(\mathsf{c}_1^2+\kappa_1^2)(\mathsf{c}_2^2+\kappa_1^2)}\Bigg).}
\end{equation}

\begin{figure}[htbp]
\begin{center}
\includegraphics[width=0.75\textwidth]{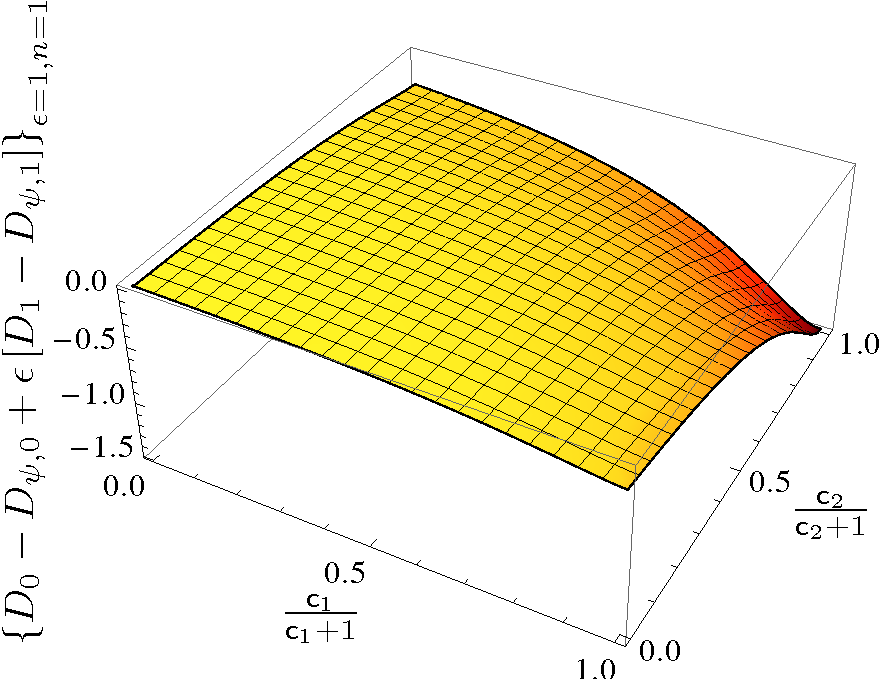}
\caption{Contribution from the lowest mode $m=1$ to the $\epsilon$~expansion of the scaling function $D(\mathsf{c}_1,\mathsf{c}_2)$ for $n=1$, evaluated at $\epsilon=1$. This quantity is given by $\{D_{0}(\mathsf{c}_1,\mathsf{c}_2)-D_{\psi,0}(\mathsf{c}_1,\mathsf{c}_2)+\epsilon [D_{1}(\mathsf{c}_1,\mathsf{c}_2)-D_{\psi,1}(\mathsf{c}_1,\mathsf{c}_2)]\}_{\epsilon=1,n=1}$.\label{fig:DpsiDdiff}}
\end{center}
\end{figure}

To visualise the result, we have plotted in figure~\ref{fig:DpsiDdiff} the difference between the $\Or(\epsilon)$ expression~\eqref{eq:Depsexp} for $D(\mathsf{c}_1,\mathsf{c}_2)$ and its analogue~\eqref{eq:fpsiresdiff} for $D_\psi(\mathsf{c}_1,\mathsf{c}_2)$ one obtains in the scalar case $n=1$ upon evaluation at $\epsilon=1$. This difference corresponds to the  contribution from the lowest mode $m=1$ to the series  expansion of $D(\mathsf{c}_1,\mathsf{c}_2)$ to $\Or(\epsilon)$. It vanishes as $(\mathsf{c}_1,\mathsf{c}_2)$ approaches the origin and reaches its minimum at the fixed point $(\mathsf{c}_1,\mathsf{c}_2)=(\infty,\infty)$ corresponding to large-scale Dirichlet boundary conditions at both boundary layers $\mathfrak{B}_1$ and $\mathfrak{B}_2$.
%
%

Combining equations~\eqref{eq:Dvarphidef} and \eqref{eq:fpsiresdiff}, we can write the result of the modified RG-improved perturbation theory used in this subsection as
\begin{equation}\label{eq:Ddec}
D(\mathsf{c}_1,\mathsf{c}_2)=D_\varphi(\mathsf{c}_1,\mathsf{c}_2)+D_\psi(\mathsf{c}_1,\mathsf{c}_2),
\end{equation}
where $D_\varphi$ stands for the expression defined through the right-hand side of equation~\eqref{eq:Dvarphidef} in conjunction with equation~\eqref{eq:Omegares}, while $D_\psi$ represents the $\Or(\epsilon)$ series expansion in equation~\eqref{eq:fpsiresdiff} with the expansion coefficients given by equations~\eqref{eq:Dpsi0} and  \eqref{eq:Dpsi1}, respectively. Note that the result must be utilised with care. It is not suitable for direct evaluation at $d=3$ ($\epsilon=1$). We defer a discussion of this and related issues to section~\ref{sec:results}.

\subsubsection{Consistency with fractional $\epsilon$ expansion for $\mathsf{c}_1=\mathsf{c}_2=0$ and Ginzburg-Levanyuk criterion}

Our motivation  for working out the modified RG-improved perturbation theory described in section~\ref{sec:consistency} was the goal to achieve consistency of the theory for general nonnegative values  of the enhancement variables $c_1$ and $c_2$ with the approach used  in references~\cite{DGS06} and \cite{GD08} to study the zero-mode case $c_1=c_2=0$ of critical surface enhancements. Since the former approach reduces to the latter when $c_1=c_2$, it is clear that consistency is ensured and the fractional $\epsilon$ expansion~\eqref{eq:Delspsp} must be recovered. To see this explicitly from the results of the previous subsection, recall that $\kappa_1$ vanishes as $\mathsf{c}_1+\mathsf{c}_2\to 0$. Therefore, $D_{\psi,0}(\mathsf{c}_1,\mathsf{c}_2)$ and 
$D_{\psi,1}(\mathsf{c}_1,\mathsf{c}_2)$ approach the limiting values $D_{0}(0,0)$ and 
$D_{1}(0,0)$, respectively. Furthermore, the first term on the right-hand side of equation~\eqref{eq:Dvarphidef} yields the $\epsilon^{3/2}$ term according to equation~\eqref{eq:Dvarphi00}.

That the modified RG-improved perturbation theory is controlled for small $\epsilon$, irrespective of whether $c_1+c_2$ is positive or zero, can also be seen from a criterion of the Ginzburg type \cite{Lev59,Gin60}. Let $\hat{\gamma}^{(4)}(\bm{p}_1,\bm{p}_2,\bm{p}_3)\,(2\pi)^{d-1}\delta(\sum_{i=1}^4\bm{p}_i)$ be the Fourier transform of the effective four-point vertex $\gamma^{(4)}(\bm{y}_1,\ldots,\bm{y}_4)$. The analogue of equation~\eqref{eq:scfgammma2} tells us that $\hat{\gamma}^{(4)}(\bm{0},\bm{0},\bm{0}) \sim \,u^*\mu^{2\eta}L^{d-5+2\eta}$, where the proportionality factor is a function of the scaling variables $\mathsf{c}_1$ and $\mathsf{c}_2$.  Following a standard reasoning (used also by Sachdev in his study of finite-temperature crossovers near quantum critical points \cite{Sac97}), we can construct from $\hat{\gamma}^{(2)}(\bm{0})$, its derivative $\partial_{p^2}\hat{\gamma}^{(2)}|_{\bm{0}}$ and $\hat{\gamma}^{(4)}(\bm{0},\bm{0},\bm{0})$ the dimensionless interaction constant
\begin{equation}\label{eq:dimeffcc}
U\equiv \frac{\hat{\gamma}^{(4)}(\bm{0},\bm{0},\bm{0})}{[(\partial_{p^2}\hat{\gamma}^{(2)})_{p=0}]^{(d-1)/2}\,[\hat{\gamma}^{(2)}(\bm{0})]^{(5-d)/2}}.
\end{equation}
Since $\hat{\gamma}^{(2)}(\bm{0})$ is proportional to $\Omega(\mathsf{c}_1,\mathsf{c}_2)$, this measure of the strength of the nonlinearities in $\mathcal{H}_{\text{eff}}$  behaves for small $\epsilon$ as 
\begin{equation}
U \sim u^*\,\Omega^{-(5-d)}
\sim\cases{\epsilon^{(1-\epsilon)/2}& if $\mathsf{c}_1=\mathsf{c}_2=0$,\cr
\epsilon & if $\mathsf{c}_1+\mathsf{c}_2>0$.}
\end{equation}
Hence the Ginzburg-Levanyuk-type criterion $U\ll 1$ is satisfied when $\epsilon\ll 1$.

\section{Discussion of results and extrapolation to $d=3$ dimensions}\label{sec:results} 

\begin{figure}[htbp]
\centerline{\includegraphics[width=0.70\textwidth]{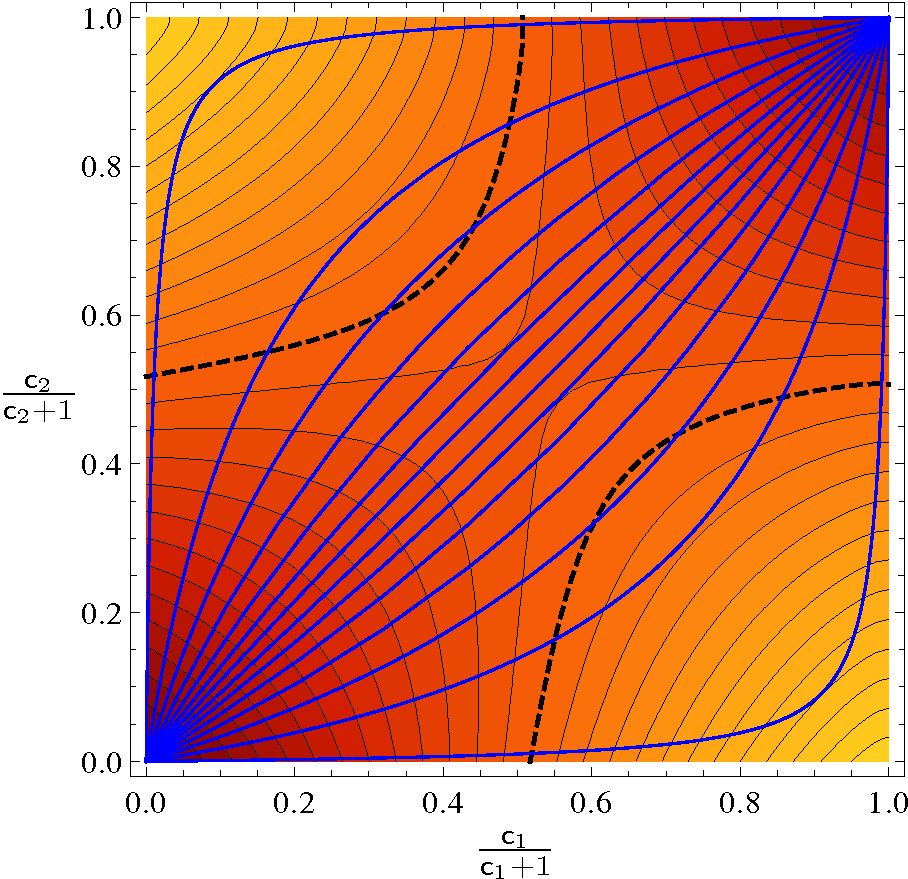}}
\caption{ Contour plot of the $\Or(\epsilon)$ result~\eqref{eq:Depsexp} for the scaling function $D_{n=1}(\mathsf{c}_1,\mathsf{c}_2)$ of the residual free-energy density $f_{\mathrm{res}}$ at the bulk critical point, evaluated at $\epsilon=1$. Along the axes the variables $x_j=\mathsf{c}_j/(\mathsf{c}_j+1)$ are plotted. For further explanations see main text. \label{fig:contourplotfres}}
\end{figure}

As our analysis in section~\ref{sec:freeen} has explicitly confirmed, the scaling functions $D(\mathsf{c}_1,\mathsf{c}_2)$ and $\mathcal{D}(\mathsf{c}_1,\mathsf{c}_2)$ of the residual free-energy density and the Casimir force have power series expansions in $\epsilon$ when $\mathsf{c}_1+\mathsf{c}_2>0$. Their coefficients of the corresponding series expansions to first order in $\epsilon$ are given by equations~\eqref{eq:D0}-\eqref{eq:D1} and follow from equation~\eqref{eq:mathcalD}, respectively. Assuming that $\mathsf{c}_1+\mathsf{c}_2>0$, the most elementary extrapolation to $d=3$ dimensions we can make is to set $\epsilon=1$ in these $\Or(\epsilon)$ expressions. In figure~\ref{fig:contourplotfres} 
a contour plot of the resulting extrapolation for the scaling function $D(\mathsf{c}_1,\mathsf{c}_2)$ is depicted as a function of the variables $\mathsf{c}_j/(\mathsf{c}_j+1)$ with $j=1,2$. A three-dimensional plot of this result can be found in figure~1 of reference~\cite{SD08}. The plotted function $D$ is negative in the vicinity of the 11 diagonal and changes sign across the thick dashed lines. The blue lines indicate paths generated by changes of $L$ at fixed values of the enhancement variables $c_1$ and $c_2$, i.e.\ RG flow lines. Note that the path associated with a given pair $(c_1,c_2)$ of enhancement variables does not depend on the value of the surface critical exponent $y_c>0$ appearing in the definition of the scaling variables $\mathsf{c}_j$ in equation~\eqref{eq:cscalvar} once the non-universal amplitude $E^*_c(u)$ has been fixed.
As is borne out by figure~\ref{fig:contourplotfres},  choices of $(c_1,c_2)$ exist for which these paths start in the region $D<0$, enter the region of positive values as $L$ increases and finally return back to the region $D<0$. Hence one expects that crossovers of the Casimir force from attractive to repulsive and back to attractive behaviours should occur as a function of $L$.

\begin{figure}[htbp]
\centerline{\includegraphics[width=0.75\textwidth]{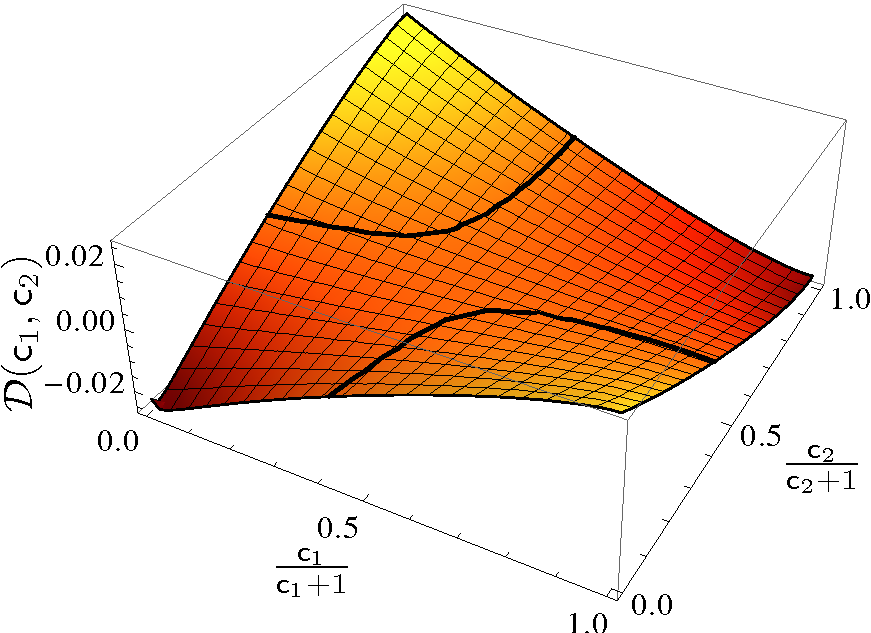}}
\caption{Three-dimensional plot of the $\Or(\epsilon)$ result for the scaling function $\mathcal{D}_{n=1}$ [equation~\eqref{eq:mathcalD}] of the Casimir force at $T_{\mathsf{c},\infty}$, evaluated at $\epsilon=1$. The axes are chosen as in figure~\ref{fig:contourplotfres}. The thick dashed lines mark the locations of zeros of $\mathcal{D}$.\label{fig:FC}}
\end{figure}

\begin{figure}[htbp]
\centerline{\includegraphics[width=0.70\textwidth]{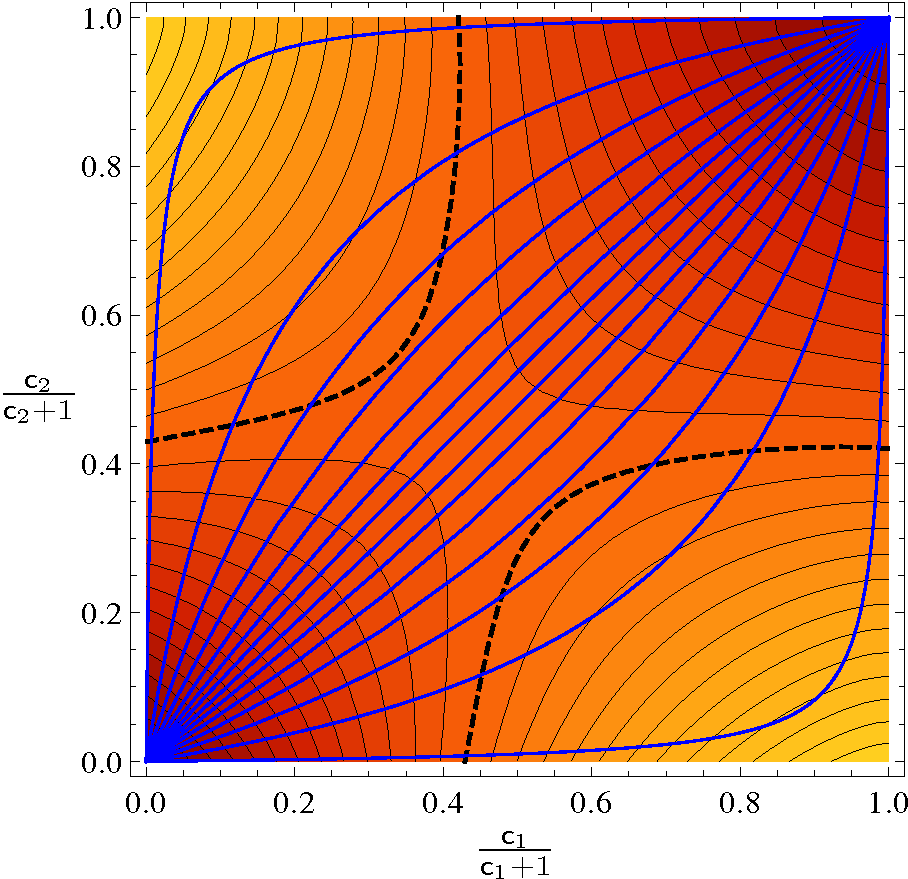}}
\caption{Contour plot of the $\Or(\epsilon)$ result for the scaling function $\mathcal{D}_{n=1}$ [equation~\eqref{eq:mathcalD}] of the Casimir force at $T_{\mathsf{c},\infty}$, evaluated at $\epsilon=1$. The axes are chosen as in figures~\ref{fig:contourplotfres}. The thick lines mark the locations of zeros of $\mathcal{D}$. The blue lines indicate the flow induced by changes of $L$ at fixed surface enhancements $(c_1,c_2)$\label{fig:contourplotFC}}
\end{figure}

The result for the scaling function $\mathcal{D}_{n=1}(\mathsf{c}_1,\mathsf{c}_2)$ of the Casimir force that follows  from this extrapolation for $D_{n=1}(\mathsf{c}_1,\mathsf{c}_2)$ via equation~\eqref{eq:mathcalD} is displayed as a three-dimensional plot in figure~\ref{fig:FC} 
%
and  as contour plot in figure~\ref{fig:contourplotFC}. 
%

To obtain these plots, we proceeded as follows. We substituted the extrapolated expression $[D_0+\epsilon D_1]_{\epsilon=1}$ for the scaling function $D$ in equation~\eqref{eq:mathcalD}.  For the pre-factor  $d-1+y_c$ we used the value $2.718$ corresponding to Hasenbusch's recent Monte Carlo estimate $y_c(n{=}1,d{=}3)\simeq  0.718(2) $  \cite{Has11}. Let us note that
some earlier Monte Carlo calculations led to significantly larger values for this exponent \cite{BL84,LB90a,VRF92}, but others and more recent  ones  \cite{RDW92,RDWW93,PS98,DBN05} yielded similar numbers.  Field-theory estimates based on the $\epsilon$ expansion to second order and the massive field-theory approach to two-loop order \cite{DS94,DS98} gave $y_c(1,3)\simeq 0.75 $ and $y_c(1,3)\simeq 0.85 $, respectively. For a detailed list of Monte Carlo and field-theory estimates for $y_c(1,3)$ the reader is referred to reference~\cite{Has11}. Choosing a somewhat larger value of $y_c(1,3)$ such as the quoted field-theory estimates would lead to small quantitative, but no qualitative changes in 
figures~\ref{fig:FC} and \ref{fig:contourplotFC}.

The behaviour of the plotted function $\mathcal{D}$ is qualitatively similar to that of $D$. As is obvious from figures~\ref{fig:FC} and \ref{fig:contourplotFC}, the critical Casimir force does exhibit the anticipated crossovers from attraction to repulsion and back to attraction for appropriate choices of $(c_1,c_2)$. According to a theorem for systems satisfying reflection positivity \cite{Bac06}, the Casimir force cannot become repulsive for equal enhancements $c_1=c_2$. This conforms with the fact that the Casimir force, by continuity, is attractive (i.e.\ $\mathcal{D}<0$)  in the vicinity of the $11$ diagonal. By choosing $(c_1,c_2)$ sufficiently away from the $11$~diagonal, one can ensure that a double sign change of $\mathcal{F}$ occurs as $L$ increases. 
A fascinating consequence is that for such choices of $(c_1,c_2)$, critical values $L=L_0(c_1,c_2)$ of the film thickness exist for which the critical Casimir force vanishes.

\begin{figure}[htbp]
\centerline{\includegraphics[width=0.70\textwidth]{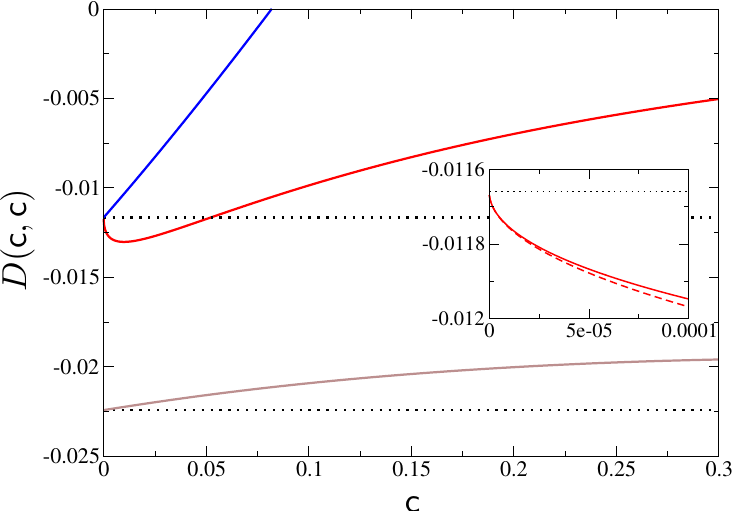}}
\caption{Extrapolated $\Or(\epsilon)$ results $[D_0+D_1]_{n=1}(\mathsf{c},\mathsf{c})$ (red curve) and $[D_{\psi,0}+D_{\psi,1}]_{n=1}(\mathsf{c},\mathsf{c})$ (blue curve); cf.\ equations~\eqref{eq:Depsexp} and \eqref{eq:fpsiresdiff}. The brown line is the function $D_{\text{app}}$ specified in equation~\eqref{eq:Dapp}. The asymptotic small-$\mathsf{c}$ form~\eqref{eq:Depsexpsmallc} of the red curve is plotted as dashed red line in the in-set. The dotted horizontal lines indicate the limiting coinciding values of the two red and blue curves, and that of the brown curve, respectively. For further explanation, see main text. \label{fig:Ddiagsmallc}}
\end{figure}

In the extrapolations for $D_{n=1}(\mathsf{c}_1,\mathsf{c}_2)$ and $\mathcal{D}_{n=1}(\mathsf{c}_1,\mathsf{c}_2)$ presented above, we tacitly assumed that $c_1+c_2>0$.  Knowing that the $\epsilon$~expansion breaks down when $c_1=c_2=0$ (becoming fractional), we may expect to see indications of this breakdown in the behaviour of these extrapolations near the origin. This is indeed the case, though only very close to it. To demonstrate this, we depict in figure~\ref{fig:Ddiagsmallc} the values $[D_0+\epsilon\,D_1]_{n=1,\epsilon=1}$ of the na{\"\i}vely extrapolated scaling function~\eqref{eq:Depsexp}  on the diagonal $\mathsf{c}_1=\mathsf{c}_2\equiv\mathsf{c}$. 

This function $[D_0+D_1]_{n=1}(\mathsf{c},\mathsf{c})$  has an infinite slope at $\mathsf{c}=0$. Its asymptotic behavior near the origin can be determined analytically in a straightforward fashion. One finds
\begin{equation}\label{eq:Depsexpsmallc}
\eqalign{[D_0+D_1]_{n=1}(\mathsf{c},\mathsf{c})&\mathop{\approx}\limits_{\mathsf{c}\to 0}-\frac{\pi^2}{1440}+a_1(1)-\frac{\pi}{144}\,\sqrt{2\mathsf{c}}+\Or(\mathsf{c})\cr &=-0.0116593-0.0308534\,\sqrt{\mathsf{c}}+\Or(\mathsf{c}).}
\end{equation}
The in-set in figure~\ref{fig:Ddiagsmallc} shows a comparison of $[D_0+D_1]_{n=1}(\mathsf{c},\mathsf{c})$ and its asymptotic form~\eqref{eq:Depsexpsmallc} for small $\mathsf{c}$. The term $\propto \mathsf{c}^{1/2}$ in equation~\eqref{eq:Depsexpsmallc} results from that part of the two-loop contribution $f^{[2]}_{\text{res}}$ whose mode summation $\sum_{m_1,m_2}$ is restricted to $m_1=1$ in either one of the two loops and to $m_2>1$ in the respective other. The easiest way to recover it is to expand $D_\varphi(\mathsf{c},\mathsf{c})|_{n=1}$ to $\Or(\epsilon)$. The coefficient of the $\Or(\epsilon)$ term is precisely the contribution $\propto \sqrt{\mathsf{c}}$ in equation~\eqref{eq:Depsexpsmallc}. It is a consequence of the (spurious) infrared singularity one encounters in this expansion due to the $m=1$ mode with $\kappa_1(\mathsf{c},\mathsf{c})\to 0$, and would imply an infrared singular derivative $\partial_{\mathsf{c}}$ of the function~\eqref{eq:Depsexpsmallc} at $\mathsf{c}=0$. This is a spurious infrared singularity since the system is not expected to exhibit critical behaviour at $T-T_{c,\infty}=\mathsf{c}_1=\mathsf{c}_2=0$ when $L<\infty$. The extrapolated scaling function $[\mathcal{D}_0+\epsilon\,\mathcal{D}_1]_{n=1,\epsilon=1}$ of the Casimir force behaves analogously near the origin, as should be clear from equation~\eqref{eq:mathcalD}.

To illustrate that the spurious behaviour of $[D_0+D_1]_{n=1}(\mathsf{c},\mathsf{c})$ near the origin is due to the zero-mode contribution, we also display the function $[D_{\psi,0}+D_{\psi,1}]_{n=1}(\mathsf{c},\mathsf{c})$ in figure~\ref{fig:Ddiagsmallc}. The latter function  is regular at $\mathsf{c}=0$ and has a finite positive slope there, since it does not involve the zero mode $m_1=1$. Its value at $\mathsf{c}=0$ agrees with that of $[D_{0}+D_{1}]_{n=1}$.

The remaining brown curve in figure~\ref{fig:Ddiagsmallc} represents  an extrapolation based on equation~\eqref{eq:Ddec}, namely the function 
\begin{equation}\label{eq:Dapp}
D_{\text{app}}(\mathsf{c},\mathsf{c})\simeq \Big\{D_{\psi,0}(\mathsf{c},\mathsf{c})+\epsilon D_{\psi,1}(\mathsf{c},\mathsf{c})-(A_3/3)\,[\Omega(\mathsf{c},\mathsf{c};\epsilon)]^{3/2}\Big\}_{n=1,\epsilon=1},
\end{equation}
This means that we have discarded the two-loop contribution to $D_\varphi$ and replaced the pre-factor $A_{d-1}/(d-1)$ of the one-loop term by its value at $d=4$. To understand the rationale for this approximation, one should note the following. If $\mathsf{c}_1=\mathsf{c}_2=0$, then this term reduces precisely to the $a_{3/2}(n)\epsilon^{3/2}$ contribution to $D(0,0)$. If we expand, on the other hand, this choice of $D_\varphi$ to linear order in $\epsilon$ when $\mathsf{c}_1+\mathsf{c}_2>0$ and substitute it into $D_{\psi,0}+\epsilon D_{\psi,1}+D_\varphi$, we recover the $\Or(\epsilon)$ result $D_{0}+\epsilon D_{1}$. 

Hence, in the limit $\mathsf{c}\to0$, $D_{\text{app}}(\mathsf{c},\mathsf{c})$ approaches the value one obtains for $\Delta^{(\mathrm{sp},\mathrm{sp})}|_{n=1,\epsilon=1} $  by evaluating the small-$\epsilon$ expansion~\eqref{eq:Delspsp} at $\epsilon=1$. Furthermore, its derivative $\partial_\mathsf{c}D_{\text{app}}$ is finite at $\mathsf{c}=0$. Despite these appealing features, we refrain from showing a plot of this function for all $\mathsf{c}\in(0,\infty)$ because away from the origin this function deviates considerably from the $\Or(\epsilon)$ extrapolation $[D_0+D_1]_{n=1}$. The reason is that  $\Omega(\mathsf{c}_1,\mathsf{c}_2;\epsilon=1)$ becomes fairly large for large $\mathsf{c}_1+\mathsf{c}_2$ (see figure~\ref{fig:Omega}) and the difference between $(A_3/3)\Omega(\mathsf{c},\mathsf{c};\epsilon)$ and its extrapolated $\Or(\epsilon)$ expansion $(1+\epsilon\partial_\epsilon)(A_3/3)\Omega(\mathsf{c},\mathsf{c};0)|_{\epsilon=1}$ is not small. Away from the origin, we therefore do not consider $D_{\text{app}}$ to be superior to the na{\"\i}ve extrapolation  $D_0+D_1$.

As we mentioned earlier, the pre-factor $A_{d-1}/{(d-1)}$, has a UV pole at $d=3$. Being guided by the aim to achieve consistency with the small-$\epsilon$ expansion for $\mathsf{c}_1=\mathsf{c}_2=0$, we expanded it in $\epsilon$ and hence could replace it by its value at $d=4$ at the order of our calculation. In an extrapolation based on equation~\eqref{eq:Ddec} one might be tempted to use the value of the UV finite one-loop term of $D_\varphi$ one obtains in the manner described below equation~\eqref{eq:Dvarphi00} by subtracting the pole term $\propto (d-3)^{-1}$ from $-\frac{A_{d-1}}{d-1}[\Omega(\mathsf{c}_1,\mathsf{c}_2,\epsilon=1)]^{(d-1)/2}$  and evaluating the difference at $d=3$. However, such an approach would mix the RG approach based on the $\epsilon$ expansion we used for $f_\psi$ with elements of a fixed-dimension RG approach. The corresponding extrapolation would again lead to substantial differences from the $\Or(\epsilon)$ extrapolation $D_0+D_1$ for large values of $\mathsf{c}_1+\mathsf{c}_2$. We feel that such attempts to evaluate $D_\varphi$ directly at $d=3$ should be based on a  RG approach in fixed dimension $d=3$. Such an approach is known to involve the study of the theory away from the bulk critical point \cite{Par80,ZJ02}. Furthermore,  the surface enhancement variables $c_1$ and $c_2$ provide two mass parameters in addition to the bulk correlation length one has to deal with. A corresponding massive RG approach in fixed dimension $d=3$ based on references \cite{DS94} and \cite{DS98} --- or combined with elements of the strategy followed in reference~\cite{SD89} ---  is technically very demanding and beyond the scope of the present work, albeit conceivable.

\begin{figure}[htbp]
\centerline{\includegraphics[width=0.70\textwidth]{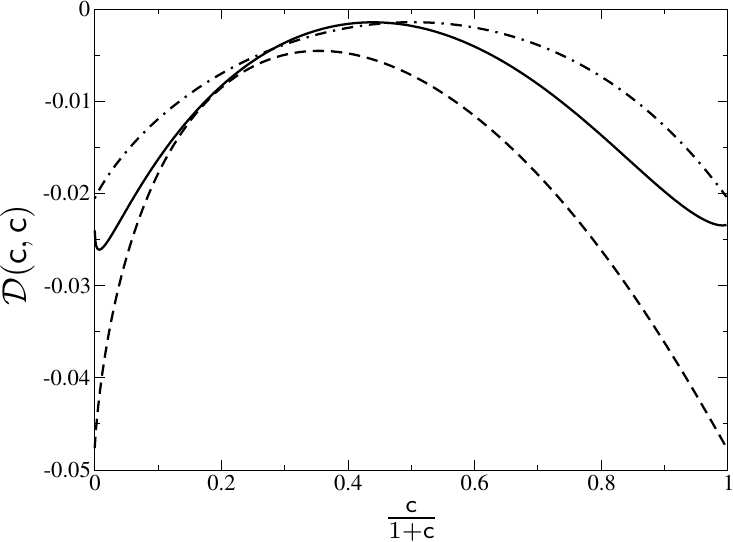}}
\caption{Comparison of the scaling functions $\mathcal{D}_{\text{Gau{\ss}}}(\mathsf{c},\mathsf{c};d)$ with $d=3$ (dashed) and $d=4$ (dash-dotted) from equation~\eqref{eq:mathcalDGauss} with the extrapolated function $\mathcal{D}_{n=1}(\mathsf{c},\mathsf{c})$ displayed in figures~\ref{fig:FC} and \ref{fig:contourplotFC} (full line).
\label{fig:forcediag}}
\end{figure}

It is instructive to compare the extrapolated scaling function $\mathcal{D}_{n=1}$ of the critical Casimir force displayed in  figures~\ref{fig:FC} and \ref{fig:contourplotFC} with their exact analogues 
\begin{equation}\label{eq:mathcalDGauss}
\mathcal{D}_{\text{Gau{\ss}}}(\mathsf{c}_1,\mathsf{c}_2;d)=(d-1+\mathsf{c}_1\partial_{\mathsf{c}_1}+\mathsf{c}_2\partial_{\mathsf{c}_2})f^{[1]}_{\text{res}}(1;0,\mathsf{c}_1,\mathsf{c}_2)/n
\end{equation}
of the Gaussian theory in $d=3$ and $d=4$ dimension. Here, the one-loop term~\eqref{eq:fres1}  must be substituted for $f^{[1]}_{\text{res}}$. A plot of these functions on the diagonal $\mathsf{c}_1=\mathsf{c}_2\equiv\mathsf{c}$ is shown in figure~\ref{fig:forcediag}. 

As one sees, the extrapolation $\mathcal{D}_{n=1}(\mathsf{c},\mathsf{c})$  lies  for most values of $\mathsf{c}$ below and above the Gaussian scaling functions~\eqref{eq:mathcalDGauss} with $d=4$ and $d=3$, respectively. Having already discussed the spurious small-$\mathsf{c}$ behaviour of the extrapolation, let us add a comment about the limiting behaviour at large $\mathsf{c}$. It has been known for long that deviations of the surface enhancement variables $c_j$ from their fixed point values $c_j=\infty$ associated with the ordinary surface transition correspond to irrelevant surface scaling fields (``extrapolation lengths'') whose RG eigenvalues  are given by the momentum dimension $-1$ \cite{DDE83,Die86a}. Taking into account the corrections to scaling $\propto L^{-1}$ implied by them is crucial for a proper analysis of Monte Carlo simulation results for  Casimir forces \cite{Huc07,VGMD07,VGMD09,GMHNB09,Has10b}. Our small-$\epsilon$ results enable us to explicitly verify the presence of such corrections to scaling. In order to comply with a correction to scaling to the residual free energy that is down by a factor $1/L$, the scaling function $D(\mathsf{c}_1,\mathsf{c}_2)$ ought to exhibit the asymptotic behaviours 
\begin{equation}\label{eq:Dlargec1c2}
nD(\mathsf{c}_1,\mathsf{c}_2)\mathop{\approx}\limits_{\mathsf{c}_1,\mathsf{c}_2\to\infty}\Delta^{(\mathrm{O},\mathrm{O})}+\Big(\mathsf{c}_1^{-1/y_c}+\mathsf{c}_2^{-1/y_c}\Big)\,\Delta^{(\mathrm{O},\mathrm{O})}_1
\end{equation}
and
\begin{equation}\label{eq:largec2}
nD(0,\mathsf{c}_2)\mathop{\approx}\limits_{\mathsf{c}_2\to\infty}\Delta^{(\mathrm{sp},\mathrm{O})}+\mathsf{c}_2^{-1/y_c}\,\Delta^{(\mathrm{sp},\mathrm{O})}_1.
\end{equation}
Inspection of our results given in equations~\eqref{eq:D0}--\eqref{eq:D1} reveals that contributions $\sim \epsilon\ln(\mathsf{c}_j)/\mathsf{c}_j$ originate from $D_1$ which can be exponentiated in accordance with these limiting forms. The universal amplitudes $\Delta^{(\mathrm{O},\mathrm{O})}_1$ and $\Delta^{(\mathrm{sp},\mathrm{O})}_1$ are found to have the $\epsilon$ expansions
\begin{equation}
\eqalign{\Delta^{(\mathrm{O},\mathrm{O})}_1=&\frac{\pi^2}{480}+\epsilon \frac{\pi^2}{960}\Bigg[3\gamma_E-\frac{17}{3}+\ln\frac{4}{\pi} -\frac{180}{\pi^{4}}\,\zeta'(4)\cr &\strut - \frac {n + 2} {n + 8}\Big (\gamma_E-\frac{1}{2}+\ln 4\Big)\Bigg]+\Or(\epsilon^2)}
\end{equation}
and
\begin{equation}
\eqalign{\Delta^{(\mathrm{sp},\mathrm{O})}_1=&-\frac{7\pi^2}{3840}+\epsilon\frac{\pi^2}{7680}\Bigg[\frac{35}{3}-7\gamma_E-12\ln 2-7\ln\pi+\frac{1260}{\pi^4}\zeta'(4)\cr &\strut +\frac{n+2}{n+8}(7\gamma_E-19+14\ln 2)\Bigg]+\Or(\epsilon^2),}
\end{equation}
respectively. Let us recall that the behaviour of the extrapolation $\mathcal{D}_{n=1}$ plotted in figure~\ref{fig:forcediag} was obtained by substituting $[D_0+\epsilon D_1]_{\epsilon=1}$ into equation~\eqref{eq:mathcalD}. Therefore, the plotted extrapolated function  does not approach its limiting value at $\mathsf{c}=\mathsf{\infty}$ as $ \mathsf{c}^{-1/y_c}$. This is because its asymptotic form contains the logarithmic term $\sim \epsilon\ln(\mathsf{c})/\mathsf{c}$ mentioned above, evaluated at $\epsilon=1$. Hence, the deviation of the plotted extrapolation $\mathcal{D}_{n=1}$ from its value at $\mathsf{c}=\infty$  varies (incorrectly) as $\sim \ln(\mathsf{c})/\mathsf{c}$. For conciseness, we have refrained from designing alternative extrapolations into which the correct asymptotic large-$\mathsf{c}$ asymptotics is incorporated via  appropriate exponentiation ansatzes.

\section{Summary and Conclusions}\label{sec:sumconcl}
Considering an $O(n)$-symmetric  $\phi^4$ model in film geometry, we investigated the fluctuation-induced forces produced by thermal fluctuations between the boundary planes of a film that is held at its bulk critical point. We restricted our attention to the case of generic non-symmetry-breaking boundary conditions. The interaction constants $\cb_j$ of the $O(n)$ symmetric boundary terms $\cb_j \phi^2/2$  included in the Hamiltonian were chosen to correspond to non-supercritical surface enhancements but otherwise allowed to take arbitrary values. Together with the restriction to the bulk critical temperature $T_{c,\infty}$ (or, more generally, to temperatures $T\ge T_{\mathsf{c},\infty}$), this requirement guarantees that the $O(n)$ symmetry is neither explicitly nor spontaneously broken.

Whenever the symmetry $\bm{\phi}\to-\bm{\phi}$ is broken, the order-parameter profile $\langle\bm{\phi}\rangle$ does not vanish. As discussed in the introduction, this implies that a nonzero Casimir force is obtained even in mean-field theory --- i.e., in the absence of fluctuations. Thus in these cases the Casimir force is not fully fluctuation induced but contains a  non-fluctuating component. This applies, in particular, to binary liquid mixtures, for which beautiful direct measurements of the Casimir forces were accomplished recently \cite{HHGDB08,GMHNB09,NHB09}. A key motivation for excluding such symmetry breakdowns in the present investigation  was the intention to keep the thermodynamic Casimir forces entirely fluctuation induced, making them share this crucial feature with the original QED Casimir force \cite{BMM01,Cas48}.

Our analysis was based on the field-theoretic RG approach in $4-\epsilon$ dimensions. The use of the $\epsilon$ expansion in studies of critical and near-critical Casimir forces has shortcomings that go beyond the familiar ones known from its application to critical behaviour in bulk and semi-infinite systems. The series expansions it yields for bulk and surface critical exponents as well as for other properties at bulk critical points --- Taylor expansions in $\epsilon$ --- are asymptotic. Naive extrapolations of such expansions to low orders in $\epsilon$ are not normally quantitatively reliable. However, quantitatively accurate results can be obtained by extending the expansions to sufficiently high orders, combining them with information from appropriate large-order results and using appropriate Borel-Pad\'e resummation techniques \cite{GZJ98,ZJ02,PV02,KS-F01}. The additional complication one encounters in the case of films is that $\epsilon=4-d$ ceases to be an appropriate expansion parameter when dealing with dimensional crossovers. (The adequate analogue of this parameter  for the study of critical behaviour at the transition temperature $T_{\mathsf{c},L}$ of a film of finite thickness $L$ and infinite extension in $d-1$ Cartesian directions would be $\epsilon_5\equiv 5-d$.) The limitations of the approach become particularly apparent when zero modes appear at zero-loop order at the bulk critical point, as happens for critical enhancement $c_1=c_2=0$ of both boundary planes. Whenever this occurs, the $\epsilon$ expansion breaks down at the bulk critical temperature $T_{\mathsf{c},\infty}$ and becomes fractional, involving also powers of $\ln\epsilon$ \cite{GD08,DGS06,DG09,Sac97}. By contrast, the $\epsilon$ expansion remains (formally) intact at $T_{\mathsf{c},\infty}$ provided at least one of the (non-supercritical) enhancements $c_j$ is subcritical, although a similar breakdown clearly must occur at a temperature below $T_{\mathsf{c},\infty}$.

Despite these deficiencies, the field-theoretic RG approach in $4-\epsilon$ dimensions has a number of appealing and valuable features. First of all, it lends itself to nontrivial checks of the scaling forms  the residual free energy and the Casimir force are expected to have according to finite-size scaling theory and the RG approach. Since we performed a two-loop calculation which gave $f_{\mathrm{res}}$ to order $u$, we were able to identify the $ \ln(\mu L)$ contributions  implied by the nontrivial $\Or(\epsilon)$ term of the surface crossover exponent $\Phi$ in the scaling arguments~\eqref{eq:scalvarc} and thus verify the scaling forms~\eqref{eq:scalvarc} and \eqref{eq:CFscf} to first order in $\epsilon$. Second, the fact that general properties such as the exponential decay in the disordered phase (see references~\cite[Sec.~VI]{KD92a} and \cite[Sec.~IV.C]{GD08}) and analyticity properties of the free-energy density $f$ (see  references~\cite[Sec.~VII]{KD92a} and \cite[Sec.~IV.C]{GD08}) must hold at any order of the small-$\epsilon$ expansion provides nontrivial checks of them. For the sake of clarity, it will be helpful to recall these analyticity properties. For given $n$ and sufficiently large $d$, the film undergoes a sharp transition to an ordered phase at a temperature $T_{\mathrm{c},L}<T_{\mathrm{c},\infty}$. In the case of non-supercritical surface enhancements considered here, there should also be no surface transitions of the film at temperatures $T\ge T_{\mathrm{c},\infty}$ (i.e.\ its surface layers should not exhibit long-range order). Consequently, the total free-energy density $f$ must be analytic in temperature at $T_{\mathrm{c},\infty}$. This requirement imposes conditions on the behaviour of the temperature dependent analogue of the scaling function of $f_{\mathrm{res}}$. We refer the reader to references ~\cite[Sec.~VII]{KD92a} and \cite[Sec.~IV.C]{GD08} for details. 

These analyticity requirements were found to be fulfilled by the two-loop $\epsilon$-expansion results for anti-periodic, $(\mathrm{O},\mathrm{O})$, and $(\mathrm{O},\mathrm{sp})$ boundary conditions, but violated at $\Or(\epsilon)$ for periodic and $(\mathrm{sp},\mathrm{sp})$ boundary conditions \cite{KD92a}. Extension of the small-$\epsilon$ expansion to  $\Or(\epsilon^{3/2})$ in the latter two cases via the effective-action approach outlined above increased the order at which a violation of  analyticity requirements occurred to $\epsilon^{3/2}$ \cite{GD08}. Third, our analyses in the latter reference and above yield evidence of a shift of the point $c_1=c_2=0$ at which the zero-mode field $\bm{\varphi}$ becomes critical at $T_{\mathrm{c},\infty}$ to an $L$-dependent location with $\mathsf{c}_1+\mathsf{c}_2< 0$. By analogy with the shift $T_{\mathrm{c},\infty}\to T_{\mathrm{c},L}$ of the transition temperature, such a shift is expected and in conformity with the required analyticity of $f$ at bulk criticality.

One crucial problem one is faced with in studies of critical behaviour and dimensional crossover in films is the difficulty of obtaining accurate results for such shifts in $d=3$ bulk dimensions by analytical methods. The reason is that fluctuations on all length scales between the microscopic one (lattice constant) and the bulk correlation length may contribute, as a result of which these shifts may acquire contributions that are non-analytic in the interaction constant \cite{Sym73,Sym73b,DS94,BBZ-J00}. Since we used RG-improved perturbation theory in $4-\epsilon$ dimension at $T_{\mathrm{c},\infty}$ to determine the effective action $\mathcal{H}_{\mathrm{eff}}$, our approach is not capable to capture such non-analytic contributions. It gave us a shift along the $c_1+c_2$ axis of the point where the $\bm{\varphi}$ component of the order parameter becomes critical that was proportional to $u^*$. This in turn implied a contribution to the critical Casimir force for critical surface enhancements $c_1=c_2=0$ of the form $(u^*)^{(3-\epsilon)/2}$. To make our results for the scaling functions $D(\mathsf{c}_1,\mathsf{c}_2)$ and $\mathcal{D}(\mathsf{c}_1,\mathsf{c}_2)$ consistent with the fractional $\epsilon$ expansions at $(\mathsf{c}_1,\mathsf{c}_2)=(0,0)$, we extended the effective-action approach of references~\cite{DGS06} and \cite{GD08} to nonzero values of $\mathsf{c}_1+\mathsf{c}_2$. We succeeded in achieving consistency with the results of these references for the case $\mathsf{c}_1=\mathsf{c}_2=0$ and the fractional $\epsilon$ expansions. Furthermore, the effective-action approach  turned out to be  well-behaved for small $\epsilon$. Nevertheless, its results must be taken with a grain of salt: they did not lend themselves easily to simple  extrapolations to $d=3$ dimensions that could be judged as clearly superior to those based on the conventional $\epsilon$ expansion.

Let us emphasise that our results exhibit a number of interesting and important \emph{qualitative} features which may be expected to persist in quantitatively more accurate investigations, irrespective of the accuracy of our extrapolation to $d=3$.
\begin{enumerate}
\item Generalising previous work \cite{KD91,KD92a,RS02,GD08,DG09}, they clearly demonstrate that  critical Casimir forces can be attractive or repulsive depending on the values of the surface enhancement variables $c_j$, even when the internal symmetry ($Z_2$ for $n=1$ and  $O(n)$ for $n>1$) is neither spontaneously nor explicitly broken. 

\item They are in full accordance with --- and hence corroborate --- the scaling forms~\eqref{eq:fressf} and \eqref{eq:CFscf} of the residual free energy and Casimir force, respectively. This means that the Casimir amplitude becomes scale dependent. 

\item As a further important observation we found that  crossovers from attractive to repulsive forces and vice versa can occur  if the surface interaction constants take appropriate values. In conjunction with (ii) this means that the critical Casimir force goes through zero at a certain thickness $L_0$.

\item  For the case of primary interest --- the Ising bulk universality class with short-range interactions and $d=3$ --- we saw that for properly chosen surface enhancement values $c_1$ and $c_2$, crossovers from attractive to repulsive and back to attractive behaviours should occur as $L$ increases. It would certainly be worthwhile to check this prediction along with (iii) by Monte Carlo calculations.

\item Finally, let us mention that crossover behaviours analogous to  (iii) and (iv) should also be possible for three-dimensional $O(n)$ systems with easy-axis spin anisotropies.  Semi-infinite systems of this kind are known to have anisotropic analogues of the isotropic special transition \cite{DE82,DE84}, characterised by the coincidence of the transition temperature at which the easy-axis component of the order parameter at the surface becomes critical with the transition temperature $T_{c,\infty}$ of the $n>1$ bulk system. Hence for finite thicknesses $L$,  the surface interaction constants associated with a given easy axis on one or both boundary planes can be critically enhanced. This suggests that the mentioned analogues of (iii) and (iv) should occur. (Films involving different directions of easy axes on the two boundary planes would require separate analyses.) 
\end{enumerate}

\appendix 

\section{Calculation of the one-loop free-energy density $f^{[1]}$ }
\label{app:f1}
\subsection{Calculation based on the Abel-Plana summation formula}\label{app:f1ap}
To compute $f^{[1]}$, we start from equation~\eqref{eq:f1loop}, differentiate this expression with respect to $\tb$ and compute the integral over $\bm{p}$. This gives
 \begin{equation}\label{eq:ftaumodesum}
\eqalign{ \partial_{\tb}f^{[1]}(L;\tb,\cb_1,\cb_2)&=\frac{n}{2}\sum_{m=1}^\infty\int_{\bm{p}}^{(d-1)}\frac{1}{\tb+p^2+k_m^2}\cr &=-L^{-(d-3)}\frac{n\pi\,K_{d-1}}{4\cos(d\pi/2)}\sum_{m=1}^\infty (\tb L^2+\kappa_m^2)^{(d-3)/2},}
 \end{equation}

The expression in the second line of equation~\eqref{eq:ftaumodesum} varies $ \sim L\int_0^\infty \rmd{k}\,(\mathring\tau+k^2)^{(d-3)/2}$ in the large-$L$ limit. We could subtract this bulk term to make the difference well defined. However, we are ultimately interested in the free-energy density $f^{[1]}(L;\tb,\cb_1,\cb_2)$. Term-by-term integration of the series for $\partial_{\tb}f^{[1]}$ increases the exponent $(d-3)/2$ of the series coefficients by one. Hence, subtracting the limiting bulk contribution $Lf_{\mathrm{b}}^{[1]}$ would not suffice to render UV finite expressions for the free-energy density $f^{[1]}(L;\tb,\cb_1,\cb_2)$ in dimensions $d\lesssim 4$. To obtain well-defined results for this quantity, we prefer to use analytical continuation in $d$, rather than making additional subtractions. We start by integrating the series in equation~\eqref{eq:ftaumodesum} term-wise, choosing the integration constant such that 
\begin{equation}\label{eq:f1modesum}
f^{[1]}(L;\tb,\cb_1,\cb_2)=-\frac{L^{-(d-1)}\,n\pi\,K_{d-1}}{2(d-1)\cos(d\pi/2)} \sum_{m=1}^\infty (\tb L^2+\kappa_m^2)^{(d-1)/2}.
\end{equation}

That the integration constant has been chosen correctly may not be obvious at this stage. However, once we will have analytically continued the above series~\eqref{eq:f1modesum} in $d$, one can confirm this choice by verifying that the resulting free-energy expressions reduce to known dimensionally regularised results at $\tb=0+$ in the limits $\cb_1,\cb_2\to 0 $ and $\cb_1,\cb_2\to \infty$. Furthermore, there are two other ways to arrive at expression~\eqref{eq:f1modesum}, both of which indicate its consistency with general rules of dimensional regularisation. The first is to use the representation 
\begin{equation}
\ln v=\int_0^\infty\frac{\rmd{t}}{t}\,\left(\rme^{-t}-\rme^{-vt}\right)
\end{equation}
for the logarithm in the second line of equation~\eqref{eq:f1loop}, perform the integrations and take into account that integrals of pure powers such as $\int_{\bm{p}}^{(d-1)}1$ vanish in dimensional regularisation. 
Another way is to insert the partial-$\bm{p}$ identity $(d-1)^{-1}\nabla_{\bm{p}}\cdot\bm{p}$ into the integrand of the integral $\int_{\bm{p}}^{(d-1)}$ in the last line of equation~\eqref{eq:f1loop} and integrate by parts, dropping the boundary terms at $p=\infty$. 

According to the result~\eqref{eq:f1modesum} we  must calculate --- and analytically continue --- a series of the form
\begin{equation}\label{eq:Sabdef}
S_{\mathcal{C}_1,\mathcal{C}_2}(a;b)\equiv \sum_{m=1}^\infty(\kappa_m^2+b^2)^a
\end{equation}
with $b^2=\tb L^2$, where $\kappa_m$ are the positive solutions to equation~\eqref{eq:transceq}. Such series are generalisations of Epstein-Hurwitz $\zeta$ functions (see e.g.\ reference~\cite{Eli95}). Let us assume that both $\mathcal{C}_j\ge 0 $ and  $\mathcal{C}_1+\mathcal{C}_2\ne0$. Then all $\kappa_m$, $m=1,2,\ldots,\infty$,  are positive, and as we showed in section~\ref{sec:backg}, non-degenerate. Hence the function $\partial_\kappa \ln R_{\mathcal{C}_1,\mathcal{C}_2}(\kappa)$ has a simple pole with residue $1$ at each of these $\kappa_m$. Since  $(\kappa^2+b^2)^a
$ is regular at $\kappa=\kappa_m$, the value of this function at $\kappa_m$ is given by the residue $\res_{\kappa=\kappa_m}[(\kappa^2+b^2)^a\, R'_{\mathcal{C}_1,\mathcal{C}_2}(\kappa)/ R_{\mathcal{C}_1,\mathcal{C}_2}(\kappa)]$. But at $\kappa=\kappa_m$ $ R'_{\mathcal{C}_1,\mathcal{C}_2}$ can be factorized as
\begin{equation}
 R'_{\mathcal{C}_1,\mathcal{C}_2}(\kappa_m)=2N_{\mathcal{C}_1,\mathcal{C}_2}(\kappa_m)/\normf_m=N_{\mathcal{C}_1,\mathcal{C}_2}(\kappa_m)\,g_{\mathcal{C}_1,\mathcal{C}_2}(\kappa_m),
\end{equation}
where 
\begin{equation}
N_{\mathcal{C}_1,\mathcal{C}_2}(\kappa)\equiv(\mathcal{C}_1 \mathcal{C}_2 - \kappa^2) \cos\kappa - (\mathcal{C}_1 + \mathcal{C}_2) \kappa\sin\kappa ,
\end{equation}
while
\begin{equation}\label{eq:gdef}
g_{\mathcal{C}_1,\mathcal{C}_2}(\kappa)\equiv 1+\frac{\mathcal{C}_1}{\mathcal{C}_1^2+\kappa^2}+\frac{\mathcal{C}_2}{\mathcal{C}_2^2+\kappa^2}.
\end{equation}
Thus we have  
\begin{equation}
\left(\kappa_m^2+b^2\right)^a=\res_{\kappa=\kappa_m}[(\kappa^2+b^2)^a\,\Upsilon_{\mathcal{C}_1,\mathcal{C}_2}(\kappa)]
\end{equation}
with
\begin{equation}
\Upsilon_{\mathcal{C}_1, \mathcal{C}_2}(\kappa)\equiv \frac{N_{\mathcal{C}_1, \mathcal{C}_2}(\kappa)\,g_{\mathcal{C}_1,\mathcal{C}_2}(\kappa)}{R_{\mathcal{C}_1,\mathcal{C}_2}(\kappa)}.
\end{equation}

\begin{figure}[htbp]
\centerline{\includegraphics[width=15em]{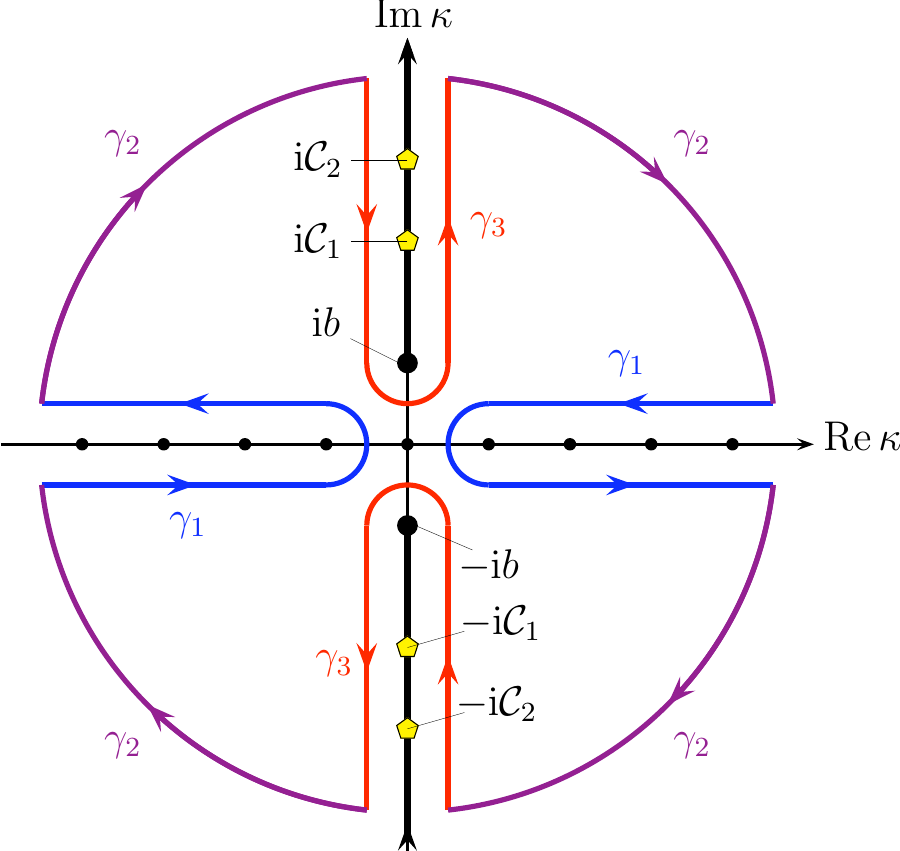}}
\caption{Paths in the complex $\kappa$-plane. }
\label{fig:Kleeblatt}
\end{figure}

Since this function and the one appearing in $S_{\mathcal{C}_1,\mathcal{C}_2}(a;b)$ are even in $\kappa$, we can extend the summation in equation~\eqref{eq:Sabdef} to all integer-valued $m\ne 0$, using $\kappa_{-|m|}=-\kappa_{|m|}$, and divide by $2$. Since $(\kappa^2+b^2)^a$ is regular at $\kappa_m$, each term of this series can be expressed as $(2\pi\rmi)^{-1}$ times an integral $\oint\rmd\kappa\, (\kappa^2+b^2)^a\,\Upsilon_{\mathcal{C}_1, \mathcal{C}_2}(\kappa)$ along a contour that passes once around the pole at the respective $\kappa_m$ in a counter-clockwise fashion and contains no other singularities. If $a <-1/2$, the series $S_{\mathcal{C}_1,\mathcal{C}_2}(a;b)$ converges and the integrand  of the contour integral decays sufficiently fast as $ \kappa\to \pm \infty\pm\rmi 0$, so that the union of all these contour integrals can be deformed into a path $\gamma_1$ encircling all poles $\kappa_m$ with $0\ne m\in\mathbb{Z}$ (see figure~\ref{fig:Kleeblatt}). We now add and subtract integrals along the paths  $\gamma_2$ and $\gamma_3$ depicted in this figure. Since the integrand is regular at all nonzero $\kappa$ inside the region bounded by the closed path $\gamma_1\cup\gamma_2\cup\gamma_3$, the integral along it is $-2\pi\rmi$ times the residue of the integrand at $\kappa=0$. Furthermore, the integral along $\gamma_2$ approaches zero as the radius of the circle on which  $\gamma_2$ is located tends to infinity. Hence we have
\begin{equation}\label{eq:Sabzwres}
\eqalign{S_{\mathcal{C}_1,\mathcal{C}_2}(a;b)&=\left(\frac{-1}{2}\res_{\kappa=0}-\frac{1}{4\pi\rmi}\int_{\gamma_3}\rmd{\kappa}\right)\,(\kappa^2+b^2
)^a\,\Upsilon_{\mathcal{C}_1, \mathcal{C}_2}(\kappa)\cr &=
\strut-\frac{1}{2}\,b^{2a}-\frac{1}{4\pi\rmi}\int_{\gamma_3}\rmd{\kappa}\,(\kappa^2+b^2
)^a\,\Upsilon_{\mathcal{C}_1, \mathcal{C}_2}(\kappa).}
\end{equation}

The factor $(\kappa^2+b^2)^a$ implies that the integrand has branch cuts from $\rmi b$ to $\rmi\infty$ and from $-\rmi b$ to $-\rmi \infty$. In addition, the function $g_{\mathcal{C}_1,\mathcal{C}_2}(\kappa)$ and hence $\Upsilon_{\mathcal{C}_1,\mathcal{C}_2}(\kappa)$ have poles at $\pm\rmi \mathcal{C}_1$ and $\pm\rmi \mathcal{C}_2$, marked by yellow pentagons in figure~\ref{fig:Kleeblatt}.  We move the path $\gamma_3$ infinitesimally close to the imaginary axes, passing around the poles $\pm\mathcal{C}_j$ along semicircles of radius $\delta\to0$. The portions of $\gamma_3$ with $\Im\,\kappa>0$ and $\Im\,\kappa<0$ give identical contributions. Taking into account that the limiting values of the power on the right and left rim of the branch cut are
\begin{equation}
\lim_{{\Re\, \kappa\to 0\pm\atop \Im\,\kappa>0}}(\kappa^2+b^2)^a=\left[(\Im\,\kappa\right)^2-b^2]^a\,\rme^{\pm\rmi\pi a},
\end{equation}
one arrives at
\begin{equation}\label{eq:Sabzwres2}
\eqalign{S_{\mathcal{C}_1,\mathcal{C}_2}(a;b)=&\strut-\frac{1}{2}\,b^{2a}-2\sin(\pi a)\,\mathcal{P}\int_b^\infty\frac{\rmd{t}}{\pi}\,(t^2-b^2)^a\,\Upsilon_{\mathcal{C}_1, \mathcal{C}_2}(\rmi t)\cr
&\strut-4\pi\rmi\sum_{j=1}^2\left(\mathcal{C}_j^2-b^2\right)^a\cos(a\pi)\res_{\kappa=\rmi\mathcal{C}_j}\Upsilon_{\mathcal{C}_1,\mathcal{C}_2}(\kappa).
}
\end{equation}
Here $\mathcal{P}\int_b^\infty\rmd{t}$ means the principal value $\lim_{\delta\to 0+}\left(\int_b^{\mathcal{C}_<-\delta}+\int_{\mathcal{C}_<+\delta}^{\mathcal{C}_>-\delta}+\int_{\mathcal{C}_>+\delta}^\infty\right)\rmd{t}$, where $\mathcal{C}_<$ and $\mathcal{C}_>$ are the smaller and larger one of $\mathcal{C}_1$ and $\mathcal{C}_2$, respectively, and $b$ is assumed to be smaller than $\mathcal{C}_<$. The contribution in the second line results from the integrals along the semicircles.

In order that the integral  $\mathcal{P}\int_b^\infty\rmd{t}$  exists, we must require  $a>-1$ (to ensure convergence at the lower limit of integration) in addition to the original condition $a<-1/2$.
In view of the behaviour of the integrand near the upper integration limit, we must have $a<-1/2$ (to guarantee convergence at the upper limit of integration). However, we can split off the limiting value $\lim_{\kappa\to\pm\rmi\infty} N_{\mathcal{C}_1,\mathcal{C}_2}(\kappa)/R_{\mathcal{C}_1,\mathcal{C}_2}(\kappa)=\mp\rmi$, obtaining
\begin{equation}\label{eq:NRdec}
\frac{N_{\mathcal{C}_1,\mathcal{C}_2}(\kappa)}{R_{\mathcal{C}_1,\mathcal{C}_2}(\kappa)}=\mp 2\rmi\left[\rme^{\mp2\rmi\kappa}\,\frac{\mathcal{C}_1\mp\rmi\kappa}{\mathcal{C}_1\pm\rmi\kappa}\,\frac{\mathcal{C}_2\mp\rmi\kappa}{\mathcal{C}_2\pm\rmi\kappa}-1\right]^{-1}\mp\rmi \quad\mbox{ for } \Im\,\kappa\gtrless 0.
\end{equation}
Since the first term vanishes at $\kappa=\rmi\mathcal{C}_j$, the contribution it yields  to the term in the second line of equation~\eqref{eq:Sabzwres2} vanishes, i.e.\ $\res_{\kappa=\rmi\mathcal{C}_j}[\Upsilon_{\mathcal{C}_1,\mathcal{C}_2}(\kappa)+\rmi\,g_{\mathcal{C}_1,\mathcal{C}_2}(\kappa)]=0$. The two contributions to $S_{\mathcal{C}_1,\mathcal{C}_2}(a;b)$  implied by the term $-\rmi$  in equation~\eqref{eq:NRdec} can be represented as an integral along the original path $\gamma_1$, namely
\begin{equation}\label{eq:Iabdef}\fl
I_{\mathcal{C}_1,\mathcal{C}_2}(a;b)=\frac{1}{4\pi} \int_{\gamma_1}\rmd{\kappa}\,\mbox{sgn}(\Im\,\kappa)\,g_{\mathcal{C}_1,\mathcal{C}_2}(\kappa)\,(\kappa^2+b^2)^a=\int_0^\infty\frac{\rmd{\kappa}}{\pi}\,g_{\mathcal{C}_1,\mathcal{C}_2}(\kappa)\left(\kappa^2+b^2\right)^a,
\end{equation}
where $\mbox{sgn}(x)$ means the sign function.
For $a<-1/2$, the latter integral is well defined and can be analytically computed. This gives
\begin{eqnarray}\label{eq:Iab}
I_{\mathcal{C}_1,\mathcal{C}_2}(a;b) &=&  b^{1+2a}\,\frac{\Gamma(-a-1/2)}{2\sqrt{\pi}\,\Gamma(-a)}+\frac{1}{2\cos(\pi a)}\sum_{j=1}^2\Bigg[\left(\mathcal{C}_j^2-b^2\right)^a\nonumber \\& &\strut-\frac{b^{1+2a}\sqrt{\pi}}{\Gamma(-a)\,\mathcal{C}_j}\;{}_2\tilde{F}_1{\left(1/2,1;a+3/2;b^2/\mathcal{C}_j^2\right)}\Bigg],
\end{eqnarray}
where ${}_2\tilde{F}_1$, the regularised hypergeometric function, is an entire function related to the standard hypergeometric function ${}_2F_1$ via
\begin{equation}
{}_2\tilde{F}_1(\alpha,\beta;\gamma;z)={}_2F_1(\alpha,\beta;\gamma;z)/\Gamma(\gamma).
\end{equation}
The expression on the right-hand side of equation~\eqref{eq:Iab} provides the analytic continuation of interest of the integral~\eqref{eq:Iabdef} to positive values of the exponent $a$. Our result for the series $S_{\mathcal{C}_1,\mathcal{C}_2}(a;b)$ 
thus becomes
\begin{equation}\fl\label{eq:Sabres}
S_{\mathcal{C}_1,\mathcal{C}_2}(a;b)=\strut-\frac{1}{2}\,b^{2a}+I_{\mathcal{C}_1,\mathcal{C}_2}(a;b)
-2\sin(\pi a)\int_b^\infty\frac{\rmd{t}}{\pi}\,\frac{g_{\mathcal{C}_1,\mathcal{C}_2}(\rmi t)\,(t^2-b^2)^a}{\frac{t+\mathcal{C}_1}{t-\mathcal{C}_1}\,\frac{t+\mathcal{C}_2}{t-\mathcal{C}_2}\,\rme^{2t}-1},
\end{equation}
where $I_{\mathcal{C}_1,\mathcal{C}_2}(a;b)$ represents the meromorphic function defined by the right-hand side of equation~\eqref{eq:Iab}.

Equation~\eqref{eq:Sabres} provides the appropriate generalisations of the results of reference~\cite{KD92a} for the special cases $(\mathcal{C}_1,\mathcal{C}_2)=(0,0)$, $(\infty,\infty)$ and $(0,\infty)$ to general nonnegative values of $\mathcal{C}_1$ and $\mathcal{C}_2$. To check the consistency with Krech and Dietrich's results, let us work out the asymptotic behaviours in these limits. Straightforward analysis yields
\begin{equation}\label{eq:SNN}\fl
S_{0,0}(a;b)=\frac{\Gamma(-a-1/2)}{2\sqrt{\pi}\,\Gamma(-a)}\,b^{2a+1}+\frac{1}{2}b^{2a}-2\sin(\pi a)\int_b^\infty\frac{\rmd{t}}{\pi}\,\frac{(t^2-b^2)^a}{\rme^{2t}-1},
\end{equation}
\begin{equation}\label{eq:SDN}
\eqalign{\lim_{\mathcal{C}_2\to\infty}\left[S_{0,\mathcal{C}_2}(a;b)-\frac{(\mathcal{C}_2^2-b^2)^a}{2\cos(\pi a)}\right]=&\frac{\Gamma(-a-1/2)}{2\sqrt{\pi}\,\Gamma(-a)}\,b^{2a+1}\cr &\strut +2\sin(\pi a)\int_b^\infty\frac{\rmd{t}}{\pi}\,\frac{(t^2-b^2)^a}{\rme^{2t}+1}}
\end{equation}
and
\begin{equation}\label{eq:SDD}
\lim_{\mathcal{C}_1,\mathcal{C}_2\to\infty}\left[S_{\mathcal{C}_1,\mathcal{C}_2}(a;b)-\sum_{j=1}^2\frac{(\mathcal{C}_j^2-b^2)^a}{2\cos(\pi a)}\right]=S_{0,0}(a;b)-b^{2a}.
\end{equation}

The right-hand sides of equations~\eqref{eq:SDD} and \eqref{eq:SDN} correspond to equations~(C5) and (C7) of reference~\cite{KD92a}, respectively, and our result~\eqref{eq:SNN} complies with the remark about the Neumann-Neumann case below its equation~(C5). To see this, one should identify  $b^2$ with $\tau L^2$ in their notation, multiply our result with the factor $L^{-2a}$ ($=L^{1-d}$) and take into account the relation  $\sin(\pi a)=-\pi/[\Gamma(-a)\,\Gamma(1+a)]$. The terms we have subtracted on the left-hand sides of equations~\eqref{eq:SDN} and \eqref{eq:SDD} approach infinity in the considered limits $\mathcal{C}_j\to\infty$. These do not appear if one sets $\mathcal{C}_2=\infty$ and $\mathcal{C}_1=\mathcal{C}_2=\infty$ from the outset because the operations of analytic continuation and of taking these limits do not commute. (As long as $a<0$, the subtracted terms approach zero as $\mathcal{C}_j\to\infty$.) Note also that these terms contribute only to the excess surface densities $f^{[1]}_{\mathrm{s},j}$, but not to $f^{[1]}_{\mathrm{res}}$.

We now insert the above results~\eqref{eq:Iab} and \eqref{eq:Sabres} into equation~\eqref{eq:f1modesum}, set $a=(d-1)/2$, and take the limit $b\to 0$ (i.e.\ $\tb\to0$). The $t$-integral of equation~\eqref{eq:Sabres} becomes
\begin{equation}\fl
\int_0^\infty\frac{\rmd{t}}{\pi}\,\frac{g_{\mathcal{C}_1,\mathcal{C}_2}(\rmi t)\,t^{d-1}}{\frac{t+\mathcal{C}_1}{t-\mathcal{C}_1}\,\frac{t+\mathcal{C}_2}{t-\mathcal{C}_2}\,\rme^{2t}-1}=\frac{1}{2}\int_0^\infty\frac{\rmd{t}}{\pi}\,t^{d-1}\frac{\rmd}{\rmd{t}}\ln\left[ 1-\frac{(t -\mathcal{C}_1) (t -\mathcal{C}_2)\, \rme^{-2 t}}{(\mathcal{C}_1+t ) (\mathcal{C}_2+t )}\right]
\end{equation}
and can be integrated by parts. An elementary change of variable $t\to p=t/L$ finally  leads us to the result
\begin{equation}\fl \label{eq:f1}
\eqalign{f^{[1]}(L;0,\cb_1,\cb_2)&=-\frac{n}{2}\,\frac{\pi\,K_{d-1}}{(d-1)\sin(\pi d)}\,(\cb_1^{d-1}+\cb_2^{d-1})\cr &\strut +
\frac{n}{2}\,K_{d-1}\int_0^\infty\rmd{p}\,p^{d-2}\ln\left[1-\frac{(p -\cb_1) (p -\cb_2)\, \rme^{-2 L p
   }}{(\cb_1+p ) (\cb_2+p )}\right],}
\end{equation}
which is in conformity with equations~\eqref{eq:fbc}, \eqref{eq:fsc1} and \eqref{eq:fres1}.

\subsection{Calculation based on the free propagator}\label{app:f1alt}

In this appendix we describe an alternative calculation of the dimensionally regularised one-loop free-energy density $f^{[1]}(L;0,\cb_1,\cb_2)$, which uses the result \eqref{eq:GLPc1c2explform} for the free propagator. We start from the one-loop expression 
\begin{equation}
\partial_{\tb}f^{[1]}(L;\tb,\cb_1,\cb_2)=\frac{n}{2}\int_{\bm{p}}^{(d-1)}\int_0^L\rmd{z}\,\hat{G}_L(\bm{p};z,z|\tb,\cb_1,\cb_2)
\end{equation}
for the energy density. Upon inserting the free propagator with $z=z'$ from equation~\eqref{eq:GLPc1c2explform}, the integration over $z$ can be performed using {\sc Mathematica} \cite{Mathematica7}. This gives 
 \begin{eqnarray}\fl
\partial_{\tb}f^{[1]}&= \frac{n}{2}\int_{\bm{p}}^{(d-1)}\frac{\sinh (L \varkappa_p ) [\varkappa_p ^2 (L (\cb_1+\cb_2)+1)-\cb_1
   \cb_2]+L \varkappa_p  (\cb_1 \cb_2+\varkappa_p ^2) \cosh (L \varkappa_p
   )}{2 \varkappa_p ^2 [(\cb_1 \cb_2+\varkappa_p ^2) \sinh (L \varkappa_p )+\varkappa_p
    (\cb_1+\cb_2) \cosh (L \varkappa_p )]}\nonumber\\ \fl
    &=\frac{n}{2}\int_{\bm{p}}^{(d-1)}\left[L\,I_{\mathrm{b}}(p,\tb) +I_{\mathrm{s}}(p,\tb,\cb_1)+I_{\mathrm{s}}(p,\tb,\cb_2)+I_{\mathrm{res}}(L;p,\tb,\cb_1,\cb_2)\right]
 \end{eqnarray}
 with the bulk, surface and $L$-dependent terms 
 \begin{equation}
 I_{\mathrm{b}}(p,\tb)=\frac{1}{2\varkappa_p},
 \end{equation}
  \begin{equation}
 I_{\mathrm{s}}(p,\tb,\cb_j)=\frac{1}{4\varkappa_p^2}\,\frac{\varkappa_p-\cb_j}{\varkappa_p+\cb_j}
 \end{equation}
 and
 \begin{equation}\label{eq:Ires}\fl
I_{\mathrm{res}}(L;p,\tb,\cb_1,\cb_2)= \frac{\cb_1 \cb_2
   (\cb_1 \cb_2 L+\cb_1+\cb_2)+L \varkappa_p ^4-\varkappa_p ^2 [
   (\cb_1^2+\cb_2^2)L+\cb_1+\cb_2]}{\varkappa_p  (\cb_1+\varkappa_p )^2
   (\cb_2+\varkappa_p )^2 \rme^{2 L \varkappa_p }-\varkappa_p(\varkappa_p ^2-\cb_1^2)
  (\varkappa_p ^2-\cb_2^2)}
 \end{equation}
of the integrand. Indefinite integrals $J_{\mathrm{b},\mathrm{s},\mathrm{res}}$ satisfying $\partial_{\tb}J_{\mathrm{b},\mathrm{s},\mathrm{res}}=I_{\mathrm{b},\mathrm{s},\mathrm{res}}$ can be determined in a straightforward fashion. One finds
\begin{equation}
J_{\mathrm{b}}(p,\tb)=\varkappa_p,\qquad J_{\mathrm{s}}(p,\tb,\cb_j)=\ln\big[\varkappa_p^{-1/2}(\cb_j+\varkappa_p)\big],
\end{equation}
and
\begin{equation}
J_{\mathrm{res}}(L;p,\tb,\cb_1,\cb_2)=\ln \left[1-\frac{(\varkappa_p -\cb_1) (\varkappa_p -\cb_2)\, \rme^{-2 L \varkappa_p
   }}{(\cb_1+\varkappa_p ) (\cb_2+\varkappa_p )}\right].
\end{equation}

Since $ J_{\mathrm{b}}$ reduces to $p$ for $\tb=0$,  the integral $\int_{\bm{p}}^{(d-1)} J_{\mathrm{b}}$ would yield the usual bulk divergence $\propto\Lambda^d$ if we regularised by cutting off the integral at $p=\Lambda$. It vanishes in the dimensional regularisation scheme because integrals of powers of $p$ are zero. Likewise, $\int_{\bm{p}}^{(d-1)} J_{\mathrm{s}}(p,0,0)$ would yield a surface divergence $\propto \Lambda^{d-1}\pmod{\ln \Lambda}$ if cut-off regularisation were used, but vanishes in our dimensionally regularised theory where only the $\cb_j$-dependent part contributes. This gives
\begin{equation}
\int_{\bm{p}}^{(d-1)}J_{\mathrm{s}}(p,0,\cb_j)=-K_{d-1}\,\frac{\pi}{(d-1)\sin(\pi d)}\,\cb_j^{d-1},
\end{equation}
where $K_{d-1}$ is defined in equation~\eqref{eq:Kddef}.

A straightforward combination of the foregoing results gives back equation~\eqref{eq:f1}.

\section{Computation of two-loop free-energy terms}\label{app:2loops}

Starting from equations~\eqref{eq:f2loop} and \eqref{eq:delta}, we first consider the contribution arising from the part proportional to $\delta_{m_1,m_2}$ of $\Delta_{m_1,m_1,m_2,m_2}(\mathcal{C}_1,\mathcal{C}_2)$. This involves the integral
\begin{equation}
\int_{\bm{p}}^{(d-1)}\frac{1}{p^2+k_m^2+\tb}=-A_{d-1}\,(k_m^2+\tb)^{(d-3)/2},
\end{equation}
where $A_d$ is defined in equation~\eqref{eq:Addef}. From the result we see that a first series we must evaluate is $\sum_{m=1}^\infty\normf_m(\kappa_m^2+b^2)^{d-3}$. This can be done along lines analogous to those followed in \ref{app:f1ap}. Since the coefficient $\normf_m$ can be expressed as $\normf_m=2/g_{\mathcal{C}_1,\mathcal{C}_2}(\kappa_m)$ according to equations~\eqref{eq:normf} and \eqref{eq:gdef}, the function $g_{\mathcal{C}_1,\mathcal{C}_2}$ drops out of the integrand of the contour integral. Thus the analogues of equations~\eqref{eq:Sabzwres} and \eqref{eq:Sabres} become
\begin{equation}\label{eq:S2ab}\fl
\eqalign{\sum_{m=1}^\infty\normf_m(\kappa_m^2+b^2)^{d-3}&=\left(-\res_{\kappa=0}-\frac{1}{2\pi\rmi}\int_{\gamma_3}\rmd\kappa\right)\frac{(\kappa^2+b^2)^{d-3}N_{\mathcal{C}_1,\mathcal{C}_2}(\kappa)}{R_{\mathcal{C}_1,\mathcal{C}_2}(\kappa)}\cr
&=-\frac{b^{2(d-3)}}{1+\mathcal{C}_1^{-1}+\mathcal{C}_2^{-1}}-2\int_0^\infty\frac{\rmd{\kappa}}{\pi}\,(\kappa^2+b^2)^{d-3}\cr
&\phantom{=\strut}+4\sin(\pi d)\int_b^\infty\frac{\rmd{t}}{\pi}\,(t^2-b^2)^{d-3}\left[\frac{t+\mathcal{C}_1}{t-\mathcal{C}_1}\,\frac{t+\mathcal{C}_2}{t-\mathcal{C}_2}\,\rme^{2t}-1\right]^{-1}.
}
\end{equation}
The integral over $\kappa$  exists when $b>0$ and $d<5/2$. It can be computed. The result
\begin{equation}\label{eq:kappaint}
2\int_0^\infty\frac{\rmd{\kappa}}{\pi}\,(\kappa^2+b^2)^{d-3}=\frac{\Gamma(5/2-d)}{\sqrt{\pi}\,\Gamma(3-d)}\,b^{2d-5}
\end{equation}
provides the desired analytic continuation to values $d>5/2$. For values of $d=4-\epsilon$ close to the upper critical dimension $d^*=4$, both the term $\propto b^{2(d-3)}$ in equation~\eqref{eq:S2ab} and the integral~\eqref{eq:kappaint} approach zero as $b\to 0$. Hence
\begin{equation}
\sum_{m=1}^\infty\normf_m\kappa_m^{2(d-3)}=4\sin(\pi d)\int_b^\infty\frac{\rmd{t}}{\pi}\,t^{2d-6}\left[\frac{t+\mathcal{C}_1}{t-\mathcal{C}_1}\,\frac{t+\mathcal{C}_2}{t-\mathcal{C}_2}\,\rme^{2t}-1\right]^{-1}
\end{equation}
for such values of $d$.

A glance at equations~\eqref{eq:f2loop} and \eqref{eq:delta} shows that the two-loop term of the free-energy density also involves the series
\begin{equation*}
\sum_{m=1}^\infty\normf_m\,\frac{\kappa_m^{2\sigma}\,(\kappa_m^2+b^2)^{(d-3)/2}}{(\kappa_m^2+\mathcal{C}_1^2)(\kappa_m^2+\mathcal{C}_2^2)}
\end{equation*}
with $\sigma=0,1,2$. These can be evaluated in the same manner as the series~\eqref{eq:Sabdef} and \eqref{eq:S2ab}. We therefore simply quote the result one obtains after taking the limit $b\to0$. It reads
\begin{equation}
\sum_{m=1}^\infty\normf_m\,\frac{\kappa_m^{2\sigma+d-3}}{(\kappa_m^2+\mathcal{C}_1^2)(\kappa_m^2+\mathcal{C}_2^2)}=2\,X_{\mathcal{C}_1,\mathcal{C}_2}^{(d,\sigma)}+2\,Y_{\mathcal{C}_1,\mathcal{C}_2}^{(d,\sigma)},
\end{equation}
where $X_{\mathcal{C}_1,\mathcal{C}_2}^{(d,\sigma)}$ and $Y_{\mathcal{C}_1,\mathcal{C}_2}^{(d,\sigma)}$ are the quantities defined by equations~\eqref{eq:Xdef} and \eqref{eq:Ydef}, respectively.

From the above results the expressions for the bulk, surface and residual free-energy densities at $\tb=0$ given in equations~\eqref{eq:fbc}, \eqref{eq:fsc2} and \eqref{eq:fres2}, respectively, follow in a straightforward fashion.

\section{Calculation of the effective two-point vertex $\gamma^{(2)}$}\label{app:gamma2eff}

In this appendix we compute the eigenvalue $k_1^2$ and the shift $\delta\tb_L(\cb_1,\cb_2)$ for arbitrary nonnegative values of $\cb_1$ and $\cb_2$ and derive the result~\eqref{eq:Omegares} for the  scaling function $\Omega(\mathsf{c}_1,\mathsf{c}_2)$. According to equations~\eqref{eq:deltauLc1c2} and \eqref{eq:Heff} the effective two-point vertex $\gamma^{(2)}$ introduced in equation~\eqref{eq:gamma2eff} is approximately given by
\begin{equation}
 k_1^2+\delta\mathring{\tau}_L=L^{-2}\left(\kappa_1^2+\frac{n+2}{3}\frac{\mathring u}{2}L^\epsilon\sum_{m=2}^{\infty}\int_{\bm{p}}^{(d-1)}\frac{\Delta_{1,1,m,m}(\mathcal{C}_1,\mathcal{C}_2)}{p^2+\kappa_m^2}\right),
\end{equation}
where the eigenvalues $\kappa_m^2$ are functions of the  dimensionless bare surface enhancement variables $\mathcal{C}_1$ and $\mathcal{C}_2$. The momentum integration is  straightforward. To compute the series $\sum_{m=2}^\infty$ we substitute expression~\eqref{eq:delta} for $\Delta_{1,1,m,m}$, add and subtract the $(m=0)$-term and then use the summation formula derived in \ref{app:f1ap}. We thus obtain
\begin{eqnarray}\label{eq:gamma2effres}\fl
\lefteqn{L^{2}\left(k_1^2+\delta\mathring{\tau}_L\right)=\kappa_1^2-\frac{n+2}{3}\frac{\mathring{u}}{2}L^\epsilon A_{d-1}\left(\sum_{m=1}^{\infty}\Delta_{1,1,m,m}(\mathcal{C}_1,\mathcal{C}_2)\kappa_m^{d-1}-\Delta_{1,1,1,1}(\mathcal{C}_1,\mathcal{C}_2)\kappa_1^{d-3}\right)}\nonumber\\
&=&\kappa_1^2-\frac{n+2}{3}\frac{\mathring{u}}{8}L^\epsilon A_{d-1}\Bigg(\sum_{\sigma,\rho=0}^2\frac{\normf_1\kappa_1^{2\sigma}}{\left(\mathcal{C}_1^2+\kappa_1^2\right)\left(\mathcal{C}_2^2+\kappa_1^2\right)}P_{\mathcal{C}_1,\mathcal{C}_2}^{(\sigma,\rho)}\left(X_{\mathcal{C}_1,\mathcal{C}_2}^{(d,\rho)}+Y_{\mathcal{C}_1,\mathcal{C}_2}^{(d,\rho)}\right)\nonumber\\
&&{}+\normf_1\kappa_1^{d-3}-4\Delta_{1,1,1,1}(\mathcal{C}_1,\mathcal{C}_2)\kappa_1^{d-3}\Bigg).
\end{eqnarray}

The right-hand side has a logarithmic UV singularity at $\epsilon=0$ (simple pole) originating from the contributions proportional to the functions $X_{\mathcal{C}_1,\mathcal{C}_2}^{(d,1)}$ and  $X_{\mathcal{C}_1,\mathcal{C}_2}^{(d,2)}$. Upon expressing $\ub$ in terms of the renormalized coupling constant $u$, its Laurent expansion becomes
\begin{equation}\fl
 \eqalign{\text{rhs\eqref{eq:gamma2effres}}=&-\frac{n+2}{6}\frac{u}{\epsilon}\sum_{\sigma,\rho=0}^2\frac{\normf_1\kappa_1^{2\sigma}}{\left(\mathcal{C}_1^2+\kappa_1^2\right)\left(\mathcal{C}_2^2+\kappa_1^2\right)}\left[P_{\mathcal{C}_1,\mathcal{C}_2}^{(\sigma,1)}-P_{\mathcal{C}_1,\mathcal{C}_2}^{(\sigma,2)}(\mathcal{C}_1^2+\mathcal{C}_2^2)\right]+\Or(\epsilon^0)\cr
&=-\frac{n+2}{3\epsilon}u\normf_1\frac{(\mathcal{C}_1+\mathcal{C}_2)\kappa_1^2(\mathcal{C}_1\mathcal{C}_2+\kappa_1^2)}{(\mathcal{C}_1^2+\kappa_1^2)(\mathcal{C}_2^2+\kappa_1^2)}+\Or(\epsilon^0).
}
\end{equation}
Its pole term gets cancelled by the counter-term provided by the contribution  $\propto u/\epsilon$ to $\kappa_1^2(\mathcal{C}_1,\mathcal{C}_2)=\kappa_1^2(L\mathring{c}_1,L\mathring{c}_2)=\kappa_1^2(L\mu Z_c c_1,L\mu Z_c c_2)$. To show this we write
\begin{equation}\fl
 \kappa_1^2(L\mu Z_c c_1,L\mu Z_c c_2)=\kappa_1^2(L\mu c_1,L\mu c_2)+2\kappa_1\frac{n+2}{3\epsilon}\sum_{j=1}^2 L\mu c_j\frac{\partial\kappa_1(L \mu c_1,L \mu c_2)}{\partial(L \mu c_j)}
\end{equation}
and calculate the partial derivatives using the implicit function theorem. This gives
\begin{equation}\fl
 \eqalign{\sum_{j=1}^2 \mathcal{C}_j\frac{\partial\kappa_1(\mathcal{C}_1,\mathcal{C}_2)}{\partial\mathcal{C}_j}&=-\left.\sum_{j=1}^2 \mathcal{C}_j\frac{\partial R_{\mathcal{C}_1,\mathcal{C}_2}(\kappa)/\partial \mathcal{C}_j}{\partial R_{\mathcal{C}_1,\mathcal{C}_2}(\kappa)/\partial \kappa}\right|_{\kappa=\kappa_1}\cr
&=\frac{(\mathcal{C}_1+\mathcal{C}_2)\kappa_1\cos\kappa_1+2\mathcal{C}_1\mathcal{C}_2\sin\kappa_1}{(2+\mathcal{C}_1+\mathcal{C}_2)\kappa_1\sin\kappa_1-(\mathcal{C}_1+\mathcal{C}_2+\mathcal{C}_1\mathcal{C}_2-\kappa_1^2)\cos\kappa_1}\cr
&=\frac{(\mathcal{C}_1+\mathcal{C}_2)\kappa_1(\mathcal{C}_1\mathcal{C}_2+\kappa_1^2)}{\mathcal{C}_1\mathcal{C}_2(\mathcal{C}_1+\mathcal{C}_2+\mathcal{C}_1\mathcal{C}_2)+(\mathcal{C}_1+\mathcal{C}_2+\mathcal{C}_1^2+\mathcal{C}_2^2)\kappa_1^2+\kappa_1^4}\cr
&=\frac{\normf_1 (\mathcal{C}_1+\mathcal{C}_2)\kappa_1(\mathcal{C}_1\mathcal{C}_2+\kappa_1^2)}{2(\mathcal{C}_1^2+\kappa_1^2)(\mathcal{C}_2^2+\kappa_1^2)}.
}
\end{equation}
Hence to the order of our calculation, $k_1^2+\delta\mathring{\tau}_L$ is UV-finite when expressed in terms of renormalized variables.  From the implied UV-finite part of \eqref{eq:gamma2effres} the result for the scaling function $\Omega(\mathsf{c}_1,\mathsf{c}_2)$ given in equation~\eqref{eq:Omegares} follows in a straightforward fashion.

\section{Behaviour of the shift $\delta\tb_L(\mathcal{C}_1,\mathcal{C}_2)$ for small $\mathcal{C}_j$}\label{app:shiftprime}

In this appendix we  calculate the shift $\delta\tb_L(\mathcal{C}_1,\mathcal{C}_2)$ to linear order in $\mathcal{C}_1$ and $\mathcal{C}_2$. We start from equation~\eqref{eq:deltauLc1c2}, use equation~\eqref{eq:delta} for $\Delta_{1,1,m,m}$, substitute the small-$\mathcal{C}_j$ expressions~\eqref{eq:smallCkappa} for $\kappa^2_1$ and $\kappa_{m\ge 2}^2$ and expand in $\mathcal{C}_j$. This gives
\begin{equation}\fl
\eqalign{\frac{\Delta_{1,1,m,m}(\mathcal{C}_1,\mathcal{C}_2)}{p^2+\kappa_m^2(\mathcal{C}_1,\mathcal{C}_2)}=&
\frac{1}{p^2+(m-1)^2\pi^2}-(\mathcal{C}_1+\mathcal{C}_2)\Bigg\{\frac{1}{2 (m-1)^2\pi^2}\,\frac{1}{ p^2+(m-1)^2\pi^2}\cr &\strut+\frac{2}{\left[p^2+(m-1)^2\pi^2\right]^2}+\Or(\mathcal{C}_1,\mathcal{C}_2)\Bigg\}.}
\end{equation}
To determine the shift from these results, we interchange the integration over $\bm{p}$ in equation~\eqref{eq:delta} with the summation over $m$. The  required momentum integrals are of the form
\begin{equation}
\int_{\bm{p}}^{(d-1)}\frac{1}{{[p^2+\kappa^2]}^s}=
-A_{d-1}\cases{\kappa^{d-3}&for $s=1$,\cr
\frac{3-d}{2}\,\kappa^{d-5}&for $s=2$.}
\end{equation}
Upon summing over $m$ we encounter the series
\begin{equation}
\sum_{m=1}^\infty (m\pi)^{d-3}=\pi^{d-3}\,\zeta(3-d)=\frac{2^{3-d}\,\Gamma(d-2)\,\zeta(d-2)}{\Gamma(3/2-d/2)\,\Gamma(d/2-1/2)}
\end{equation}
and
\begin{equation}
\sum_{m=1}^\infty (m\pi)^{d-5}=\pi^{d-5}\,\zeta(5-d).
\end{equation}
From the above results, equations~\eqref{eq:deltauLsp} and \eqref{eq:shiftprime} follow in a straightforward manner.

\section{Calculation of $f_\psi-f$ to two-loop order}\label{app:fpsidiff}

In this appendix we calculate the the difference $f_\psi-f$ of free-energy densities to two-loop order. Its one-loop contribution is given in equation~\eqref{eq:fpsi1l}. To determine its two-loop term, we must compute the two graphs  appearing inside square brackets on the right-hand side of equation~\eqref{eq:fdiff2l}. The rightmost one of these (red) can be read off from equations~\eqref{eq:fvarphi} and \eqref{eq:fvarphi1} by setting $\delta\mathring{\tau}_L=0$. Upon adding the contribution from the remaining graph  \raisebox{-0.75em}{\includegraphics[width=1.2em]{./figure10}} in square brackets, we arrive at
\begin{eqnarray}\fl
L^{d-1}(f_{\psi}^{[2]}-f^{[2]})\nonumber\\ \fl =\frac{\mathring{u}n(n+2)}{4!}L^\epsilon A_{d-1}^2 \Bigg\{\kappa_1^{2(d-3)}\Delta_{1,1,1,1}(\mathcal{C}_1,\mathcal{C}_2)-2\kappa_1^{d-3}\sum_{m=1}^{\infty}\kappa_m^{d-3}\Delta_{1,1,m,m}(\mathcal{C}_1,\mathcal{C}_2)\Bigg\}
 \nonumber\\ \fl
=\frac{\mathring{u}{n(n+2)}}{4!}L^\epsilon A_{d-1}^2\frac{\kappa_1^{d-3}}{2}\Bigg\{\kappa_1^{d-3}\left[2\Delta_{1,1,1,1}(\mathcal{C}_1,\mathcal{C}_2)-\normf_1\right]\nonumber\\ \fl
\phantom{=}\strut -
\sum_{\sigma,\rho=0}^2\frac{\normf_1\kappa_1^{2\sigma}}{(\mathcal{C}_1^2+\kappa_1^2)(\mathcal{C}_2^2+\kappa_1^2)}P_{\mathcal{C}_1,\mathcal{C}_2}^{(\sigma,\rho)}\left(X_{\mathcal{C}_1,\mathcal{C}_2}^{(d,\rho)}+Y_{\mathcal{C}_1,\mathcal{C}_2}^{(d,\rho)}\right)\Bigg\}\label{eq:fpsi2ldiff}
\end{eqnarray}
where the definition of $\mathcal{C}_j$ should be recalled from equation~\eqref{eq:dimvardef}.
The result contains pole terms; its Laurent expansion reads
\begin{equation}\fl
 \eqalign{\text{rhs\eqref{eq:fpsi2ldiff}} &=\strut-n\frac{n+2}{48\pi}\frac{u}{\epsilon}\kappa_1\sum_{\rho=0}^2\frac{\normf_1\kappa_1^{2\sigma}}{\left(\mathcal{C}_1^2+\kappa_1^2\right)\left(\mathcal{C}_2^2+\kappa_1^2\right)}\left[P_{\mathcal{C}_1,\mathcal{C}_2}^{(\sigma,1)}-P_{\mathcal{C}_1,\mathcal{C}_2}^{(\sigma,2)}(\mathcal{C}_1^2+\mathcal{C}_2^2)\right]+\Or(\epsilon^0)\cr
&=\strut-n\frac{n+2}{24\pi}\frac{u}{\epsilon}\normf_1\frac{(\mathcal{C}_1+\mathcal{C}_2)\kappa_1^3(\mathcal{C}_1\mathcal{C}_2+\kappa_1^2)}{(\mathcal{C}_1^2+\kappa_1^2)(\mathcal{C}_2^2+\kappa_1^2)}+\Or(\epsilon^0).}
\end{equation} 
The pole terms are cancelled by those provided by the  one-loop term  $f_\psi^{[1]}-f$ upon expressing the bare variables $\cb_j$ in terms of their renormalized analogues $c_j$  via equations~\eqref{eq:surfrep} and \eqref{eq:Z1orderu}. Upon adding the one-loop term one arrives at the two-loop result for $f_{\psi,\text{res},\text{R}}\,L^{d-1}$ given in \eqref{eq:fpsiresdiff} in a straightforward fashion. 

\ack
Partial support by Deutsche Forschungsgemeinschaft under Grant No.\ Di-378/5 is gratefully acknowledged.


\section*{References}
\begin{thebibliography}{10}


\bibitem{BMM01}
For a review of the Casimir effect in QED and an extensive list of references, see\\
M.~Bordag, U.~Mohideen, and V.~M. Mostepanenko.
\newblock New developments in the {C}asimir effect.
\newblock {\em Phys. Rep.}, 353:1--205, 2001.

\bibitem{Kre94}
M.~Krech.
\newblock {\em {C}asimir Effect in Critical Systems}.
\newblock World Scientific, Singapore, 1994.

\bibitem{KG99}
M.~Kardar and R.~Golestanian.
\newblock The ``friction'' of vacuum, and other fluctuation-induced forces.
\newblock {\em Rev. Mod. Phys.}, 71(4):1233--1245, 1999.

\bibitem{Cas48}
H.~B.~G. Casimir.
\newblock On the attraction between two perfectly conducting plates.
\newblock {\em Proc. K. Ned. Akad. Wet.}, B51:793--795, 1948.

\bibitem{SWM95}
F.~M. Serry, D.~Walliser, and G.~J. Maclay.
\newblock The anharmonic {C}asimir oscillator (aco) - {T}he {C}asimir effect in a
  model microelectromechanical system.
\newblock {\em Journal of Microelectromechanical Systems}, 4(4):193 --205, 1995.

\bibitem{BR01}
E.~Buks and M.~L. Roukes.
\newblock Stiction, adhesion energy, and the {C}asimir effect in
  micromechanical systems.
\newblock {\em Phys. Rev. B}, 63:033402, 2001.

\bibitem{BR01b}
E.~Buks and M.~L. Roukes.
\newblock Metastability and the {C}asimir effect in micromechanical systems.
\newblock {\em Europhys. Lett.}, 54(2):220, 2001.

\bibitem{CAKBC01}
H.~B. Chan, V.~A. Aksyuk, R.~N. Kleiman, D.~J. Bishop, and F.~Capasso.
\newblock Quantum mechanical actuation of microelectromechanical systems by the
  {C}asimir force.
\newblock {\em Science}, 291:1941--1943, March 2001.
\newblock Erratum: Science \textbf{27} 2001: Vol. 293. no. 5530, p. 607.

\bibitem{Emi10}
T.~Emig.
\newblock Casimir physics: Geometry, shape and material.
\newblock {\em Int. J. Mod. Phys. A}, 25(11):2177--2195, 2010.

\bibitem{Kre99}
M.~Krech.
\newblock Fluctuation-induced forces in critical fluids.
\newblock {\em J. Phys.: Condens. Matter}, 11(37):R391, 1999.

\bibitem{BDT00}
J.~G. Brankov, D.~M. Dantchev, and N.~S. Tonchev.
\newblock {\em Theory of Critical Phenomena in Finite-Size Systems --- Scaling
  and Quantum Effects}.
\newblock World Scientific, Singapore, 2000.

\bibitem{FdG78}
M.~E. Fisher and P.-G. de~Gennes.
\newblock Ph{\'e}nom{\`e}nes aux parois dans un m{\'e}lange binaire critique.
\newblock {{\em C.\ R.\ Acad.\ Sci.} B}, 287:207--209,
  1978.

\bibitem{GC99}
R.~Garcia and M.~H.~W. Chan.
\newblock Critical fluctuation-induced thinning of $^4${He} films near the
  superfluid transition.
\newblock {\em Phys. Rev. Lett.}, 83:1187, 1999.

\bibitem{FYP05}
M.~Fukuto, Y.~F. Yano, and P.~S. Pershan.
\newblock Critical {C}asimir effect in three-dimensional {I}sing systems:
  measurements on binary wetting films.
\newblock {\em Phys. Rev. Lett.}, 94:135702--1--4, 2005.

\bibitem{RBM07}
S.~Rafai, D.~Bonn, and J.~Meunier.
\newblock Repulsive and attractive critical {C}asimir forces.
\newblock {\em Physica A}, 386(1):31--35, 2007.

\bibitem{HHGDB08}
C.~Hertlein, L.~Helden, A.~Gambassi, S.~Dietrich, and C.~Bechinger.
\newblock Direct measurement of critical {C}asimir forces.
\newblock {\em Nature}, 451:172--175, 2008.

\bibitem{Gam09}
A.~Gambassi.
\newblock The {C}asimir effect: From quantum to critical fluctuations.
\newblock {\em J. Phys.: Conference Series}, 161(1):012037, 2009.

\bibitem{NI85}
M.~P. Nightingale and J.~O. Indekeu.
\newblock Effect of criticality on wetting layers.
\newblock {\em Phys. Rev. Lett.}, 54(16):1824--1827, 1985.

\bibitem{KD91}
M.~Krech and S.~Dietrich.
\newblock Finite-size scaling for critical films.
\newblock {\em Phys. Rev. Lett.}, 66:345--348, 1991.
\newblock [Erratum {\bf 67}, 1055 (1991)].

\bibitem{KD92a}
M.~Krech and S.~Dietrich.
\newblock Free energy and specific heat of critical films and surfaces.
\newblock {\em Phys. Rev. A}, 46(4):1886--1921, 1992.

\bibitem{SD08}
F.~M. Schmidt and H.~W. Diehl.
\newblock Crossover from attractive to repulsive {C}asimir forces and vice
  versa.
\newblock {\em Phys. Rev. Lett.}, 101(10):100601, 2008.

\bibitem{Die86a}
H.~W. Diehl.
\newblock Field--theoretical approach to critical behaviour at surfaces.
\newblock In C.~Domb and J.~L. Lebowitz, editors, {\em Phase Transitions and
  Critical Phenomena}, volume~10, pages 75--267. Academic, London, 1986.

\bibitem{Die97}
H.~W. Diehl.
\newblock The theory of boundary critical phenomena.
\newblock {\em Int.\ J.\ Mod.\ Phys.\ B}, 11:3503--3523, 1997.
\newblock cond-mat/9610143.

\bibitem{Die94a}
H.~W. Diehl.
\newblock Critical adsorption of fluids and the equivalence of extraordinary
  and normal surface transitions.
\newblock {\em Ber.\ Bunsenges.\ Phys.\ Chem.}, 98:466--471, 1994.

\bibitem{Kre97}
M.~Krech.
\newblock {C}asimir forces in binary liquid mixtures.
\newblock {\em Phys. Rev. E}, 56(2):1642--1659, 1997.

\bibitem{Huc07}
A.~Hucht.
\newblock Thermodynamic {C}asimir effect in $^4${He} films near ${T}_\lambda$:
  {M}onte {C}arlo results.
\newblock {\em Phys. Rev. Lett.}, 99(18):185301, 2007.

\bibitem{VGMD07}
O.~Vasilyev, A.~Gambassi, A.~Macio{\l}ek, and S.~Dietrich.
\newblock {M}onte {C}arlo simulation results for critical {C}asimir forces.
\newblock {\em Europhys. Lett.}, 80(6):60009 (6pp), 2007.

\bibitem{VGMD09}
O.~Vasilyev, A.~Gambassi, A.~Macio\l{}ek, and S.~Dietrich.
\newblock Universal scaling functions of critical {C}asimir forces obtained by
  {M}onte {C}arlo simulations.
\newblock {\em Phys. Rev. E}, 79(4):041142, 2009.

\bibitem{GMHNB09}
A.~Gambassi, A.~Macio\l{}ek, C.~Hertlein, U.~Nellen, L.~Helden, C.~Bechinger,
  and S.~Dietrich.
\newblock Critical {C}asimir effect in classical binary liquid mixtures.
\newblock {\em Phys. Rev. E}, 80(6):061143, 2009.

\bibitem{Has10b}
M.~Hasenbusch.
\newblock Thermodynamic {C}asimir effect for films in the {3D} {I}sing
  universality class: Symmetry breaking boundary conditions, 2010.
\newblock {\em Phys. Rev. B}, 82(10):104425, 2010.

\bibitem{MMD10}
T.~F. Mohry, A.~Macio\l{}ek, and S.~Dietrich.
\newblock Crossover of critical {C}asimir forces between different surface
  universality classes.
\newblock {\em Phys. Rev. E}, 81(6):061117, 2010.

\bibitem{Doh08}
V.~Dohm.
\newblock Diversity of critical behavior within a universality class.
\newblock {\em Phys. Rev. E}, 77(6):061128, 2008.

\bibitem{DC09}
H.~W. Diehl and H.~Chamati.
\newblock Dynamic critical behavior of model ${A}$ in films: Zero-mode boundary
  conditions and expansion near four dimensions.
\newblock {\em Phys. Rev. B}, 79(10):104301, 2009.

\bibitem{DD81a}
H.~W. Diehl and S.~Dietrich.
\newblock Field-theoretical approach to static critical phenomena in
  semi-infinite systems.
\newblock {\em Z.\ Phys.\ B: Condens. Matter}, 42:65--86, 1981.
\newblock Erratum: {\bf 43}, 281 (1981).

\bibitem{DD81b}
H.~W. Diehl and S.~Dietrich.
\newblock Field-theoretical approach to multicritical behavior near free
  surfaces.
\newblock {\em Phys. Rev. B}, 24(5):2878--2880, 1981.

\bibitem{DD83a}
H.~W. Diehl and S.~Dietrich.
\newblock Multicritical behaviour at surfaces.
\newblock {\em Z.\ Phys.\ B: Condens. Matter}, 50:117--129, 1983.

\bibitem{RS02}
A.~Romeo and A.~A. Saharian.
\newblock {C}asimir effect for scalar fields under {R}obin boundary conditions
  on plates.
\newblock {\em J. Phys. A}, 35:1297--1320, 2002.

\bibitem{CH68}
R.~Courant and D.~Hilbert.
\newblock {\em Methoden der mathematischen {Physik I}}, volume~30 of {\em
  {Heidelberger} {Taschenb\"ucher}}.
\newblock Springer-Verlag, Berlin, 3nd edition, 1968.

\bibitem{Jae01}
K.~J{\"a}nich.
\newblock {\em Analysis f{\"u}r Physiker und Ingenieure}.
\newblock Springer-Lehrbuch. Springer, Berlin, Heidelberg, 2001.

\bibitem{Pey70b}
A.~Peyerimhoff.
\newblock {\em Gew{\"o}hnliche Differentialgleichungen II}.
\newblock Studien-text. Akademische Verlagsgesellschaft, Frankfurt am Main,
  1970.

\bibitem{CC97}
E.~A. Coddington and R.~Carlson.
\newblock {\em Linear ordinary differential equations}.
\newblock Siam, Society for Industrial and Applied Mathematics, Philadelphia,
  USA, 1997.

\bibitem{Sch08}
F.~M. Schmidt.
\newblock {Kritischer Casimir Effekt bei Robin-Randbedingungen}.
\newblock Master's thesis, Fachbereich Physik, Universit{\"a}t Duisburg-Essen,
  Duisburg, 2008.

\bibitem{GD08}
D.~Gr{\"u}neberg and H.~W. Diehl.
\newblock Thermodynamic {C}asimir effects involving interacting field theories
  with zero modes.
\newblock {\em Phys. Rev. B}, 77(11):115409, 2008.
\newblock {arXiv:0710.4436}.

\bibitem{DD80}
H.~W. Diehl and S.~Dietrich.
\newblock Scaling laws and surface exponents from renormalization--group
  equations.
\newblock {\em Phys.\ Lett.}, 80A:408--412, 1980.

\bibitem{Die82}
H.~W. Diehl.
\newblock Critical behavior of semi--infinite magnets (invited).
\newblock {\em J.\ Appl.\ Phys.}, 53:7914--7919, 1982.

\bibitem{DGS06}
H.~W. Diehl, Daniel Gr{\"u}neberg, and M.~A. Shpot.
\newblock Fluctuation-induced forces in periodic slabs: Breakdown of $\epsilon$
  expansion at the bulk critical point and revised field theory.
\newblock {\em Europhys. Lett.}, 75:241--247, 2006.
\newblock cond-mat/0605293.

\bibitem{DG09}
H.~W. Diehl and Daniel Gr{\"u}neberg.
\newblock Critical {C}asimir amplitudes for $n$-component $\phi^4$ models with
  ${O}(n)$-symmetry breaking quadratic boundary terms.
\newblock {\em Nucl. Phys. B [FS]}, 822:517--542, 2009.

\bibitem{Sac97}
S.~Sachdev.
\newblock Theory of finite-temperature crossovers near quantum critical points
  close to, or above, their upper-critical dimension.
\newblock {\em Phys. Rev. B}, 55(1):142--163, 1997.

\bibitem{KD10}
B.~Kastening and V.~Dohm.
\newblock Finite-size effects in film geometry with nonperiodic boundary
  conditions: Gaussian model and renormalization-group theory at fixed
  dimension.
\newblock {\em Phys. Rev. E}, 81(6):061106, 2010.

\bibitem{Lev59}
A.~P. Levanyuk.
\newblock Contribution to the theory of light scattering near the second-order
  phase transition points.
\newblock {\em Sov. Phys.-JETP}, 9:571, 1959.
\newblock [Zh. Eksp. Teor. Fiz. \textbf{36}, 810 (1959)].

\bibitem{Gin60}
V.~L. Ginzburg.
\newblock Some remarks on phase transitions of the 2nd kind and the microscopic
  theory of ferroelectric materials.
\newblock {\em Fiz. Tverd. Tela}, 2:2031, 1960.
\newblock [Sov. Phys.-Solid St. \textbf{2}, 1824 (1961)].

\bibitem{Has11}
M.~Hasenbusch.
\newblock A {M}onte {C}arlo study of surface critical phenomena: The special
  point, 2011.
\newblock {\em Phys. Rev. B}, 84:134405, 2011.

\bibitem{BL84}
K.~Binder and D.~P. Landau.
\newblock Crossover scaling and critical behavior at the "surface-bulk"
  multicritical point.
\newblock {\em Phys. Rev. Lett.}, 52:318--321, 1984.

\bibitem{LB90a}
D.~P. Landau and K.~Binder.
\newblock {M}onte {C}arlo study of surface phase transitions in the
  three-dimensional {I}sing model.
\newblock {\em Phys.\ Rev.\ B}, 41:4633--4645, 1990.

\bibitem{VRF92}
M.~Vendruscolo, M.~Rovere, and A.~Fasolino.
\newblock Magnetic-phase transitions of {I}sing surfaceswith modified
  surface-bulk coupling: a {M}onte {C}arlo study.
\newblock {\em Europhys. Lett.}, 20(6):547, 1992.

\bibitem{RDW92}
C.~Ruge, S.~Dunkelmann, and F.\ Wagner.
\newblock New method for determination of critical parameters.
\newblock {\em Phys.\ Rev.\ Lett.}, 69:2465--2468, 1992.

\bibitem{RDWW93}
C.~Ruge, S.~Dunkelmann, F.\ Wagner, and J.~Wulf.
\newblock Study of the three-dimensional {I}sing model on film geometry with
  the cluster {M}onte {C}arlo method.
\newblock {\em J.\ Stat.\ Phys.}, 73:293--317, 1993.

\bibitem{PS98}
M.~Pleimling and W.~Selke.
\newblock Critical phenomena at perfect and non-perfect surfaces.
\newblock {\em Eur.\ Phys.\ J.\ B}, 1:385--391, 1998.

\bibitem{DBN05}
Y.~Deng, H.~W.~J. Bl{\"o}te, and M.~P. Nightingale.
\newblock Surface and bulk transitions in three-dimensional o(n) models.
\newblock {\em Phys. Rev. E}, 72(1):016128, 2005.

\bibitem{DS94}
H.~W. Diehl and M.~Shpot.
\newblock Surface critical behavior in fixed dimensions $d<4$: {N}onanalyticity
  of critical surface enhancement and massive field theory approach.
\newblock {\em Phys. Rev. Lett.}, 73(25):3431--3434, 1994.

\bibitem{DS98}
H.~W. Diehl and M.~Shpot.
\newblock Massive field-theory approach to surface critical behavior in
  three-dimensional systems.
\newblock {\em Nucl. Phys. B}, 528(3):595--647, 1998.
\newblock cond-mat/9804083.

\bibitem{Bac06}
C.~P. Bachas.
\newblock Comment on the sign of the {C}asimir force.
\newblock {\em J. Phys. A: Math. Theor.}, 40(30): 9089--9096, 2006.

\bibitem{Par80}
G.~Parisi.
\newblock Field-theoretic approach to second-order phase transitions in two-
  and three-dimensional systems.
\newblock {\em J. Stat. Phys.}, 23(1):49--81, 1980.

\bibitem{ZJ02}
J.~Zinn-Justin.
\newblock {\em Quantum Field Theory and Critical Phenomena}.
\newblock International series of monographs on physics. Oxford University
  Press, Oxford, 4th edition, 2002.

\bibitem{SD89}
R.~Schloms and V.~Dohm.
\newblock Minimal renormalization without $\epsilon$-expansion: critical
  behavior in three dimensions.
\newblock {\em Nucl. Phys. B}, 328:639--663, 1989.

\bibitem{DDE83}
H.~W. Diehl, S.~Dietrich, and E.~Eisenriegler.
\newblock Universality, irrelevant surface operators, and corrections to
  scaling in systems with free surfaces and defect planes.
\newblock {\em Phys. Rev. B}, 27(5):2937--2954, 1983.

\bibitem{NHB09}
U.~Nellen, L.~Helden, and C.~Bechinger.
\newblock Tunability of critical {C}asimir interactions by boundary conditions.
\newblock {\em Europhys. Lett.}, 88(2):26001, 2009.

\bibitem{GZJ98}
R.~Guida and J.~Zinn-Justin.
\newblock Critical exponents of the {$N$}-vector model.
\newblock {\em J. Phys. A}, 31(40):8103--8121, 1998.

\bibitem{PV02}
A.~Pelissetto and E.~Vicari.
\newblock Critical phenomena and renormalization-group theory.
\newblock {\em Phys. Rep.}, 368:549--727, 2002.

\bibitem{KS-F01}
H.~Kleinert and V.~Schulte-Frohlinde.
\newblock {\em Critical properties of $\phi^4$-theories}.
\newblock World Scientific, Singapore, 2001.

\bibitem{Sym73}
K.~Symanzik.
\newblock Massless $\phi^4$ theory in $4-\epsilon$ dimensions.
\newblock {\em Lett. Nuovo Cimento}, 8:771--774, 1973.

\bibitem{Sym73b}
K.~Symanzik.
\newblock Massless $\phi^4$ theory in $4-\epsilon$ dimensions.
\newblock technical report 73/58, DESY, Hamburg, 1973.

\bibitem{BBZ-J00}
G.~Baym, J.-P. Blaizot, and J.~Zinn-Justin.
\newblock The transition temperature of the dilute interacting {B}ose gas for
  $n$ internal states.
\newblock {\em Europhys. Lett.}, 49(2):150--155, 2000.

\bibitem{DE82}
H.~W. Diehl and E.~Eisenriegler.
\newblock Irrelevance of surface anisotropies for critical behavior near free
  surface.
\newblock {\em Phys. Rev. Lett.}, 48(25):1767, 1982.

\bibitem{DE84}
H.~W. Diehl and E.~Eisenriegler.
\newblock Effects of surface exchange anisotropies on magnetic critical and
  multicritical behavior at surfaces.
\newblock {\em Phys. Rev. B}, 30(1):300--314, 1984.

\bibitem{Eli95}
E.~Elizalde.
\newblock {\em Ten applications of spectral zeta functions}, volume m35 of {\em
  Lecture Notes in Physics}.
\newblock Springer, Berlin-Heidelberg, Germany, 1995.

\bibitem{Mathematica7}
Wolfram Research, Computer code {\sc Mathematica}, version 7.

\end{thebibliography}

\end{document}